\documentclass[prd,amssymb,amsmath,showpacs]{revtex4}

\allowdisplaybreaks

\begin{document}

\title{Light deflection in the post-linear gravitational field of bounded point-like masses}

\author{Michael H. Br\"ugmann}
\affiliation{Theoretisch-Physikalisches Institut, Friedrich-Schiller-Universit\"at Jena,
                Max-Wien-Platz 1, 07743 Jena, Germany}
\date{\today}

\begin{abstract}
Light deflection in the post-linear gravitational field of two bounded point-like masses is treated.  
Both the light source and the observer are assumed to be located at infinity in an asymptotically flat space. 
The equations of light propagation are explicitly integrated to the second order in $G/c^2$. 
Some of the integrals are evaluated by making use of an expansion in powers of the ratio of the relative 
separation distance to the impact parameter $(r_{12}/\xi)$. A discussion of which orders must be retained to be 
consistent with the expansion in terms of $G/c^2$ is given.  
It is shown that the expression obtained in this paper for the angle of light deflection
is fully equivalent to the expression obtained by Kopeikin and Sch\"afer up to the order given there.
The deflection angle takes a particularly simple form for a light ray originally propagating
orthogonal to the orbital plane of a binary with equal masses.
Application of the formulae for the deflection angle to the double pulsar PSR J0737-3039 
for an impact parameter five times greater than the relative separation distance of the binary's components
shows that the corrections to the Epstein-Shapiro light deflection angle of about $10^{-6}$ arcsec
lie between $10^{-7}$ and $10^{-8}$ arcsec.
\end{abstract}


\maketitle

\section{Introduction}\label{introduction}
Light deflection by a gravitational field \cite{EIN1916} is one of the observational cornerstones 
of general relativity. The observational confirmation
of Einstein's prediction that light would be deflected by the
gravitational field of the Sun \cite{DYED1920} brought general relativity to the attention of 
the general public in the 1920's.
Today, technology has reached a level at which the extremely high precision
of current ground-based radio interferometric astronomical observations approaches 1
$\mu$arcsec and within the next decade the accuracy of space-based astrometric positional
observations is also expected to reach this accuracy.
At this level of accuracy we can no longer treat the gravitational field of 
a system of moving bodies as static and spherically symmetric.
This fact is one of the principal reasons for the necessity of a more accurate solution
to the problem of the propagation of electromagnetic waves in non-stationary gravitational fields
of celestial bodies. 
To reach the accuracy of 1 $\mu$arcsec, many subtle relativistic effects must be taken into account in 
the treatment of light propagation in non-stationary gravitational fields of moving bodies.
One of the most intricate problems is the computation of the effects of translational
motion of the gravitating bodies on light propagation.

This question was treated for the first time by Hellings in 1986 \cite{HELL86}. In 1989 Klioner \cite{KLI03} solved
the problem completely to the first post-Newtonian (1PN) order (i.e. to the order $1/c^2$) for the case of bodies
moving with constant velocity. The complete solution of the problem for arbitrarily 
moving bodies in the first post-Minkowskian approximation (linear in the gravitational constant G) was found by 
Kopeikin and Sch\"afer in 1999 \cite{KSCH99}. They succeeded in integrating analytically the 
post-Minkowskian equations of light propagation in the field of arbitrarily moving masses.
In Ref. \cite{BLPT04}, Le Poncin-Lafitte et al. have recently developed 
an alternative approach to the problem of light deflection and time/frequency transfer 
in post-Minkowskian gravitational fields based on an expansion of the Synge world function for null geodesics. 
In that paper the world function and time transfer function were computed for a static spherically 
symmetric body to the second post-Minkowskian approximation.

In this paper we treat light deflection in the post-linear gravitational field 
of two bounded point-like masses (binary system).
We assume that the light source as well the observer are located at infinity in an asymptotically flat space.
To compute the light deflection we integrate the equations of light propagation explicitly to the second
order in $G/c^2$, i.e. to the order $G^2/c^4$.

The assumption that the gravitational field is weak along the light path allows us to consider the metric as a perturbation
of a flat metric represented by a power series in the gravitational constant $G$
\begin{align}
g_{\mu \nu}[x^{\sigma},G] \equiv g^{(0)}_{\mu \nu}+ \sum^{\infty}_{n=1} G^n g^{(n)}_{\mu \nu}(x^{\sigma})\label{Metric0},
\end{align}
with
\begin{align*}
g^{(0)}_{\mu \nu}= \eta_{\mu \nu} = \mathrm{diag}(-1,1,1,1).
\end{align*}
For the same reason we can consider the light trajectory as a perturbation of its trajectory in flat space (a straight line)
represented by a power series in $G$
\begin{align}
\vec{z}(t)=\vec{z}_{(0)}(t)+\sum^{\infty}_{n=1} G^n \delta \vec{\tilde{z}}_{(n)}(t)\label{Trajec0}.
\end{align}
It follows from Eq.~\eqref{Trajec0} that the vector tangent to the light trajectory takes the form
\begin{align}
\vec{l}(t)\equiv \frac{d\vec{z}(t)}{dt}=\vec{l}_{(0)}+\sum^{\infty}_{n=1} G^n \delta \vec{\tilde{l}}_{(n)}(t)\label{Tangent0},
\end{align}
where $\vec{l}_{(0)}$ is the constant vector tangent to the unperturbed light trajectory.

In order to obtain the post-linear equations for light propagation we introduce into the differential
equations for the null geodesics the metric as given in \eqref{Metric0} and the expression for the tangent vector
$\vec{l}(t)$ as given in \eqref{Tangent0}. 
As a result we get a set of ordinary coupled differential equations of first order for the perturbation
terms $\delta \vec{l}_{(n)}(t)$.
Each term $\delta \vec{l}_{(n)}$ is given in the form of a line integral along a straight line in the fictitious
metric $g^{(0)}_{\mu \nu}$, i.e. along the original unperturbed light trajectory.
We get the post-linear light deflection to the order $G^2/c^4$ after computing the perturbation terms
$\delta \vec{l}_{(1)}(t)$, $\delta \vec{l}_{(2)}(t)$ and the corrections arising from introducing the linear
perturbation of the light trajectory $\delta \vec{z}_{(1)}(t)$, the motion of the masses and the shift 
of the 1PN-centre of mass with respect to the Newtonian centre of mass in the expression for linear
light deflection.
The final result we obtain is the expression for light deflection in the post-linear gravitational
field of two bounded point-like masses.
The deflection angle takes a particularly simple form for a light ray originally propagating orthogonal to
the orbital plane of a binary with equal masses.

This paper is organized as follows.
In Section \ref{light} we derive the post-linear light propagation equations. We set out an approximation scheme
to integrate these equations. The deflection angle as a function of the perturbations of the vector tangent
to the light ray is introduced.
In Section \ref{gravfield} the post-linear and linear metric for two bounded point-like masses in harmonic coordinates are given.
The coordinate frame is chosen so that the 1PN-centre of mass is at rest at the origin.
In Section \ref{lightlingrav} we compute the perturbation of the vector tangent to the unperturbed light ray
and the corresponding light deflection in the linear gravitational field. 
In Section \ref{lightpostlingrav} we compute the light deflection in the post-linear gravitational field.
To facilitate the computations we separate the light deflection terms which are functions of the post-linear 
metric coefficients from the terms that are functions of the linear metric coefficients and the perturbations
of the first order in $G$ of the vector tangent to the unperturbed light ray.
The resulting integrals are given in Appendices \ref{ap:A} and \ref{ap:B}.
In Section \ref{lightmotion} we calculate the additional linear and post-linear light deflection terms arising from 
the introduction of the motion of the masses into the expression for the linear perturbation. 
In Section \ref{trajectory} we compute the corrections to the post-linear light deflection arising from the introduction
of the linear perturbed light trajectory into the expression for the linear light deflection.
The resulting integrals are given in Appendix \ref{ap:C}.
Section \ref{centre} is devoted to the computation of the corrections to the linear and post-linear light deflection arising from
the introduction of the shift of the 1PN-centre of mass with respect to the Newtonian centre of mass into
the expression for the linear light deflection.
The resulting expressions for the total linear perturbation, linear and post-linear light deflection by
the gravitational field of two bounded point-like masses are given in an explicit form in Secs \ref{totlightlin} and \ref{totlightpostlin}.
In Section \ref{results} we present our results and give in an explicit form the light deflection expression for some simple
cases. The derived formulae for the angle of light deflection are applied to the double pulsar PSR J0737-3039. 
In Appendix \ref{ap:D} we compute the linear light deflection terms arising from the acceleration terms in the 
metric coefficients $h^{(1)}_{00}$ and $h^{(1)}_{pq}$.
In Appendix \ref{ap:E} we show that the expression for the linear light deflection given in this paper is fully equivalent
to the expression for the light deflection computed by Kopeikin and Sch\"afer \cite{KSCH99}
in the event that the velocities of the masses are small with respect to the velocity of light 
and the retarded times in the expression given by Kopeikin and Sch\"afer are close to the time of closest approach 
of the unperturbed light ray to the origin of the coordinate system.
Finally, Section \ref{discussion} is devoted to the discussion of the results.
\subsubsection*{Notation}
Let us summarize the notation and symbols used in this paper: 
\begin{enumerate}
\item $G$ is the Newtonian constant of gravitation;
\item $c$ is the velocity of light;
\item the Greek indices $\alpha$, $\beta$, $\gamma$..., are space-time indices and run from 0 to 3;
\item the Latin indices $i$, $j$, $k$,..., are spatial indices and run from 1 to 3;
\item $g_{\mu \nu}$ is a metric tensor of curved, four-dimensional space-time, depending on spatial coordinates and
time;
\item the signature adopted for $g_{\mu \nu}$ is $(-+++)$;
\item we suppose that space-time is covered by a harmonic coordinate system system $(x^{\mu})=(x^0,x^i)$, where
$x^0=c\,t$, $t$ being the time coordinate;
\item the three-dimensional quantities (3-vectors) are denoted by $\vec{a}=a^i$;
\item the three-dimensional unit vector in the direction of $\vec{a}$ is denoted by $\vec{e}_a=e^i_a$;
\item the Latin indices are lowered and raised by means of the 
unit matrix $\delta_{ij}$ = $\delta^{ij} = \mathrm{diag}(1,1,1)$;
\item by $,_{\sigma}$ we denote the partial derivative with respect to the coordinate $x^{\sigma}$;
\item the scalar product of any two 3-vectors $\vec{a}$ and $\vec{b}$ with respect to the Euclidean metric
$\delta_{ij}$ is denoted by $\vec{a} \cdot \vec{b}$ and can be computed as
$\vec{a} \cdot \vec{b}=\delta_{ij} a^i b^j=a^ib^i$;
\item the Euclidean norm of a 3-vector $\vec{a}$ is denoted by $a=|\vec{a}|$ and can be computed as
$a=\Big[\delta_{mn}a^ma^n\Big]^{1/2}$;
\item by $\vec{l}_{(0)}$ we denote the vector tangent to the unperturbed light ray $\vec{z}(t)$ and the unit vector
$\vec{e}_{(0)}$ is defined by $\vec{e}_{(0)}=\vec{l}_{(0)}/|\vec{l}_{(0)}|$.
\end{enumerate}

\section{Light propagation and light deflection in the post-linear gravitational field}\label{light}
\subsection{The light propagation equation}\label{lightprop}
In this paper we calculate the light deflection in the post-linear gravitational field of two
bounded masses for the case when the impact parameter $|\vec{\xi}\,|$ is much larger (5 times or more) 
than the distance $r_{12}$ between the two accelerating masses, so that we can suppose that the gravitational 
field is weak along the light path.

For weak gravitational fields we can assume that the light propagation is very well governed by the laws 
of geometric optics,
whereby light rays (photons) move in curved space-time along null geodesics.
The equations of null geodesics with the time coordinate as a parameter are given by (e.g. see \cite{WEIN72})

\begin{align}
\frac{dl^{i}}{dt}+ \Gamma^{i}_{\alpha \beta} l^{\alpha} l^{\beta}=c^{-1} \Gamma^{0}_{\nu \sigma} l^\nu l^\sigma l^i,
\label{geodesic}
\end{align}
where
\begin{align}
\Gamma^{\mu}_{\rho \sigma}=\frac{1}{2} g^{\mu \lambda}\big[g_{\rho \lambda,\sigma}+g_{\sigma \lambda,\rho}
-g_{\rho \sigma,\lambda}\big] \label{Christoffel}
\end{align}
are the Christoffel symbols of the second kind and $l^\mu=dz^\mu/dt$ denotes the 4-vector $l^\mu=(c,l^i)$. 
Here, it is important to recall that $z^{\mu}=(z^0,z^i)$, where $z^0=c\,t$.
Notice that $l^{\mu}$ is not exactly a 4-vector because we differentiate with respect to the time coordinate $t$. 
So $l^{\mu}$ is a 4-vector up to a factor.
The space component of $l^{\mu}$ given by $l^i=dz^i/dt$ is the 3-vector tangent to the light ray $z^i(t)$.
In the present case of null geodesics, $l^\mu$ has to fulfil the condition
\begin{align}
  l^2 \equiv g_{\mu \nu}[z^0,z^i(t),G]l^\mu l^\nu=0. \label{nullvector}
\end{align}

Now we consider a light ray $z^{i}(t)$ that is propagating in a curved space-time $g_{\mu \nu}[z^0,z^i(t),G]$ with
the signature $(-+++)$.
If the gravitational field is weak, we can write the fundamental metric tensor $g_{\mu \nu}[z^0,z^i(t),G]$ as a power series
in the gravitational constant $G$
\begin{align}
g_{\mu \nu}[z^0,z^i(t),G] \equiv \eta_{\mu \nu}+\sum_{n=1}^{\infty} h_{\mu \nu}^{(n)}[z^0,z^i(t),G] ,\label{Metric1}
\end{align}
where $\eta_{\mu \nu}$ is the Minkowski metric and  $h_{\mu \nu}^{(n)}[z^0,z^i(t),G]$ is a perturbation of the order $n$
in the gravitational constant $G$ equivalent to $G^n g^{(n)}_{\mu \nu}(z^0,z^i(t))$ of Eq.~\eqref{Metric0}
(physically, this means an expansion in the dimensionless parameter $Gm/c^2d$ which is usually very small, $d$ being the 
characteristic length of the problem and $m$ a characteristic mass).

In order to obtain from equation (\ref{geodesic}) the equations of light propagation for the metric given in (\ref{Metric1}) we substitute the Christoffel symbols into the equation (\ref{geodesic}).
To save writing we denote the metric coefficients $h^{(1)}_{pq}[z^0,z^i(t),G]$, 
$h^{(2)}_{pq}[z^0,z^i(t),G]$ by $h^{(1)}_{pq}$ and $h^{(2)}_{pq}$.
Then the resulting equation of light propagation to the second order in $G/c^2$ is given by

\begin{align}
&\frac{dl^i}{dt}=\frac{1}{2}c^2 h^{(1)}_{00,i}-c^2 h^{(1)}_{0i,0}
 -c\,h^{(1)}_{0i,m}l^m
 +c\,h^{(1)}_{0m,i}l^m-c\,h^{(1)}_{mi,0}l^m 
 -h^{(1)}_{mi,n}l^ml^n\nonumber\\
&+
  \frac{1}{2}h^{(1)}_{mn,i}l^ml^n
 -\frac{1}{2}c\,h^{(1)}_{00,0}l^i
 -h^{(1)}_{00,k}l^k l^i
 +\Big(\frac{1}{2}c^{-1}h^{(1)}_{mp,0}
 -c^{-1}h^{(1)}_{op,m}\Big)l^ml^pl^i \nonumber\\
&+
  \frac{1}{2}c^2h^{(2)}_{00,i}
 -\frac{1}{2}c^2h^{(1)ik}h^{(1)}_{00,k}
 -h^{(2)}_{00,k}l^kl^i
 -\Big(h^{(2)}_{mi,n}
 -\frac{1}{2}h^{(2)}_{mn,i}\Big)l^ml^n\nonumber\\
&+
  h^{(1)ik}\Big(h^{(1)}_{mk,n}-\frac{1}{2}h^{(1)}_{mn,k}\Big)l^ml^n 
 -h^{(1)}_{00}h^{(1)}_{00,k}l^kl^i,\label{Lightpropagation}
\end{align}
where by $_{,0}$ and $_{,i}$ we denote $\partial/\partial z^0$ and $\partial/\partial z^i$ respectively.
To calculate the light deflection we need to solve equation (\ref{Lightpropagation}) for $l^i$.
In order to solve this complicated, non-linear differential equation, we turn to approximation techniques.

\subsection{The Approximation Scheme}\label{approximation}
We can write the 3-vector $l^i(t)$ as
\begin{align}
l^i(t)=l^i_{(0)}+\sum_{n=1}^{\infty}\delta l^i_{(n)}(t),\label{Wavevector}
\end{align}
where $l^i_{(0)}$ denotes the constant incoming tangent vector $l^i(-\infty)$ and
$\delta l^i_{(n)}(t)$ the perturbation of the constant tangent vector $l^i_{(0)}$ of order $n$
in $G$ equivalent to $G^n \delta \vec{\tilde{l}}_{(n)}(t)$.\\
After introducing the expression for $l^i(t)$ given by (\ref{Wavevector}) into the equation (\ref{Lightpropagation}) we
obtain differential equations for the perturbations $\delta l^i_{(1)}$ and $\delta l^i_{(2)}$.
These are given by:

\begin{align}
&\frac{d\delta l^i_{(1)}}{dt}=\frac{1}{2} c^2 h^{(1)}_{00,i}
 -c^2h^{(1)}_{0i,0}
 -c\,h^{(1)}_{0i,m}l^m_{(0)}
 +c\,h^{(1)}_{0m,i}l^m_{(0)}
 -c\,h^{(1)}_{mi,0}l^m_{(0)}
 -h^{(1)}_{mi,n}l^m_{(0)}l^n_{(0)}\nonumber\\
&+
  \frac{1}{2}h^{(1)}_{mn,i}l^m_{(0)}l^n_{(0)}  
 -\frac{1}{2}c\,h^{(1)}_{00,0}l^i_{(0)}
 -h^{(1)}_{00,k}l^k_{(0)} l^i_{(0)} 
 +\Big(\frac{1}{2}c^{-1}h^{(1)}_{mp,0}
 -c^{-1}h^{(1)}_{0p,m}\Big)l^m_{(0)}l^p_{(0)}l^i_{(0)} \label{Perturbation1}
\end{align}
and

\begin{align}
&\frac{d \delta l^i_{(2)}}{dt}=
  \frac{1}{2}c^2h^{(2)}_{00,i}
 -\frac{1}{2}c^2h^{(1)ik}h^{(1)}_{00,k}
 -h^{(2)}_{00,k}l^k_{(0)}l^i_{(0)}
 -\Big(h^{(2)}_{mi,n}
 -\frac{1}{2}h^{(2)}_{mn,i}\Big)l^m_{(0)}l^n_{(0)}\nonumber\\
&+
  h^{(1)ik}\Big(h^{(1)}_{mk,n}-\frac{1}{2}h^{(1)}_{mn,k}\Big)l^m_{(0)}l^n_{(0)}
 -h^{(1)}_{00}h^{(1)}_{00,k}l^k_{(0)}l^i_{(0)}\nonumber\\
&-
  c\,h^{(1)}_{0i,m}\delta l^m_{(1)}
 +c\,h^{(1)}_{0m,i}\delta l^m_{(1)}
 -c\,h^{(1)}_{mi,0} \delta l^m_{(1)}\nonumber\\
&-
  h^{(1)}_{mi,n} \delta l^m_{(1)}l^n_{(0)}
 -h^{(1)}_{mi,n} l^m_{(0)}\delta l^n_{(1)}
 +h^{(1)}_{mn,i} \delta l^m_{(1)}l^n_{(0)}\nonumber\\ 
&-
  \frac{1}{2}ch^{(1)}_{00,0} \delta l^i_{(1)}
 -h^{(1)}_{00,k}\delta l^k_{(1)} l^i_{(0)} 
 -h^{(1)}_{00,k} l^k_{(0)} \delta l^i_{(1)}\nonumber\\
&+
  c^{-1}h^{(1)}_{mp,0} \delta l^m_{(1)}l^p_{(0)}l^i_{(0)}
 -c^{-1}h^{(1)}_{op,m}\delta l^m_{(1)}l^p_{(0)} l^i_{(0)} 
 -c^{-1}h^{(1)}_{op,m} l^m_{(0)} \delta l^p_{(1)} l^i_{(0)}\nonumber\\
&+
  \Big(\frac{1}{2}c^{-1}h^{(1)}_{mp,0}
 -c^{-1}h^{(1)}_{0p,m}\Big)l^m_{(0)}l^p_{(0)}\delta l^i_{(1)}. \label{Perturbation2}
\end{align}

In order to calculate the perturbations $\delta l^i_{(1)}(t)$ and $\delta l^i_{(2)}(t)$ we have to integrate equations
(\ref{Perturbation1}) and (\ref{Perturbation2}) along the light ray trajectory to the appropriate order.

Before performing the integration it is convenient to introduce a new independent parameter $\tau$ along the photon's
trajectory as defined by Kopeikin and Sch\"afer \cite{KSCH99}. The relationship between the parameter $\tau$ and the time
coordinate $t$ is given by
\begin{align}
\tau=t-t^{\ast},\label{tau}
\end{align}
where $t^{\ast}$ is the time of closest approach of the unperturbed trajectory of the photon to the origin in an
asymptotically flat harmonic coordinate system.
Then the equation of the unperturbed light ray can be represented as
\begin{align}
z^i(\tau)=\tau \,l^i_{(0)}+\xi^i,
\end{align}
where $\xi^i$ is a vector directed from the origin of the coordinate system towards the point of closest approach.
The vector $\xi^i$ is often called the impact parameter and is orthogonal to the vector $l^i_{(0)}$.
The distance $r(\tau)=|\vec{z}(\tau)|$, of the photon from the origin of the coordinate system reads
\begin{align}
r(\tau)=\sqrt{c^2 \tau^2+\xi^2}.
\end{align}
\begin{widetext}
It follows from equation \eqref{tau} that the differential identity $dt=d \tau$ is valid, so that we can always replace the
integration along the unperturbed light ray with respect to $t$ by the integration with respect to the variable $\tau$.

Then the resulting expression for $\delta l^i_{(1)}$ is given by
\begin{align}
&\delta l^i_{(1)}(\tau)=\frac{1}{2} \int^{\tau}_{-\infty} d \sigma \, l^{\alpha}_{(0)} l^{\beta}_{(0)} h^{(1)}_{\alpha \beta,i}
|_{(\rightarrow)}
 -c\,h^{(1)}_{0i}- h^{(1)}_{mi} l^m_{(0)}
 -h^{(1)}_{00} l^i_{(0)}\nonumber\\
&+
  \frac{1}{2}\,c\int^{\tau}_{-\infty} d \sigma \, h^{(1)}_{00,0} l^i_{(0)}
  |_{(\rightarrow)} 
 +\int^{\tau}_{-\infty} d \sigma \, l^m_{(0)} l^p_{(0)} \Big[\frac{1}{2} c^{-1} h^{(1)}_{mp,0}-c^{-1}h^{(1)}_{0p,m}
  \Big]l^i_{(0)}\mid_{(\rightarrow)}. 
\label{Perturbation11}
\end{align}

On the right-hand side of equation (\ref{Perturbation11}) after evaluating the partial derivatives of the metric
coefficients with respect to the photon's coordinates (i.e. $(z^0,z^i(t))$), we replace in the integrals the photon 
trajectory by its unperturbed approximation $z^i(\sigma)_{\mathrm{unpert.}}= \sigma l^i_{(0)}+\xi^i$ and the time coordinate $z^0$ by 
$\sigma+t^{\ast}$. In this paper we denote this operation by the symbol $|_{(\rightarrow)}$. Then we perform the integration with respect to $\sigma$.\\
After substituting the expression obtained for $\delta l^i_{(1)}$ into equation (\ref{Perturbation2}) we can
integrate it to get $\delta l^i_{(2)}$.
To calculate the perturbation $\delta l^i_{(2)}$ we separate the part of $\delta l^i_{(2)}$ that depends on
the post-linear metric coefficients from the part that depends on the linear metric coefficients.
We denote these parts of $\delta l^i_{(2)}$ by $\delta l^i_{(2)\rm{I}}$ and $\delta l^i_{(2) \rm{II}}$ respectively.
As in the case of equation (\ref{Perturbation11}) we replace the photon trajectory by its unperturbed approximation and the time
coordinate $z^0$ by $\sigma+t^{\ast}$ after evaluating the partial derivatives of the metric coefficients with respect to
the photon coordinates. The expressions for $\delta l^i_{(2)\rm{I}}$ and $\delta l^i_{(2) \rm{II}}$ are given by

\begin{align}
&\delta l^i_{(2)\rm{I}}(\tau)=\int^{\tau}_{-\infty} d \sigma \,\Big[\frac{1}{2} c^2 h^{(2)}_{00,i}
 -h^{(2)}_{00,k} l^k_{(0)}l^i_{(0)} \Big]|_{(\rightarrow)}
 +\int^{\tau}_{-\infty} d \sigma \,\Big[\frac{1}{2} h^{(2)}_{mn,i}-h^{(2)}_{mi,n}\Big] l^m_{(0)}l^n_{(0)}
  |_{(\rightarrow)} \label{Perturbation21}
\end{align}
and

\begin{align}
&\delta l^i_{(2)\rm{II}}(\tau)=-\int^{\tau}_{-\infty} d \sigma \,\Big[\frac{1}{2} c^2 h^{(1)ik}h^{(1)}_{00,k}
 +h^{(1)}_{00}h^{(1)}_{00,k}l^k_{(0)}l^i_{(0)}\Big]|_{(\rightarrow)}  \nonumber\\
&+
  \int^{\tau}_{-\infty} d \sigma \,\Big[h^{(1)ik}\big(h^{(1)}_{mk,n}-\frac{1}{2}h^{(1)}_{mn,k}\big)\Big]l^m_{(0)}l^n_{(0)} 
  |_{(\rightarrow)} \nonumber\\
&+ 
  c \int^{\tau}_{-\infty} d \sigma \,\Big[h^{(1)}_{0m,i}-h^{(1)}_{0i,m}-h^{(1)}_{mi,0}\Big] \delta l^m_{(1)}(\sigma)
  |_{(\rightarrow)}\nonumber\\
&+
  \int^{\tau}_{-\infty} d \sigma \,\Big[h^{(1)}_{mn,i} \delta l^m_{(1)}(\sigma)l^n_{(0)}
 -h^{(1)}_{mi,n} \delta l^m_{(1)}(\sigma)l^n_{(0)}
 -h^{(1)}_{mi,n}l^m_{(0)} \delta l^n_{(1)}(\sigma)\Big]|_{(\rightarrow)}\nonumber\\
&-
  \int^{\tau}_{-\infty} d \sigma \,\Big[\frac{1}{2}ch^{(1)}_{00,0} \delta l^i_{(1)}(\sigma)
 +h^{(1)}_{00,k} \delta l^k_{(1)}(\sigma)l^i_{(0)}
 +h^{(1)}_{00,k}l^k_{(0)} \delta l^i_{(1)}(\sigma)\Big]|_{(\rightarrow)}\nonumber\\
&+
  c^{-1} \int^{\tau}_{-\infty} d \sigma \,\Big[h^{(1)}_{mp,0} \delta l^m_{(1)}(\sigma)l^p_{(0)}
 -h^{(1)}_{0p,m} \delta l^m_{(1)}(\sigma)l^p_{(0)}
 -h^{(1)}_{0p,m}l^m_{(0)}\delta l^p_{(1)}(\sigma)\Big]l^i_{(0)}|_{(\rightarrow)}\nonumber\\
&+
  c^{-1} \int^{\tau}_{-\infty} d \sigma \,\Big[\frac{1}{2} h^{(1)}_{mp,0}
 -h^{(1)}_{0p,m} \Big]l^m_{(0)}l^p_{(0)} 
  \delta l^i_{(1)}(\sigma)|_{(\rightarrow)}. \label{Perturbation22}
\end{align}

\subsection{Light deflection}\label{lightdeflection}
The dimensionless vector $\alpha^i_{(n)}$ of order $n$ in $G$ describing the angle of total deflection of the light ray measured
at the point of observation and calculated with respect to the vector $l^i_{(0)}$ (see \cite{KSCH99}) is given by
\begin{align}
\alpha^i_{(n)}(\tau)=P^i_q \frac{\delta l^q_{(n)}(\tau)}{|\vec{l}_{(0)}|}, \label{deflection}
\end{align}
where $\delta l^i_{(n)}$ is the perturbation of the constant tangent vector of order $n$ in $G$.
Here, $P^i_q=\delta^i_q-e^i_{(0)}e_{(0)q}$ is the projection tensor onto the plane orthogonal to the vector $l^i_{(0)}$.
In the case of light rays (photons) $|\vec{l}_{(0)}|=c$.

\section{The post-linear gravitational field of two bounded point-like masses}\label{gravfield}
In the computation of the metric generated by a system of two bounded point-like masses
we distinguish between 3 zones:
the near zone, the intermediate zone and the wave zone or far zone.\\
In Refs. \cite{DF98} and \cite{KSCHGE99} it was shown that leading order terms for the effect
of light deflection in the case of small impact parameter depend neither on the radiative part $(\sim 1/\xi)$
of the gravitational field nor on the intermediate $(\sim 1/\xi^2)$ zone terms. The main effect rather comes
from the near-zone $(\sim 1/\xi^3)$ terms.
Taking into account this property of strong suppression of the influence of gravitational waves on the light
propagation, we can assume in the present work that the light deflection in the post-linear gravitational
field of two point-like masses is mainly determined by the near-zone metric. 

\subsection{The metric in the near zone}\label{metric}
In Ref. \cite{BFP98}, Blanchet et al. calculated the conservative 2 PN harmonic-coordinate metric 
for the near-zone of a system of
two bounded point-like masses as a function of the position $\vec{z}$ and of the positions and velocities 
of the masses $\vec{x}_{a}(t)$ and $\vec{v}_{a}(t)$ respectively, with $a=1,2$.
We shall use their metric in this paper.
The post-linear metric for two bounded point-like masses, at the 2PN-order, is given by

\begin{align}
h^{(2)}_{00}&= \frac{1}{c^4}\Bigg\{-2 \frac{G^2 m_{1}^2}{r_{1}^2}+G^2 m_{1} m_{2} \left(-\frac{2}{r_{1} r_{2}}- \frac{r_{1}}{2
r_{12}^3}+\frac{r_{1}^2}{2 r_{2} r_{12}^3}-\frac{5}{2 r_{2} r_{12}}
\right)\Bigg\}\nonumber\\
&+\frac{1}{c^4}(1\leftrightarrow 2), \label{Metric2}\\
h^{(2)}_{pq}&=\frac{1}{c^4}\Bigg\{\delta^{pq}\Big[\frac{G^2 m_{1}^2}{r_{1}^2}
 + G^2 m_{1} m_{2}
\Big(\frac{2}{r_{1}r_{2}}-\frac{r_{1}}{2r_{12}^3}+\frac{r_{1}^2}{2r_{2}r_{12}^3}-\frac{5}{2r_{1}r_{12}}
+\frac{4}{r_{12}S}\Big)\Big]\nonumber\\
&+\frac{G^2  m_{1}^2}{r_{1}^2}n_{1}^{p}n_{1}^{q} -4 G^2 m_{1}m_{2}
n_{12}^{p}n_{12}^{q}\left(\frac{1}{S^2}+\frac{1}{r_{12}S}\right)
+\frac{4G^2m_{1}m_{2}}{S^2}\left(n_{1}^{(p}n_{2}^{q)}+2n_{1}^{(p}n_{12}^{q)}\right)\Bigg \}\nonumber\\
&+ \frac{1}{c^4} (1 \leftrightarrow 2)  \label{Metric3}, 
\end{align}
where the symbol $(1 \leftrightarrow 2)$ refers to the preceding term in braces but with the labels 1 and 2 exchanged;
by S we denote $S=r_1+r_2+r_{12}$, where $r_1=|\vec{z}-\vec{x}_{1}(t)|$, 
$r_2=|\vec{z}- \vec{x}_{2}(t)|$ and $r_{12}=|\vec{x}_{1}(t)-\vec{x}_{2}(t)|$.
The vectors $n^{p}_{1}$, $n^{p}_{2}$ and $n^{p}_{12}$ are unit vectors defined by
$n^{p}_{1}=r^{p}_{1}/r_1$, $n^{p}_{2}=r^{p}_{2}/r_2$ and $n^{p}_{12}=r^{p}_{12}/r_{12}$.

In our computations we also need a part of the linear gravitational field of two bounded point-like
masses. The part that is relevant to our calculation is given by
\begin {align}
&h^{(1)}_{00}= 2 \frac{G}{c^2}\sum^{2}_{a=1}\frac{m_a}{r_a}+ \frac{G}{c^4}\sum^{2}_{a=1}\frac{m_a}{r_a} 
\Big[-(\vec{n}_{a} \cdot \vec{v}_{a})^2+4 v^2_a\Big],\label{Metric4}\\
&h^{(1)}_{0p}=-4 \frac{G}{c^3}\sum^{2}_{a=1}\frac{{m_a}}{r_a} v^{p}_{a}, \label{Metric5}\\
&h^{(1)}_{pq}=2 \frac{G}{c^2}\sum^{2}_{a=1}\frac{m_a}{r_a} \delta^{pq}
+\frac{G}{c^4}\sum^{2}_{a=1}\frac{m_a}{r_a}\Big[-(\vec{n}_a \cdot \vec{v}_a)^2\,\delta^{pq}+4 v^p_a v^q_a\Big], \label{Metric6} 
\end{align}
where $v^p_a$ denotes the velocity of the mass $m_a$ and $n^p_a$ is the unit vector defined by $n^p_a=r^p_a/r_a$.

Here, it is worthwhile to point out that the parts of the linear gravitational field in $h^{(1)}_{00}$ and $h^{(1)}_{pq}$ that
contain the accelerations of the masses were introduced into the part of the gravitational field quadratic in $G$ after substituting
the accelerations by explicit functions of the coordinate positions of the masses by means of the Newtonian equations of motion.

\subsection{The barycentric coordinate system}\label{barycentric}

We use a harmonic coordinate system, the origin of which coincides with the 1PN-centre of mass. 
Using the 1PN-accurate centre of mass theorem of Ref. \cite{TDND81}, we can express the individual centre of mass 
frame positions of the two masses in terms of the relative position $\vec{r}_{12} \equiv \vec{x}_1-\vec{x}_2$ 
and the relative velocity $\vec{v}_{12} \equiv \vec{v}_1-\vec{v}_2$ \\
as
\begin{align}
&\vec{x}_1=\Big[X_2+\frac{1}{c^2} \epsilon_{\mathrm{1PN}} \Big]\vec{r}_{12} \label{pos1},\\
&\vec{x}_2=\Big[-X_1+\frac{1}{c^2} \epsilon_{\mathrm{1PN}} \Big] \vec{r}_{12} \label{pos2},
\end{align}
where $X_1$, $X_2$ and  $\epsilon_{1PN}$ are given by
\begin{align}
&X_1 \equiv \frac{m_1}{M},\\
&X_2 \equiv \frac{m_2}{M},\\
&\epsilon_{\mathrm{1PN}}= \frac{\nu (m_1-m_2)}{2 M} \Big[v^2_{12}-\frac{GM}{r_{12}}\Big].
\end{align}
Here, we have introduced
\begin{align}
M \equiv m_1+m_2,\,\, v_{12}=|\vec{v}_{12}|
\end{align}
and
\begin{align}
\nu \equiv \frac{m_1 m_2}{M^2}.
\end{align}
Here, it is important to remark that in our computation of the post-linear light deflection
up to the order $G^2/c^4$, we need only to consider the 1PN-corrections to the Newtonian centre of mass, because, as we shall see in Sec.\,\ref{centre}, the 2PN-corrections to the Newtonian centre of mass are related to post-linear light deflection
terms of order higher than $G^2/c^4$.
\section{Light deflection in the linear gravitational field of two bounded point-like masses}\label{lightlingrav}
In this section we compute the perturbation term $\delta l^i_{(1)}(\tau)$ and the corresponding angle of
light deflection for an observer situated at infinity in an asymptotically flat space.
To compute the perturbation term  $\delta l^i_{(1)}(\tau)$ we have to introduce the linear
metric coefficients given by Eqs~\eqref{Metric4}-\eqref{Metric6} into Eq.~\eqref{Perturbation11}.
After introducing the linear metric coefficients into Eq.~\eqref{Perturbation11}, we find
\begin{align}
&\delta l^i_{(1)}(\tau)= -2\,G \sum^{2}_{a=1} m_a \int^{\tau}_{-\infty} d \sigma \,\frac{1}{r^3_a} \big[z^i-x^i_a(t)\big]
|_{(\rightarrow)}\nonumber\\
&+
  4\,\frac{G}{c}\sum^{2}_{a=1} m_a \int^{\tau}_{-\infty} d \sigma \,\frac{1}{r^3_a} (\vec{e}_{(0)}\cdot\vec{v}_a(t)) 
\big[z^i-x^i_a(t)\big]|_{(\rightarrow)}\nonumber\\
&-
  2\,\frac{G}{c^2}\sum^{2}_{a=1} m_a \int^{\tau}_{-\infty} d \sigma \,\frac{1}{r^3_a}\,\Big[v^2_a(t)
 +
  (\vec{e}_{(0)}\cdot\vec{v}_a(t))^2\Big]\big[z^i-x^i_a(t)\big]|_{(\rightarrow)}\nonumber\\
&+
  3\,\frac{G}{c^2}\sum^{2}_{a=1} m_a \int^{\tau}_{-\infty} d \sigma \,\frac{1}{r^5_a}(\vec{r}_a\cdot\vec{v}_{a}(t))^2\big[z^i-x^i_a(t)\big]|_{(\rightarrow)}\nonumber\\
&-
  4\,\frac{G}{c^2}\sum^{2}_{a=1} m_a \int^{\tau}_{-\infty} d \sigma \,\frac{1}{r^3_a}(\vec{e}_{(0)}\cdot\vec{v}_{a}(t))
\Big[c\,\sigma-(\vec{e}_{(0)}\cdot\vec{x}_{a}(t))\Big]l^i_{(0)}|_{(\rightarrow)}\nonumber\\ 
&-
  2\,\frac{G}{c^2}\sum^{2}_{a=1} m_a \int^{\tau}_{-\infty} d \sigma \, \frac{1}{r^3_a} 
\Bigg\{(\vec{e}_{(0)}\cdot\vec{v}_{a}(t))c\, \sigma+\Big[\vec{\xi}\cdot\vec{v}_a(t)-\vec{x}_a(t)\cdot\vec{v}_a(t)\Big]\Bigg\}
 v^i_a(t)|_{(\rightarrow)}\nonumber\\
&+
  4\,\frac{G}{c^2}\sum^{2}_{a=1}\frac{m_a}{r_a}v^i_a(t)
-
  4\,\frac{G}{c^2}\sum^{2}_{a=1}\frac{m_a}{r_a}l^i_{(0)} 
-
  4\,\frac{G}{c^3}\sum^{2}_{a=1}\frac{m_a}{r_a}(\vec{e}_{(0)}\cdot\vec{v}_{a}(t))v^i_a(t)\nonumber\\
&+
  2\, \frac{G}{c^4}\sum^{2}_{a=1}m_a\Bigg\{\frac{(\vec{r}_a\cdot\vec{v}_a(t))^2}{r^3_a}-2 \frac{1}{r_a}v^2_a(t)\Bigg\}l^i_{(0)}, \label{LinPert.}
\end{align}
where $\vec{r}_a=\vec{z}-\vec{x}_a(t)$ and $r_a=|\vec{r}_a|$.

Because the linear metric coefficients are functions of the positions and velocities of the masses
$\vec{x}_a(t)$ and $\vec{v}_a(t)$ respectively, the expression for $\delta l^i_{(1)}(\tau)$
given in Eq.~\eqref{LinPert.} is a function of these quantities. This means that we have to take into
account the motion of the masses when we are going to compute the integrals in Eq.~\eqref{LinPert.}.
Considering that the influence of the gravitational field on the light propagation is strongest
near the barycentre of the binary and that the velocities of the masses are small with respect to the velocity of light, 
we are allowed to make the following
approximations:
\begin{enumerate}
\item We may assume that the linear gravitational field is determined by the positions and velocities of the masses
taken at the time of closest approach ($t=t^{\ast}$) of the unperturbed light ray to the barycentre of the binary (i.e. to the origin
of the asymptotically flat harmonic coordinate system).
The expression, resulting from Eq.~\eqref{LinPert.} after setting $t=t^{\ast}$ for the positions and 
velocities and computing the integrals, is denoted by $\delta l^i_{(1)\mathrm{I}}(\tau)$;
\item We treat the effect of the motion of the masses on light propagation as a correction to
the expression of $\delta l^i_{(1)\mathrm{I}}(\tau)$, which we denote by $\delta l^i_{(1)\mathrm{II}}(\tau)$.
We shall compute this correction in Section \ref{lightmotion}.
\end{enumerate}  
After fixing the values of the quantities $\vec{x}_a(t)$ and $\vec{v}_a(t)$ to $\vec{x}_a(t^{\ast})$ and
$\vec{v}_a(t^{\ast})$ and evaluating the integrals in Eq.~\eqref{LinPert.}, we find

\begin{align}
\delta l^{i}_{(1)\mathrm{I}}(\tau)&=-2\,\frac{G}{c} \sum^{2}_{a=1} m_a B_a\big[\xi^i-x^i_a(t^{\ast})\big]
 -
  \frac{G}{c^2}\sum^{2}_{a=1} m_a \Big\{2\,A_a
 +
  \frac{4}{r_a}\Big\} l^i_{(0)}\nonumber\\
&+
  4\,\frac{G}{c^2}\sum^{2}_{a=1} m_a (\vec{e}_{(0)} \cdot \vec{v}_a(t^{\ast}))B_a\big[\xi^i-x^i_a(t^{\ast})\big]
 +
  4\,\frac{G}{c^2}\sum^{2}_{a=1} \frac{m_a}{r_a} v^i_a(t^{\ast})\nonumber\\
&+
  4\,\frac{G}{c^3}\sum^{2}_{a=1} m_a (\vec{e}_{(0)} \cdot \vec{v}_a(t^{\ast}))\Big\{A_a
 +
  \frac{1}{r_a}\Big\}l^i_{(0)}\nonumber\\
&+
  \frac{G}{c^3}\sum^{2}_{a=1} m_a(\vec{e}_{(0)} \cdot \vec{v}_a(t^{\ast}))^2 \Big\{3\,F_{a2}
 -
  2\,B_a\Big\}\big[\xi^i-x^i_a(t^{\ast})\big]\nonumber\\
&+
  6\,\frac{G}{c^3}\sum^{2}_{a=1} m_a(\vec{e}_{(0)} \cdot \vec{v}_a(t^{\ast}))\Big[\vec{\xi}\cdot\vec{v}_a(t^{\ast})
 -
  \vec{x}_a(t^{\ast})\cdot \vec{v}_a(t^{\ast})\Big]F_{a3}\big[\xi^i-x^i_a(t^{\ast})\big]\nonumber\\
&-
  2\,\frac{G}{c^3}\sum^{2}_{a=1} m_a v^2_a(t^{\ast})B_a\big[\xi^i-x^i_a(t^{\ast})\big]\nonumber\\
&+
  3\,\frac{G}{c^3}\sum^{2}_{a=1} m_a \Big[\vec{\xi}\cdot\vec{v}_a(t^{\ast})
 -
  \vec{x}_a(t^{\ast})\cdot \vec{v}_a(t^{\ast})\Big]^2 F_{a4}\big[\xi^i-x^i_a(t^{\ast})\big]\nonumber\\
&-
  \frac{G}{c^3}\sum^{2}_{a=1} m_a\Bigg\{(\vec{e}_{(0)} \cdot \vec{v}_a(t^{\ast}))\bigg[2\,A_a
 +
  \frac{4}{r_a}\bigg]
 + 
  2\,\Big[\vec{\xi}\cdot\vec{v}_a(t^{\ast})-\vec{x}_a(t^{\ast})\cdot \vec{v}_a(t^{\ast})\Big]B_a
\Bigg\}v^i_a(t^{\ast})\nonumber\\
&+
  6\,\frac{G}{c^4}\sum^{2}_{a=1} m_a(\vec{e}_{(0)}\cdot\vec{v}_{a}(t^{\ast}))\Big[\vec{\xi}\cdot\vec{v}_a(t^{\ast})
 -
  \vec{x}_a(t^{\ast})\cdot \vec{v}_a(t^{\ast})\Big]F_{a2}l^i_{(0)}\nonumber\\
&+
  \frac{G}{c^4}\sum^{2}_{a=1} m_a(\vec{e}_{(0)} \cdot \vec{v}_a(t^{\ast}))^2\Big\{3\,F_{a1}
 -
  2\,A_a\Big\}l^i_{(0)}\nonumber\\
&-
  \frac{G}{c^4}\sum^{2}_{a=1} m_a v^2_a(t^{\ast})\Big\{2\,A_a
 +
  \frac{4}{r_a}\Big\}l^i_{(0)}
 +
  2\,\frac{G}{c^4}\sum^{2}_{a=1} \frac{m_a}{r^3_a}(\vec{r}\cdot\vec{v}_a(t^{\ast}))^2l^i_{(0)}\nonumber\\ 
&+
  3\,\frac{G}{c^4}\sum^{2}_{a=1}m_a\Big[\vec{\xi}\cdot\vec{v}_a(t^{\ast})
 -
  \vec{x}_a(t^{\ast})\cdot \vec{v}_a(t^{\ast})\Big]^2F_{a3}l^i_{(0)},\label{Perturbed1}
\end{align}
where the functions $A_a$, $B_a$, $F_{a1}$, $F_{a2}$, $F_{a3}$, and $F_{a4}$ are given by
\begin{align}
&A_a=\frac{1}{r_a\,R_a}\Big[-r^2_a(0,t^{\ast})
 +
  (\vec{e}_{(0)}\cdot \vec{x}_a(t^{\ast}))\big(c\,\tau+r_a\big)\Big],\label{Aa}\\
&B_a=\frac{1}{r_a\,R_a}\Big[-(\vec{e}_{(0)}\cdot \vec{x}_a(t^{\ast}))
 +
  c\,\tau+r_a\Big],\label{Ba}\\
&F_{a1}=\frac{1}{3\,r^3_a\,R^2_a}\Bigg\{-r^2_a(0,t^{\ast})\Bigg[2\,r^4_a(0,t^{\ast})
 -
  (\vec{e}_{(0)}\cdot \vec{x}_a(t^{\ast}))\Big(3\,r^2_a(0,t^{\ast})
 -
  (\vec{e}_{(0)}\cdot \vec{x}_a(t^{\ast}))^2\Big)\,r_a\Bigg]\nonumber\\
&+
  2\,(\vec{e}_{(0)}\cdot \vec{x}_a(t^{\ast}))\Bigg[3\,r^4(0,t^{\ast})
 -
  (\vec{e}_{(0)}\cdot \vec{x}_a(t^{\ast}))\Big(3\,r^2_a(0,t^{\ast})
 -
  (\vec{e}_{(0)}\cdot \vec{x}_a(t^{\ast}))^2\Big)\,r_a\Bigg] c\,\tau \nonumber\\
&-
  \Bigg[3\,r^4_a(0,t^{\ast})
 +
  3\,(\vec{e}_{(0)} \cdot \vec{x}_a(t^{\ast}))^2r^2_a(0,t^{\ast})
 -
  (\vec{e}_{(0)}\cdot \vec{x}_a(t^{\ast}))\Big(3\,r^2_a(0,t^{\ast})
 -
  (\vec{e}_{(0)}\cdot \vec{x}_a(t^{\ast}))^2\Big)\,r_a \Bigg] c^2 \tau^2 \nonumber\\
&+
  (\vec{e}_{(0)}\cdot \vec{x}_a(t^{\ast}))\Bigg[3\,r^2_a(0,t^{\ast})
 -
  (\vec{e}_{(0)}\cdot \vec{x}_a(t^{\ast}))^2\Bigg] c^3 \tau^3\Bigg\},\label{Fa1}\\
&F_{a2}=\frac{1}{3\,r^3_a\,R^2_a}\Bigg\{-2\,(\vec{e}_{(0)}\cdot \vec{x}_a(t^{\ast}))\,r^4_a(0,t^{\ast})
 +
  \Big[r^2_a(0,t^{\ast})
 +
  (\vec{e}_{(0)}\cdot \vec{x}_a(t^{\ast}))^2\Big]\,r^2_a(0,t^{\ast})\,r_a \nonumber\\
&-
  2\,(\vec{e}_{(0)}\cdot \vec{x}_a(t^{\ast}))\Big[\big(r^2_a(0,t^{\ast})
 +
  (\vec{e}_{(0)}\cdot \vec{x}_a(t^{\ast}))^2\big)\,r_a
 -
  3\,(\vec{e}_{(0)}\cdot \vec{x}_a(t^{\ast}))\,r^2_a(0,t^{\ast})\Big]c\,\tau\nonumber\\
&+
  \Big[r^2_a(0,t^{\ast})
 +
  (\vec{e}_{(0)}\cdot \vec{x}_a(t^{\ast}))^2\Big]\Big[r_a
 -
  3\,(\vec{e}_{(0)}\cdot \vec{x}_a(t^{\ast}))\Big] c^2 \tau^2 \nonumber\\
&+
  \Big[r^2_a(0,t^{\ast})
 +
  (\vec{e}_{(0)}\cdot \vec{x}_a(t^{\ast}))^2\Big]c^3 \tau^3\Bigg\}, \label{Fa2}\\
&F_{a3}=\frac{1}{3\,r^3_a\,R^2_a} \Bigg\{-r^2_a(0,t^{\ast})\Big[r^2_a(0,t^{\ast})
 +
  (\vec{e}_{(0)}\cdot \vec{x}_a(t^{\ast}))^2
 -
  2\,(\vec{e}_{(0)}\cdot \vec{x}_a(t^{\ast}))\,r_a\Big]\nonumber\\
&+
  (\vec{e}_{(0)}\cdot \vec{x}_a(t^{\ast}))\Big[3\,r^2_a(0,t^{\ast})
 +
  3\,(\vec{e}_{(0)}\cdot \vec{x}_a(t^{\ast}))^2
 -
  4(\vec{e}_{(0)}\cdot \vec{x}_a(t^{\ast}))\,r_a\Big]c\,\tau \nonumber\\ 
&+
  2\,(\vec{e}_{(0)}\cdot \vec{x}_a(t^{\ast}))\Big[r_a
 -
  3\,(\vec{e}_{(0)}\cdot \vec{x}_a(t^{\ast}))\Big]c^2 \tau^2
 +
  2\,(\vec{e}_{(0)}\cdot \vec{x}_a(t^{\ast}))c^3 \tau^3\Bigg\},\label{Fa3}\\
&F_{a4}=\frac{1}{3\,r^3_a\,R^2_a}\Bigg\{(\vec{e}_{(0)}\cdot \vec{x}_a(t^{\ast}))\Big[-3\,r^2_a(0,t^{\ast})
 +
  (\vec{e}_{(0)}\cdot \vec{x}_a(t^{\ast}))^2\Big]
 +2\,r^2_a(0,t^{\ast})\,r_a\nonumber\\
&+
  \Big[3\,r^2_a(0,t^{\ast})
 +
  3\,(\vec{e}_{(0)}\cdot \vec{x}_a(t^{\ast}))^2
 -
  4\,(\vec{e}_{(0)}\cdot \vec{x}_a(t^{\ast}))r_a\Big]c\,\tau\nonumber\\
&+
  2\,\Big[r_a
 -
  3\,(\vec{e}_{(0)}\cdot\vec{x}_a(t^{\ast}))\Big]c^2\tau^2
 +
  2\,c^3\tau^3\Bigg\}.\label{Fa4}
\end{align}
Here, the suffix $a$ labels the masses and $r_a$ is the distance between the position of the photon 
along its unperturbed trajectory and the position of the mass $m_a$ at the time $t^{\ast}$. Explicitly the distance $r_a$ 
is given by
\begin{align}
r_a=r_a(\tau,t^{\ast})=\bigg[c^2 \tau^2+\xi^2 - 2\,c\,\tau \vec{e}_{(0)}\cdot \vec{x}_{a}(t^{\ast})
- 2\,\vec{\xi}\cdot \vec{x}_{a}(t^{\ast})
+x^{2}_{a}(t^{\ast})\bigg]^{1/2}.\label{distanceA}
\end{align}
It follows from the expression for $r_a$ that $r_a(0,t^{\ast})$ is the distance between the point
of closest approach of the unperturbed light ray to the origin of the coordinate system and the position of the
mass $m_a$ at the time $t^{\ast}$. The quantity $R_a$ appearing in Eqs \eqref{Aa}-\eqref{Fa4} is given by 
\begin{align}
R_a=r^2_a(0,t^{\ast})-(\vec{e}_{(0)} \cdot \vec{x}_a(t^{\ast}))^2.\label{distanceB}
\end{align}

To get the expression for the angle of light deflection we have to introduce 
$\delta l^i_{(1)\mathrm{I}}(\tau)$ into Eq. \eqref{deflection} and compute the limit $\tau \rightarrow \infty$.
We have to compute the limit $\tau \rightarrow \infty$, because we assume that the observer is located at infinity in
an asymptotically flat space.

After introducing the perturbation $\delta l^i_{(1)\mathrm{I}}(\tau)$ into Eq. \eqref{deflection}
and computing the limit for $\tau \rightarrow \infty$, we find

\begin{align}
\alpha^i_{(1)\mathrm{I}}&=\lim_{\tau\to\infty}\Big[\frac{1}{c}P^i_q \delta l^q_{(1)\mathrm{I}}(\tau)\Big]\nonumber\\
&=-4\frac{G}{c^2} \sum_{a=1}^{2} \frac{m_a}{R_a} \big[\xi^i-P^i_qx^q_a(t^{\ast})\big]\nonumber\\
&+
  8\frac{G}{c^3} \sum_{a=1}^{2} \frac{m_a}{R_a}(\vec{e}_{(0)}\cdot \vec{v}_a(t^{\ast}))
  \big[\xi^i-P^i_qx^q_a(t^{\ast})\big]\nonumber\\
&-
  4\frac{G}{c^4} \sum_{a=1}^{2} \frac{m_a}{R_a} v^2_a(t^{\ast})\big[\xi-P^i_qx^q_a(t^{\ast})\Big] 
 -
  4\frac{G}{c^4} \sum_{a=1}^{2} \frac{m_a}{R_a}(\vec{e}_{(0)}\cdot \vec{v}_a(t^{\ast}))^2 
  \big[\xi^i-P^i_qx^q_a(t^{\ast})\big]\nonumber\\
&-
  4\frac{G}{c^4} \sum_{a=1}^{2} \frac{m_a}{R_a}\Big\{(\vec{e}_{(0)}\cdot \vec{x}_a(t^{\ast}))
  (\vec{e}_{(0)}\cdot \vec{v}_a(t^{\ast}))+\vec{\xi}\cdot\vec{v}_a(t^{\ast})
 -
  \vec{x}_a(t^{\ast})\cdot\vec{v}_a(t^{\ast})\Big\}P^i_q v^q_a(t^{\ast})\nonumber\\
&+
  2\frac{G}{c^4} \sum_{a=1}^{2} \frac{m_a}{R_a}(\vec{e}_{(0)}\cdot\vec{v}_a(t^{\ast}))^2\big[\xi^i-P^i_qx^q_a(t^{\ast})\big]
\nonumber\\
&+
  4\frac{G}{c^4}\sum_{a=1}^{2} \frac{m_a}{R^2_a}(\vec{e}_{(0)}\cdot\vec{v}_a(t^{\ast}))^2\,(\vec{e}_{(0)}\cdot\vec{x}_a(t^{\ast}))^2 \big[\xi^i-P^i_qx^q_a(t^{\ast})\big]\nonumber\\
&+
  8\frac{G}{c^4} \sum_{a=1}^{2} \frac{m_a}{R^2_a}(\vec{e}_{(0)}\cdot\vec{x}_a(t^{\ast}))
  (\vec{e}_{(0)}\cdot\vec{v}_a(t^{\ast}))\Big[\vec{\xi}\cdot \vec{v}_a(t^{\ast})
 -
  \vec{x}_a(t^{\ast})\cdot \vec{v}_a(t^{\ast})\Big]\big[\xi^i-P^i_qx^i_a(t^{\ast})\big]\nonumber\\
&+
  4\frac{G}{c^4} \sum_{a=1}^{2} \frac{m_a}{R^2_a}\Big[(\vec{\xi}\cdot \vec{v}_a(t^{\ast}))^2
 -
  2(\vec{\xi}\cdot \vec{v}_a(t^{\ast})(\vec{x}_a(t^{\ast})\cdot\vec{v}_a(t^{\ast}))
 +
  (\vec{x}_a(t^{\ast})\cdot\vec{v}_a(t^{\ast}))^2\Big]\nonumber\\
&\big[\xi^i-P^i_qx^q_a(t^{\ast})\big].
\end{align}

\section{The post-linear light deflection in the post-linear gravitational field of two bounded 
point-like masses}\label{lightpostlingrav}
In this section we present our computations for light deflection in the post-linear gravitational field of
two bounded point-like masses.
In our computations we assume that both the light source and the observer are at infinity in an asymptotically
flat space so that the effects of $h^{(1)}_{\mu \nu}$ and $h^{(2)}_{\mu \nu}$ near the light source and near the observer 
are negligible.   
We take into account  only the terms of the order $G^2/c^4$.
From equations \eqref{Perturbation21}, \eqref{Perturbation22} and \eqref{deflection} we see that $\alpha^{i}_{(2)}$ is 
a function of the  post-linear metric coefficients $h^{(2)}_{\mu \nu}$ and of the linear metric coefficients
$h^{(1)}_{\mu \nu}$. To facilitate the computations we separate the light deflection terms that are functions of the post-linear 
metric coefficients from the terms that are functions of the linear metric coefficients and the perturbations
of the first order in $G$ of the vector tangent to the unperturbed light ray.
First we compute the terms of $\alpha^{i}_{(2)}$ that are functions of the post-linear metric coefficients, which we
denote by $\alpha^{i}_{(2)\mathrm{I}}$.

\subsection{The post-linear light deflection terms that depend  on the metric
coefficients quadratic in $G$}\label{postlinlightquad}
It follows from equations \eqref{Perturbation21} and \eqref{deflection} that a part of the 
post-linear light deflection is given by:
\begin{align}         
&\alpha^{i}_{(2)\mathrm{I}}=\frac{1}{c}P^{i}_{q} \Bigg[\frac{1}{2} c^2 \int^{\infty}_{-\infty} d \tau h^{(2)}_{00,q}
\mid_{(\rightarrow)}
 +\int^{\infty}_{-\infty} d \tau \Big[\frac{1}{2} h^{(2)}_{mn,q}-h^{(2)}_{qm,n}\Big]l^m_{(0)}l^n_{(0)}
\mid_{(\rightarrow)}\Bigg]\label{deflection1}.
\end{align}
Upon introducing the post-linear metric \eqref{Metric2} and \eqref{Metric3} into \eqref{deflection1} we obtain
integrals whose integrands are functions of the distances $r_1$, $r_2$, $S$ and their inverses.
Through the distances $r_1$, $r_2$ and $S$, the resulting integrals from Eq. \eqref{deflection1} are functions of the
positions of the masses $\vec{x}_a(t)$.

For the same reason as in the case of the linear light deflection we are here allowed to fix the values of the positions
of the masses $\vec{x}_a(t)$ to their values at the time $t^{\ast}$ before performing the integration.
The resulting integrals are given explicitly in Appendix \ref{ap:A}.

To evaluate the integrals that cannot be represented by elementary functions we resort as usual to a series
expansion of the integrands.
To perform the series expansion we consider the integrands as functions of the distances
$r_1$, $r_2$ and $S$. Then, we expand these functions in a Taylor series about the origin of the coordinate system
$\vec{x}_1=\vec{x}_2=0$ to the second order.
We need only to perform the Taylor expansion up to second order, since with an expansion to this order we obtain
a result which is sufficiently accurate for the applications that we shall consider in this paper.

The positions of the masses in the centre of mass frame defined in Subsection~\ref{barycentric} 
are given by equations \eqref{pos1} and \eqref{pos2}.
Here, we do not need to take into account the 1PN-corrections in the positions of the masses, because if we introduce these
into Eq.~\eqref{deflection1} we will obtain terms of higher order than $G^2/c^4$.

In Section \ref{centre} we shall compute the post-linear light deflection terms resulting from the introduction of the
1PN-corrections in the positions of the masses into the equation for the linear light deflection.

Also, we do not need to consider here the correction terms arising from introducing the motions of the masses into 
Eq.~\eqref{deflection1}, because these terms are of higher order than $G^2/c^4$.
The correction to the post-linear light deflection arising from introducing the motion of the masses into the
expression for the linear perturbation is denoted by $\alpha^i_{(2)\mathrm{III}}$ and we shall compute it in Section \ref{lightmotion}.

\subsection{The post-linear light deflection terms that are dependent on the 
metric coefficients linear in $G$}\label{postlinlightlin}
The post-linear light deflection terms which are functions of the linear metric coefficients and the linear perturbations
$\delta \vec{l}_{(1)}(\tau)$ we denote by $\alpha^{i}_{(2)\mathrm{II}}$.
It follows from equations \eqref{Perturbation11} and \eqref{deflection} that the resulting expression for
the post-linear light deflection $\alpha^{i}_{(2)\mathrm{II}}$ is given by  

\begin{align}
&\alpha^{i}_{(2)\rm{II}}=\frac{1}{c}P^{i}_{q}\,\, \bigg[ -\frac{1}{2} c^2 \int^{\infty}_{-\infty} 
d \tau \, h^{(1)qm}h^{(1)}_{00,m}|_{(\rightarrow)}\nonumber\\
&+
  \int^{\infty}_{-\infty} d \tau \,\Big[h^{(1)qp}\big(h^{(1)}_{mp,n}
 -\frac{1}{2} h^{(1)}_{mn,p}\big)\Big]l^m_{(0)}l^n_{(0)}|_{(\rightarrow)}\nonumber\\
&+
  c \int^{\infty}_{-\infty} d \tau \,\Big[h^{(1)}_{0m,q}-h^{(1)}_{0q,m}\Big] \delta l^m_{(1)}(\tau)
  |_{(\rightarrow)}\nonumber\\
&+
  \int^{\infty}_{-\infty} d \tau \,\Big[h^{(1)}_{mn,q} \delta l^m_{(1)}(\tau)l^n_{(0)}
 -h^{(1)}_{mq,n} \delta l^m_{(1)}(\tau)l^n_{(0)}
 -h^{(1)}_{mq,n}l^m_{(0)} \delta l^n_{(1)}(\tau)\Big]|_{(\rightarrow)}\nonumber\\
&-
  \int^{\infty}_{-\infty}d \tau \, h^{(1)}_{00,k}l^k_{(0)}\delta l^q_{(1)}(\tau)|_{(\rightarrow)}\nonumber\\
&-
  \frac{1}{c}\int^{\infty}_{-\infty}d \tau \, 
  h^{(1)}_{0p,m}l^m_{(0)}l^p_{(0)} \delta    l^q_{(1)}(\tau)|_{(\rightarrow)}\bigg].\label{deflection2}
\end{align}

To compute $\alpha^{i}_{(2)\rm{II}}$, we introduce the expressions for the perturbations $\delta l^i_{(1)}(\tau)$
given by Eq.~\eqref{Perturbed3} and the metric functions \eqref{Metric4}, \eqref{Metric6} and \eqref{Metric5} 
into the expression for $\alpha^{i}_{(2)\rm{II}}$. 
Here, we may use the same approximations as before, i.e. we can fix the values of the positions and velocities
of the masses to their values at the time $t^{\ast}$ before performing the integrals.
The resulting integrals are given in Appendix \ref{ap:B}.
As explained in the preceding section, with the help of a Taylor expansion of the integrands we can evaluate the 
integrals, which cannot be represented by elementary functions.

\section{Light deflection and the motion of the masses}\label{lightmotion}
In this section we compute the correction terms to the linear and post-linear light deflection arising from the
effect of the motion of the masses on light propagation.
The correction terms to the linear and post-linear light deflection  can be found by means of Taylor expansions 
of the linear perturbation \eqref{LinPert.} in which the coefficients depend on the sources' coordinates $x^{i}_{a}$ 
and their successive derivatives with regard to $t$, namely
\begin{align*}
\frac{dx^{i}_{a}}{dt}= v^{i}_{a}(t), \frac{d^2 x^{i}_{a}}{dt^2}=\frac{dv^i_a}{dt}= a^{i}_{a}(t),..., 
\end{align*}
taken at the time $t^{\ast}$.

\subsection{The linear light deflection and the motion of the masses}\label{linlightmotion}

The correction terms to the linear perturbation arising from the Taylor expansion of
Eq.~\eqref{LinPert.} are given by

\begin{align}
&\delta l^i_{(1)\mathrm{II}}(\tau)=G\sum^2_{a=1} m_a \int^{\tau}_{-\infty} d\sigma \, 
  \Bigg\{\bigg[-\frac{6}{r^5_a}(\vec{r}_a\cdot\vec{v}_a(t^{\ast}))r^i_a
 +
  \frac{2}{r^3_a}v^i_a(t^{\ast})\bigg]\sigma \nonumber\\
&+
  \bigg[-\frac{15}{r^7_a}(\vec{r}_a\cdot\vec{v}_a(t^{\ast}))^2r^i_a
 +
  \frac{6}{r^5_a}(\vec{r}_a\cdot\vec{v}_a(t^{\ast}))v^i_a(t^{\ast})
 +
  \frac{3}{r^5_a}v^2_a(t^{\ast})r^i_a\bigg]\sigma^2\Bigg\}_{|(\rightarrow)}\nonumber\\
&+
  \frac{G}{c}\sum^2_{a=1} m_a (\vec{e}_{(0)}\cdot\vec{v}_a(t^{\ast}))\int^{\tau}_{-\infty} d\sigma \, 
\Bigg\{\bigg[\frac{12}{r^5_a}(\vec{r}_a\cdot\vec{v}_a(t^{\ast}))r^i_a
 -
  \frac{4}{r^3_a}v^i_a(t^{\ast})\bigg]\sigma \nonumber\\
&+
  \bigg[\frac{30}{r^7_a}(\vec{r}_a\cdot\vec{v}_a(t^{\ast}))^2r^i_a
 -
  \frac{12}{r^5_a}(\vec{r}_a\cdot\vec{v}_a(t^{\ast}))v^i_a(t^{\ast})
 -
  \frac{6}{r^5_a}v^2_a(t^{\ast})r^i_a\bigg]\sigma^2\Bigg\}_{|(\rightarrow)}\nonumber\\
&+
  \frac{G}{c^2}\sum^2_{a=1} m_a \int^{\tau}_{-\infty} d\sigma \,
\Bigg\{\Big[v^2(t^{\ast})
 +
  (\vec{e}_{(0)}\cdot\vec{v}(t^{\ast}))^2\Big]
  \bigg[\bigg(-\frac{6}{r^5_a}(\vec{r}_a\cdot\vec{v}_a(t^{\ast}))r^i_a \nonumber\\
&+
  \frac{2}{r^3_a}v^i_a(t^{\ast})\bigg)\sigma 
 +
  \bigg(-\frac{15}{r^7_a}(\vec{r}_a\cdot\vec{v}_a(t^{\ast}))^2r^i_a
 +
  \frac{6}{r^5_a}(\vec{r}_a\cdot\vec{v}_a(t^{\ast}))v^i_a(t^{\ast})
 +
  \frac{3}{r^5_a}v^2_a(t^{\ast})r^i_a\bigg)\sigma^2\bigg]\nonumber\\
&+
  \bigg[\bigg(\frac{15}{r^7_a}(\vec{r}_a\cdot\vec{v}_a(t^{\ast}))^3r^i_a
 -
  \frac{3}{r^5_a}(\vec{r}_a\cdot\vec{v}_a(t^{\ast}))^2v^i_a(t^{\ast})
 -
  \frac{6}{r^5_a}(\vec{r}_a\cdot\vec{v}_a(t^{\ast}))v^2_a(t^{\ast})r^i_a\bigg)\sigma \nonumber\\
&+
  \bigg(\frac{105}{2}\frac{(\vec{r}_a\cdot\vec{v}_a(t^{\ast}))^4}{r^9_a}r^i_a
 -
  \frac{15}{r^7_a}(\vec{r}_a\cdot\vec{v}_a(t^{\ast}))^3v^i_a(t^{\ast})
 -
  \frac{75}{2}\frac{(\vec{r}_a\cdot\vec{v}_a(t^{\ast}))^2}{r^7_a}v^2_a(t^{\ast})r^i_a \nonumber\\
&+
  \frac{6}{r^5_a}(\vec{r}_a\cdot\vec{v}_a(t^{\ast}))v^2_a(t^{\ast})v^i_a(t^{\ast})
 +
  \frac{3}{r^5_a}v^4_a(t^{\ast})r^i_a\bigg)\sigma^2\bigg]\nonumber\\
&-
  (\vec{e}_{(0)}\cdot\vec{v}_a(t^{\ast}))\Big[c \,\sigma
 -
  (\vec{e}_{(0)}\cdot\vec{x}_a(t^{\ast}))\Big]
\bigg[\frac{12}{r^5_a}(\vec{r}_a\cdot\vec{v}_a(t^{\ast}))\sigma \nonumber\\
&+
  \bigg(\frac{30}{r^7_a}(\vec{r}_a\cdot\vec{v}_a(t^{\ast}))^2
 -
  \frac{6}{r^5_a}v^2_a(t^{\ast})\bigg)\sigma^2\bigg]l^i_{(0)}\nonumber\\
&+
  (\vec{e}_{(0)}\cdot\vec{v}_a(t^{\ast}))^2\bigg[\frac{4}{r^3_a}\sigma
 +
  \frac{12}{r^5_a}(\vec{r}_a\cdot\vec{v}_a(t^{\ast}))\sigma^2\bigg]l^i_{(0)}\nonumber\\
&+
  \bigg[(\vec{e}_{(0)}\cdot\vec{v}_a(t^{\ast}))c\,\sigma
 +
  \Big(\vec{\xi}\cdot\vec{v}_a(t^{\ast})
 -
  \vec{x}_a(t^{\ast})\cdot\vec{v}_a(t^{\ast})\Big)\bigg]
\bigg[-\frac{6}{r^5_a}(\vec{r}_a\cdot\vec{v}_a(t^{\ast}))\sigma \nonumber\\
&+
  \bigg(-\frac{15}{r^7_a}(\vec{r}_a\cdot\vec{v}_a(t^{\ast}))^2
 +
  \frac{3}{r^5_a}v^2_a(t^{\ast})\bigg)\sigma^2\bigg]v^i_a(t^{\ast})\nonumber\\
&+
  v^2_a(t^{\ast})\bigg[\frac{2}{r^3_a}\sigma
 +
  \frac{6}{r^5_a}(\vec{r}_a\cdot\vec{v}_a(t^{\ast}))\sigma^2\bigg]v^i_a(t^{\ast})
\Bigg\}_{|(\rightarrow)}\nonumber\\
&+
  \frac{G}{c^2}\sum^2_{a=1} m_a \Bigg\{\frac{4}{r^3_a}(\vec{r}_a\cdot\vec{v}_a(t^{\ast}))\tau
 +
  \bigg[\frac{6}{r^5_a}(\vec{r}_a\cdot\vec{v}_a(t^{\ast}))^2
 -
  \frac{2}{r^3_a}v^2_a(t^{\ast})\bigg]\tau^2\Bigg\}\Big[v^i_a(t^{\ast})-l^i_{(0)}\Big]\nonumber\\
&+
  \frac{G}{c^3}\sum^2_{a=1} m_a (\vec{e}_{(0)}\cdot\vec{v}_a(t^{\ast}))
  \Bigg\{-\frac{4}{r^3_a}(\vec{r}_a\cdot\vec{v}_a(t^{\ast})\tau 
 +
  \bigg[-\frac{6}{r^5_a}(\vec{r}_a\cdot\vec{v}_a(t^{\ast}))^2 \nonumber\\
&+
  \frac{2}{r^3_a}v^2_a(t^{\ast})\bigg]\tau^2\Bigg\}v^i_a(t^{\ast})\nonumber\\
&+
  \frac{G}{c^4}\sum^2_{a=1} m_a\Bigg\{\bigg[\frac{6}{r^5_a}(\vec{r}_a\cdot\vec{v}_a(t^{\ast}))^3
 -
  \frac{4}{r^3_a}(\vec{r}_a\cdot\vec{v}_a(t^{\ast}))v^2_a(t^{\ast})\bigg]\tau\nonumber\\
&+
  \bigg[\frac{15}{r^7_a}(\vec{r}_a\cdot\vec{v}_a(t^{\ast}))^4
 -
  \frac{15}{r^5_a}(\vec{r}_a\cdot\vec{v}_a(t^{\ast}))^2v^2_a(t^{\ast})
 +
  \frac{2}{r^3_a}v^4_a(t^{\ast})\bigg]\tau^2\nonumber\\
&-
  4 v^2_a(t^{\ast})\bigg[\frac{(\vec{r}_a\cdot\vec{v}_a(t^{\ast}))}{r^3_a}\tau
 +
  \bigg(\frac{3}{2}\frac{(\vec{r}_a\cdot\vec{v}_a(t^{\ast}))^2}{r^5_a}
 -
  \frac{1}{2}\frac{v^2_a(t^{\ast})}{r^3_a}\bigg)\tau^2\Bigg\}l^i_{(0)}
\label{Perturbation12}
\end{align}

Considering that in the present work we compute the post-linear light deflection up to the order $G^2/c^4$, 
we must retain the linear light deflection terms up to the order $G/c^4$. Notice that the linear terms 
of the order $G/c^4$ are of the same order as the post-linear terms of the order $G^2/c^4$ since for a system of 
bounded point-like masses, the virial theorem applies (i.e. $v^2_a \sim G/d$) and, considering 
that the terms of the order $G/c^4$ are also terms in $v^2_a$, it is easy to see that these terms are of the same order 
as the post-linear terms of the order $G^2/c^4$.
Linear light deflection terms of course we get from the perturbation $\delta l^i_{(1)\mathrm{II}}(\tau)$ too.
 To obtain the perturbation $\delta l^i_{(1)\mathrm{II}}(\tau)$ we have to evaluate the integrals in the expression 
above.
Taking into account that we are only interested in the angle of light deflection to the order $G/c^4$, we
need only retain the terms of the order $G/c^2$ and $G/c^3$, since the expression for the
light deflection angle (Eq.~\eqref{motion1}) contains a further factor $1/c$.
After performing the integration of the expression above and retaining only terms of the order $G/c^2$ 
and $G/c^3$, we finally find
\begin{align}
&\delta l^i_{(1)\mathrm{II}}(\tau)=-6 \frac{G}{c^2} \sum^2_{a=1} m_a \bigg[(\vec{e}_{(0)}\cdot\vec{v}_{a}(t^{\ast}))F_{a2}
+
 \Big[\vec{\xi}\cdot\vec{v}_a(t^{\ast})-\vec{x}_a(t^{\ast})\cdot\vec{v}_a(t^{\ast})\Big]F_{a3}\bigg]
 [\xi^i-x^i_a(t^{\ast})\big]\nonumber\\
&-
  4\frac{G}{c^2} \sum^2_{a=1}\frac{m_a}{r^3_a} \bigg[(\vec{e}_{(0)}\cdot\vec{v}_{a}(t^{\ast}))c\,\tau
 +
  \Big[\vec{\xi}\cdot\vec{v}_a(t^{\ast})-\vec{x}_a(t^{\ast})\cdot\vec{v}_a(t^{\ast})\Big]\bigg]\tau l^i_{(0)}\nonumber\\
&+
  2 \frac{G}{c^2} \sum^2_{a=1} m_a \Bigg\{A_a+\frac{2}{r^3_a}\bigg[(\vec{e}_{(0)}\cdot\vec{v}_{a}(t^{\ast}))c\,\tau 
 +
  \Big[\vec{\xi}\cdot\vec{v}_a(t^{\ast})-\vec{x}_a(t^{\ast})\cdot\vec{v}_a(t^{\ast})\Big]\bigg]\tau\Bigg\}v^i_a(t^{\ast})\nonumber\\
&+
  4\frac{G}{c^2} \sum^2_{a=1} \frac{m_a}{r^3_a}\bigg[(\vec{e}_{(0)}\cdot\vec{v}_{a}(t^{\ast}))c\,\tau
 +
  \Big[\vec{\xi}\cdot\vec{v}_a(t^{\ast})-\vec{x}_a(t^{\ast})\cdot\vec{v}_a(t^{\ast})\Big]\bigg]\tau v^i_a(t^{\ast})\nonumber\\
&+
  12\frac{G}{c^3} \sum^2_{a=1}m_a(\vec{e}_{(0)}\cdot\vec{v}_{a}(t^{\ast}))\bigg[(\vec{e}_{(0)}\cdot\vec{v}_{a}(t^{\ast}))F_{a2}
 +
  \Big[\vec{\xi}\cdot\vec{v}_a(t^{\ast})-\vec{x}_a(t^{\ast})\cdot\vec{v}_a(t^{\ast})\Big]F_{a3}\bigg]\big[\xi^i-x^i_a(t^{\ast})\big]
  \nonumber\\
&-
  \frac{G}{c^3} \sum^2_{a=1}m_a\bigg[15(\vec{e}_{(0)}\cdot\vec{v}_{a}(t^{\ast}))^2G_{a2}
 +
  30(\vec{e}_{(0)}\cdot\vec{v}_{a}(t^{\ast}))\Big[\vec{\xi}\cdot\vec{v}_a(t^{\ast})-\vec{x}_a(t^{\ast})\cdot\vec{v}_a(t^{\ast})\Big]G_{a3}  \nonumber\\
&+
  15\Big[\vec{\xi}\cdot\vec{v}_a(t^{\ast})-\vec{x}_a(t^{\ast})\cdot\vec{v}_a(t^{\ast})\Big]^2G_{a4}- 3 v^2_a(t^{\ast})F_{a2}\bigg]
  \big[\xi^i-x^i_a(t^{\ast})\big]\nonumber\\ 
&-
  6\frac{G}{c^3} \sum^2_{a=1}m_a \bigg[(\vec{e}_{(0)}\cdot\vec{v}_a(t^{\ast}))F_{a1}
 +
  \Big[\vec{\xi}\cdot\vec{v}_a(t^{\ast})-\vec{x}_a(t^{\ast})\cdot\vec{v}_a(t^{\ast})\Big]F_{a2}\bigg]l^i_{(0)}\nonumber\\
&+
  6\frac{G}{c^3} \sum^2_{a=1}m_a \bigg[(\vec{e}_{(0)}\cdot\vec{v}_{a}(t^{\ast}))F_{a1}
 +
  \Big[\vec{\xi}\cdot\vec{v}_a(t^{\ast})-\vec{x}_a(t^{\ast})\cdot\vec{v}_a(t^{\ast})\Big]F_{a2}\bigg]v^i_a(t^{\ast})\nonumber\\ 
&-
  4\frac{G}{c^3} \sum^2_{a=1}m_a (\vec{e}_{(0)}\cdot\vec{v}_{a}(t^{\ast}))\Bigg\{A_a 
 +
  \frac{1}{r^3_a}\bigg[(\vec{e}_{(0)}\cdot\vec{v}_{a}(t^{\ast}))c\,\tau \nonumber\\ 
&+    
 \Big[\vec{\xi}\cdot\vec{v}_a(t^{\ast})-\vec{x}_a(t^{\ast})\cdot\vec{v}_a(t^{\ast})\Big]\bigg]\tau \Bigg\}v^i_a(t^{\ast}),\label{Perturbed2} 
\end{align}
where the functions $A_a$, $F_{a1}$, $F_{a2}$, $F_{a3}$ are given in Section \ref{lightlingrav}  by Eqs~ \eqref{Aa}-\eqref{Fa4}.
The functions $G_{a2}$, $G_{a3}$ and $G_{a4}$ are given by

\begin{align}
&G_{a2}=\frac{1}{15\,r^5_a\,R^3_a}\Bigg\{\bigg[3\,r^8_a(0,t^{\ast})\,r_a 
 - 
  8\,(\vec{e}_{(0)}\cdot\vec{x}_a(t^{\ast}))\,r^8_a(0,t^{\ast})\nonumber\\ 
&+ 
  6\,{(\vec{e}_{(0)}\cdot\vec{x}_a(t^{\ast}))}^2\,r^6_a(0,t^{\ast})\,r_a
 - 
  {(\vec{e}_{(0)}\cdot\vec{x}_a(t^{\ast}))}^4\,r^4_a(0,t^{\ast})\,r_a\bigg] \nonumber\\
 &- 
   \bigg[ 12\,(\vec{e}_{(0)}\cdot\vec{x}_a(t^{\ast}))\,r^6_a(0,t^{\ast})\,r_a 
  - 
   40\,{(\vec{e}_{(0)}\cdot\vec{x}_a(t^{\ast}))}^2\,r^6_a(0,t^{\ast}) \nonumber\\
 &+ 
   24\,{(\vec{e}_{(0)}\cdot\vec{x}_a(t^{\ast}))}^3\,r^4_a(0,t^{\ast})\,r_a
  - 
   4\,{(\vec{e}_{(0)}\cdot\vec{x}_a(t^{\ast}))}^5\,r^2_a(0,t^{\ast})\,r_a\bigg] \,c\,\tau \nonumber\\
&-
  \bigg[ -6\,r^6_a(0,t^{\ast})\,r_a 
 +
  20\,(\vec{e}_{(0)}\cdot\vec{x}_a(t^{\ast}))\,r^6_a(0,t^{\ast})
 - 
  24\,{(\vec{e}_{(0)}\cdot\vec{x}_a(t^{\ast}))}^2\,r^4_a(0,t^{\ast})\,r_a \nonumber\\
&+ 
  60\,{(\vec{e}_{(0)}\cdot\vec{x}_a(t^{\ast}))}^3\,r^4_a(0,t^{\ast})
 - 
  22\,{(\vec{e}_{(0)}\cdot\vec{x}_a(t^{\ast}))}^4\,r^2_a(0,t^{\ast})\,r_a 
 + 
  4\,{(\vec{e}_{(0)}\cdot\vec{x}_a(t^{\ast}))}^6\,r_a\bigg] \,c^2\,\tau^2 \nonumber\\
 &- 
  \bigg[12\,(\vec{e}_{(0)}\cdot\vec{x}_a(t^{\ast}))\,r^4_a(0,t^{\ast})\,r_a 
 - 
  60\,{(\vec{e}_{(0)}\cdot\vec{x}_a(t^{\ast}))}^2\,r^4_a(0,t^{\ast})\nonumber\\
 &+ 
  24\,{(\vec{e}_{(0)}\cdot\vec{x}_a(t^{\ast}))}^3\,r^2_a(0,t^{\ast})\,r_a
 - 
  20\,{(\vec{e}_{(0)}\cdot\vec{x}_a(t^{\ast}))}^4\,r^2_a(0,t^{\ast}) 
 - 
  4\,{(\vec{e}_{(0)}\cdot\vec{x}_a(t^{\ast}))}^5\,r_a \bigg] \,c^3\,\tau^3 \nonumber\\
&- 
  \bigg[ -3\,r^4_a(0,t^{\ast})\,r_a
 + 
  15\,(\vec{e}_{(0)}\cdot\vec{x}_a(t^{\ast}))\,r^4_a(0,t^{\ast})
 - 
  6\,{(\vec{e}_{(0)}\cdot\vec{x}_a(t^{\ast}))}^2\,r^2_a(0,t^{\ast})\,r_a \nonumber\\
&+ 
  30\,{(\vec{e}_{(0)}\cdot\vec{x}_a(t^{\ast}))}^3\,r^2_a(0,t^{\ast}) 
 + 
  {(\vec{e}_{(0)}\cdot\vec{x}_a(t^{\ast}))}^4\,r_a 
 - 
  5\,{(\vec{e}_{(0)}\cdot\vec{x}_a(t^{\ast}))}^5\bigg] \,c^4\,\tau^4 \nonumber\\
&- 
  \bigg[-3\,r^4_a(0,t^{\ast}) 
 - 
  6\,{(\vec{e}_{(0)}\cdot\vec{x}_a(t^{\ast}))}^2\,r^2_a(0,t^{\ast})
 + 
  {(\vec{e}_{(0)}\cdot\vec{x}_a(t^{\ast}))}^4\bigg] \,c^5\,\tau^5\Bigg\},\label{Ga2}
\end{align}

\begin{align}
&G_{a3}=\frac{1}{15\,r^5_a\,R^3_a}\Bigg\{-2\,r^4_{a}(0,t^{\ast})\,\bigg[r^4_{a}(0,t^{\ast}) 
- 
 {(\vec{e}_{(0)}\cdot\vec{x}_{a}(t^{\ast}))}^3\,r_a \nonumber\\
&+ 
 3\,(\vec{e}_{(0)}\cdot\vec{x}_{a}(t^{\ast}))\,r^2_{a}(0,t^{\ast})\, 
 \Big[-r_a + (\vec{e}_{(0)}\cdot\vec{x}_{a}(t^{\ast}))\,\Big]\bigg]\nonumber\\
&+ 
  2\,(\vec{e}_{(0)}\cdot\vec{x}_{a}(t^{\ast}))\,r^2_{a}(0,t^{\ast})\,
  \bigg[5\,r^4_{a}(0,t^{\ast}) 
 - 
  4\,{(\vec{e}_{(0)}\cdot\vec{x}_{a}(t^{\ast}))}^3\,r_a\nonumber\\
 &+ 
  3\,(\vec{e}_{(0)}\cdot\vec{x}_{a}(t^{\ast}))\,r^2_{a}(0,t^{\ast})\,\Big[-4\,r_a 
+ 
 5\,(\vec{e}_{(0)}\cdot\vec{x}_{a}(t^{\ast}))\Big]  \bigg]\,c\,\tau \nonumber\\ 
&- 
  \bigg[5\,r^6_{a}(0,t^{\ast}) 
 - 
  8\,{(\vec{e}_{(0)}\cdot\vec{x}_{a}(t^{\ast}))}^5\,r_a 
 + 
  6\,(\vec{e}_{(0)}\cdot\vec{x}_{a}(t^{\ast}))\,r^4_{a}(0,t^{\ast})\,\Big[-2\,r_a \nonumber\\
&+ 
  5\,(\vec{e}_{(0)}\cdot\vec{x}_{a}(t^{\ast}))\, \Big]  
 + 
  {(\vec{e}_{(0)}\cdot\vec{x}_{a}(t^{\ast}))}^3\,r^6_{a}(0,t^{\ast})\Big[-28\,r_a 
 + 
  45\,(\vec{e}_{(0)}\cdot\vec{x}_{a}(t^{\ast}))\Big]\bigg]\,c^2\tau^2 \nonumber\\
&+ 
  (\vec{e}_{(0)}\cdot\vec{x}_{a}(t^{\ast}))\,\bigg[3\,r^2_{a}(0,t^{\ast}) 
 + 
  {(\vec{e}_{(0)}\cdot\vec{x}_{a}(t^{\ast}))}^2\bigg] \,\bigg[5\,r^2_{a}(0,t^{\ast}) 
 + 
  (\vec{e}_{(0)}\cdot\vec{x}_{a}(t^{\ast}))\,\Big[-8\,r_a\nonumber\\
 &+ 
  15\,(\vec{e}_{(0)}\cdot\vec{x}_{a}(t^{\ast}))\, \Big]  \bigg] \,c^3\,\tau^3\nonumber\\
&- 
  (\vec{e}_{(0)}\cdot\vec{x}_{a}(t^{\ast}))\,\bigg[-r_a 
 + 
  5\,(\vec{e}_{(0)}\cdot\vec{x}_{a}(t^{\ast}))\, \bigg] \,\bigg[3\,r^2_{a}(0,t^{\ast}) 
 + 
  {(\vec{e}_{(0)}\cdot\vec{x}_{a}(t^{\ast}))}^2\bigg]\,c^4\,\tau^4\nonumber\\
&+ 
  2\,(\vec{e}_{(0)}\cdot\vec{x}_{a}(t^{\ast}))\,\,
     \bigg[3\,r^2_{a}(0,t^{\ast}) + 
       {(\vec{e}_{(0)}\cdot\vec{x}_{a}(t^{\ast}))}^2\bigg]\,c^5\tau^5\Bigg\},\label{Ga3}
\end{align}

\begin{align}
&G_{a4}=\frac{1}{15\,r^5_a\,R^3_a}\Bigg\{2\,r^4_{a}(0,t^{\ast})\,\bigg[ r^2_{a}(0,t^{\ast})\,\Big[ r_a 
 - 
  3\,(\vec{e}_{(0)}\cdot\vec{x}_{a}(t^{\ast}))\Big]  
 + 
  {(\vec{e}_{(0)}\cdot\vec{x}_{a}(t^{\ast}))}^2 \Big[ 3\,r_a \nonumber\\
&- 
  (\vec{e}_{(0)}\cdot\vec{x}_{a}(t^{\ast}))\Big]\bigg]\nonumber\\
&+ 
  2\,(\vec{e}_{(0)}\cdot\vec{x}_{a}(t^{\ast}))\,r^2_{a}(0,t^{\ast})\,\bigg[ {(\vec{e}_{(0)}\cdot\vec{x}_{a}(t^{\ast}))}^2\,
  \Big[ -12\,r_a 
 + 
  5\,(\vec{e}_{(0)}\cdot\vec{x}_{a}(t^{\ast}))\Big]\nonumber\\ 
&+ 
  r^2_{a}(0,t^{\ast})\,\Big[ -4\,r_a 
 + 
  15\,(\vec{e}_{(0)}\cdot\vec{x}_{a}(t^{\ast}))\Big]\bigg] \,c\,\tau\nonumber\\
&+ 
  \bigg[r^2_{a}(0,t^{\ast}) 
 + 
  3\,{(\vec{e}_{(0)}\cdot\vec{x}_{a}(t^{\ast}))}^2\bigg]\,\bigg[ r^2_{a}(0,t^{\ast})\,\Big[ 4\,r_a 
 - 
  15\,(\vec{e}_{(0)}\cdot\vec{x}_{a}(t^{\ast}))\Big]\nonumber\\  
&+ 
  {(\vec{e}_{(0)}\cdot\vec{x}_{a}(t^{\ast}))}^2 \Big[ 8\,r_a 
 - 
  5\,(\vec{e}_{(0)}\cdot\vec{x}_{a}(t^{\ast}))\Big]\bigg] \,c^2\,\tau^2 \nonumber\\ 
&+ 
  \bigg[ r^2_{a}(0,t^{\ast}) 
 + 
  3\,{(\vec{e}_{(0)}\cdot\vec{x}_{a}(t^{\ast}))}^2\bigg] \bigg[ 5\,r^2_{a}(0,t^{\ast}) 
 + 
  (\vec{e}_{(0)}\cdot\vec{x}_{a}(t^{\ast}))\,\Big[-8\,r_a \nonumber\\
 &+ 
  15\,(\vec{e}_{(0)}\cdot\vec{x}_{a}(t^{\ast}))\Big]\bigg] \,c^3\,\tau^3\nonumber\\
&+ 
  2\,\bigg[ r_a - 
       5\,(\vec{e}_{(0)}\cdot\vec{x}_{a}(t^{\ast}))\, \bigg] \,
     \bigg[ r^2_{a}(0,t^{\ast}) + 
       3\,{(\vec{e}_{(0)}\cdot\vec{x}_{a}(t^{\ast}))}^2\bigg]\,c^4\,\tau^4\nonumber\\
&+ 
  2\,\bigg[ r^2_{a}(0,t^{\ast}) 
+ 
  3\,{(\vec{e}_{(0)}\cdot\vec{x}_{a}(t^{\ast}))}^2\bigg] \,c^5\,\tau^5 \Bigg\}.\label{Ga4}
\end{align}

After introducing the perturbation $\delta l^i_{(1)\mathrm{II}}(\tau)$ into Eq.~\eqref{deflection} and computing
the limit for $\tau \rightarrow \infty$, we obtain

\begin{align}
\alpha^i_{(1)\mathrm{II}}&=\lim_{\tau\to\infty}\Big[\frac{1}{c}P^i_q \delta l^q_{(1)\mathrm{II}}(\tau)\Big]\nonumber\\
&=-4 \frac{G}{c^3} \sum^{2}_{a=1} \frac{m_a}{R_a}(\vec{e}_{(0)}\cdot\vec{v}_a(t^{\ast}))
\big[\xi^i-P^i_qx^q_a(t^{\ast})\big]\nonumber\\
&-8\frac{G}{c^3} \sum^{2}_{a=1} \frac{m_a}{R^2_a}(\vec{e}_{(0)}\cdot\vec{x}_a(t^{\ast}))
\Big[(\vec{e}_{(0)}\cdot \vec{x}_a(t^{\ast}))(\vec{e}_{(0)}\cdot \vec{v}_a(t^{\ast}))+\vec{\xi}\cdot \vec{v}_a(t^{\ast})
-\vec{x}_a(t^{\ast})\cdot \vec{v}_a(t^{\ast})\Big]\nonumber\\
&\big[\xi^i-P^i_qx^q_a(t^{\ast})\big]\nonumber\\
&+4\frac{G}{c^3} \sum^{2}_{a=1} \frac{m_a}{R_a}(\vec{e}_{(0)}\cdot\vec{x}_a(t^{\ast}))P^i_qv^q_a(t^{\ast})\nonumber\\
&+2\frac{G}{c^4} \sum^{2}_{a=1} \frac{m_a}{R_a}v^2_a(t^{\ast})[\big[\xi^i-P^i_qx^q_a(t^{\ast})\big]\nonumber\\
&+4\frac{G}{c^4} \sum^{2}_{a=1} \frac{m_a}{R^2_a}v^2_a(t^{\ast})(\vec{e}_{(0)}\cdot\vec{x}_a(t^{\ast}))^2
\big[\xi^i-P^i_qx^q_a(t^{\ast})\big]\nonumber\\
&+2\frac{G}{c^4} \sum^{2}_{a=1} \frac{m_a}{R_a}(\vec{e}_{(0)}\cdot\vec{v}_a(t^{\ast}))^2
\big[\xi^i-P^i_qx^q_a(t^{\ast})\big]\nonumber\\
&-8\frac{G}{c^4} \sum^{2}_{a=1} \frac{m_a}{R^2_a}(\vec{e}_{(0)}\cdot\vec{x}_a(t^{\ast}))^2(\vec{e}_{(0)}\cdot\vec{v}_a(t^{\ast}))^2
\big[\xi^i-P^i_qx^q_a(t^{\ast})\big]\nonumber\\
&-16\frac{G}{c^4} \sum^{2}_{a=1} \frac{m_a}{R^3_a}(\vec{e}_{(0)}\cdot\vec{x}_a(t^{\ast}))^4(\vec{e}_{(0)}\cdot\vec{v}_a(t^{\ast}))^2
\big[\xi^i-P^i_qx^q_a(t^{\ast})\big]\nonumber\\
&-8\frac{G}{c^4} \sum^{2}_{a=1} \frac{m_a}{R^2_a}(\vec{e}_{(0)}\cdot\vec{x}_a(t^{\ast}))(\vec{e}_{(0)}\cdot\vec{v}_a(t^{\ast}))
\Big[\vec{\xi}\cdot\vec{v}_a(t^{\ast})-\vec{x}_a(t^{\ast})\cdot \vec{v}_a(t^{\ast})\Big]
\big[\xi^i-P^i_qx^q_a(t^{\ast})\big]\nonumber\\
&-32\frac{G}{c^4} \sum^{2}_{a=1} \frac{m_a}{R^3_a}(\vec{e}_{(0)}\cdot\vec{x}_a(t^{\ast}))^3(\vec{e}_{(0)}\cdot\vec{v}_a(t^{\ast}))
\Big[\vec{\xi}\cdot\vec{v}_a(t^{\ast})-\vec{x}_a(t^{\ast})\cdot\vec{v}_a(t^{\ast})\Big]            
\big[\xi^i-P^i_qx^q_a(t^{\ast})\big]\nonumber\\
&-4\frac{G}{c^4} \sum^{2}_{a=1} \frac{m_a}{R^2_a}\Big[\vec{\xi}\cdot\vec{v}_a(t^{\ast})-\vec{x}_a(t^{\ast})\cdot \vec{v}_a(t^{\ast})\Big]^2
\big[\xi^i-P^i_qx^q_a(t^{\ast})\big]\nonumber\\
&-16\frac{G}{c^4} \sum^{2}_{a=1} \frac{m_a}{R^3_a}(\vec{e}_{(0)}\cdot\vec{x}_a(t^{\ast}))^2
\Big[\vec{\xi}\cdot\vec{v}_a(t^{\ast})-\vec{x}_a(t^{\ast})\cdot \vec{v}_a(t^{\ast})\Big]^2
\big[\xi^i-P^i_qx^q_a(t^{\ast})\big]\nonumber\\
&+4\frac{G}{c^4} \sum^{2}_{a=1} \frac{m_a}{R_a}(\vec{e}_{(0)}\cdot\vec{x}_a(t^{\ast}))(\vec{e}_{(0)}\cdot\vec{v}_a(t^{\ast}))P^i_qv^q_a(t^{\ast})
\nonumber\\
&+8\frac{G}{c^4} \sum^{2}_{a=1} \frac{m_a}{R^2_a}(\vec{e}_{(0)}\cdot\vec{x}_a(t^{\ast}))^2
\Bigg\{\Big[\vec{\xi}\cdot\vec{v}_a(t^{\ast})-\vec{x}_a(t^{\ast})\cdot \vec{v}_a(t^{\ast})\Big]\nonumber\\
&+(\vec{e}_{(0)}\cdot\vec{x}_a(t^{\ast}))(\vec{e}_{(0)}\cdot\vec{v}_a(t^{\ast}))\Bigg\}P^i_qv^q_a(t^{\ast})\nonumber\\
&+4\frac{G}{c^4} \sum^{2}_{a=1} \frac{m_a}{R_a}\Big[\vec{\xi}\cdot\vec{v}_a(t^{\ast})-\vec{x}_a(t^{\ast})\cdot \vec{v}_a(t^{\ast})\Big]
P^i_qv^q_a(t^{\ast}),\label{motion1}
\end{align} 
where the quantity $R_a$ is given by Eq.~\eqref{distanceB}.
\subsection{The post-linear light deflection and the motion of the masses}\label{postlinlightmotion}
The correction terms to the post-linear light deflection to the order $G^2/c^4$ resulting from the Taylor
expansion of Eq.~\eqref{LinPert.} are given by
\begin{align}
\alpha^i_{(2)\mathrm{III}}&=-3\frac{G}{c}\sum^{2}_{a=1} m_a \int_{-\infty}^{\infty} d \tau \, \frac{\tau^2}{r^5_a}\Big[c \tau (\vec{e}_{(0)}\cdot \vec{a}_a(t^{\ast}))+\vec{\xi}\cdot\vec{a}_a(t^{\ast})
-\vec{x}_a(t^{\ast})\cdot \vec{a}_a(t^{\ast})\Big]
\big[\xi^i-P^i_qx^q_a(t^{\ast})\big]\nonumber\\
&+\frac{G}{c}\sum^{2}_{a=1} m_a \int_{-\infty}^{\infty} d \tau \,\frac{\tau^2}{r^3_a}P^i_qa^q_a(t^{\ast})\nonumber\\
&+4\frac{G}{c^2}\sum^{2}_{a=1} m_a \int_{-\infty}^{\infty} d \tau \,\frac{\tau}{r^3_a}
(\vec{e}_{(0)}\cdot \vec{a}_a(t^{\ast}))\big[\xi^i-P^i_qx^q_a(t^{\ast})\big].\label{motion2}
\end{align}
In the preceding equation the second integral diverges. Therefore we resort to a Taylor expansion of its
integrand about the origin of the coordinate system $\vec{x}_a=0$, up to second order.
Then only the first term of the Taylor expansion is a divergent integral and it is given by
\begin{align*}
\frac{G}{c}\sum^2_{a=1} m_a\int^{\infty}_{-\infty} d \tau \tau^2 \frac{1}{\big[c^2 \tau^2+\xi^2\big]^{3/2}} 
P^i_q a^q_a(t^{\ast}).
\end{align*}
Because in this case we do not need to take into account the 1PN-corrections
in the positions of the masses we can assume that the origin of the coordinate system which is located at the
1PN-centre of mass coincides with the position of the Newtonian centre of mass.
Taking into account the consequence of the Newtonian centre of mass theorem 
\begin{align*}
\sum^{2}_{a=1} m_a \vec{a}_{a}=0,
\end{align*}
it is easy to see that the divergent integral vanishes.
After performing the integration with respect to the parameter $\tau$ we find:
\begin{align}
\alpha^i_{(2)\mathrm{III}}&=2\frac{G}{c^4}\sum^{2}_{a=1} \frac{m_a}{R_a} (\vec{e}_{(0)}\cdot\vec{x}_a(t^{\ast}))
(\vec{e}_{(0)}\cdot\vec{a}_a(t^{\ast}))\big[\xi^i-P^i_qx^q_a(t^{\ast})\big]\nonumber\\
&-4\frac{G}{c^4}\sum^{2}_{a=1} \frac{m_a}{R^2_a}(\vec{e}_{(0)}\cdot\vec{x}_a(t^{\ast}))^3
(\vec{e}_{(0)}\cdot\vec{a}_a(t^{\ast}))\big[\xi^i-P^i_qx^q_a(t^{\ast})\big]\nonumber\\
&-2\frac{G}{c^4}\sum^{2}_{a=1} \frac{m_a}{R_a}\Big[\vec{\xi}\cdot\vec{a}_a(t^{\ast})
-\vec{x}_a(t^{\ast})\cdot\vec{a}_a(t^{\ast})\Big]\big[\xi^i-P^i_qx^q_a(t^{\ast})\big]\nonumber\\
&-4\frac{G}{c^4}\sum^{2}_{a=1} \frac{m_a}{R^2_a}(\vec{e}_{(0)}\cdot\vec{x}_a(t^{\ast}))^2
\Big[\vec{\xi}\cdot\vec{a}_a(t^{\ast})-\vec{x}_a(t^{\ast})\cdot\vec{a}_a(t^{\ast})\Big]
\big[\xi^i-P^i_qx^q_a(t^{\ast})\big]\nonumber\\
&+\frac{G}{c^4}\sum^{2}_{a=1} m_a \Bigg\{2\frac{\vec{\xi}\cdot\vec{x}_a(t^{\ast})}{\xi^2}
-\frac{x^2_a(t^{\ast})}{\xi^2}+3\frac{(\vec{e}_{(0)}\cdot\vec{x}_a(t^{\ast}))^2}{\xi^2}
+2\frac{(\vec{\xi}\cdot \vec{x}_a(t^{\ast}))^2}{\xi^4}\Bigg\}P^i_qa^q_a(t^{\ast}),\label{Defl4a}
\end{align}
where the quantity $R_a$ is given by Eq.~\eqref{distanceB}.
Note that in Eq.~\eqref{motion2} two of the three integrals were exactly integrated, so that the resulting expression
given by Eq.~\eqref{Defl4a} is a combination of exact terms with a term which is represented as an expansion
in powers of $(x_a(t^{\ast})/\xi)$.
In view of further applications and for the sake of uniformity we perform the Taylor expansion of the exact 
terms about the origin of the coordinate system $\vec{x}_a=0$ up to second order.
After performing the Taylor expansion of the exact terms in Eq.~\eqref{Defl4a} about the origin of the coordinate system $\vec{x}_a=0$,
up to second order, we find
\begin{align}
\alpha^{i}_{(2)\rm{III}}&=2 \frac{G}{c^4} \sum^{2}_{a=1} m_a (\vec{e}_{(0)}\cdot \vec{x}_a(t^{\ast}))(\vec{e}_{(0)}\cdot \vec{a}_a(t^{\ast}))
\Bigg\{\frac{1}{\xi^2}+2 \frac{(\vec{\xi}\cdot \vec{x}_a(t^{\ast}))}{\xi^4}-\frac{x^2_a(t^{\ast})}{\xi^4}
+\frac{(\vec{e}_{(0)}\cdot \vec{x}_a(t^{\ast}))^2}{\xi^4}\nonumber\\
&+4 \frac{(\vec{\xi}\cdot \vec{x}_a(t^{\ast}))^2}{\xi^6}
\Bigg\}\big[\xi^i-P^i_qx^q_a(t^{\ast})\big]\nonumber\\
&-4 \frac{G}{c^4}\sum^{2}_{a=1} m_a (\vec{e}_{(0)}\cdot \vec{x}_a(t^{\ast}))^3(\vec{e}_{(0)}\cdot \vec{a}_a(t^{\ast}))\Bigg\{\frac{1}{\xi^4}
+4\frac{\vec{\xi}\cdot \vec{x}_a(t^{\ast})}{\xi^6}-2 \frac{x^2_a(t^{\ast})}{\xi^6}+2 \frac{(\vec{e}_{(0)}\cdot\vec{x}_a(t^{\ast}))^2}{\xi^6}\nonumber\\
&+12\frac{(\vec{\xi}\cdot\vec{x}_a(t^{\ast}))^2}{\xi^8}\Bigg\}\big[\xi^i-P^i_qx^q_a(t^{\ast})\big]\nonumber\\
&-2 \frac{G}{c^4}\sum^{2}_{a=1} m_a \Big[\vec{\xi}\cdot\vec{a}_a(t^{\ast})-\vec{x}_a(t^{\ast})\cdot\vec{a}_a(t^{\ast})\Big]
\Bigg\{\frac{1}{\xi^2}+2 \frac{(\vec{\xi}\cdot \vec{x}_a(t^{\ast}))}{\xi^4}-\frac{x^2_a(t^{\ast})}{\xi^4}
+\frac{(\vec{e}_{(0)}\cdot \vec{x}_a(t^{\ast}))^2}{\xi^4}\nonumber\\
&+4 \frac{(\vec{\xi}\cdot \vec{x}_a(t^{\ast}))^2}{\xi^6}
\Bigg\}\big[\xi^i-P^i_qx^q_a(t^{\ast})\big]\nonumber\\
&-4\frac{G}{c^4}\sum^{2}_{a=1} m_a (\vec{e}_{(0)}\cdot\vec{x}_a(t^{\ast}))^2
\Big[\vec{\xi}\cdot\vec{a}_a(t^{\ast})-\vec{x}_a(t^{\ast})\cdot\vec{a}_a(t^{\ast})\Big]
\Bigg\{\frac{1}{\xi^4}
+4\frac{\vec{\xi}\cdot \vec{x}_a(t^{\ast})}{\xi^6}-2 \frac{x^2_a(t^{\ast})}{\xi^6}\nonumber\\
&+2 \frac{(\vec{e}_{(0)}\cdot\vec{x}_a(t^{\ast}))^2}{\xi^6}+12\frac{(\vec{\xi}\cdot\vec{x}_a(t^{\ast}))^2}{\xi^8}\Bigg\}
\big[\xi^i-P^i_qx^q_a(t^{\ast})\big]\nonumber\\
&+\frac{G}{c^4}\sum^{2}_{a=1} m_a \Bigg\{2\frac{\vec{\xi}\cdot\vec{x}_a(t^{\ast})}{\xi^2}-\frac{x^2_a(t^{\ast})}{\xi^2}
+3 \frac{(\vec{e}_{(0)}\cdot \vec{x}_a(t^{\ast}))^2}{\xi^2}+2\frac{(\vec{\xi}\cdot\vec{x}_a(t^{\ast}))^2}{\xi^4}\Bigg\}P^i_qa^q_a(t^{\ast}).\label{Defl4}
\end{align}
To replace the accelerations in the preceding equation by functions of the positions
we shall use the Newtonian equations of motion. After replacing the accelerations in Eq.~\eqref{Defl4} and expressing
the positions of the masses by their centre of mass frame coordinates without considering the 1PN-corrections we find,

\begin{align}
\alpha^{i}_{(2)\mathrm{III}}&=
\frac{G^2 m_1 m_2}{c^4 r^2_{12}} \Bigg\{2\,(\vec{e}_{\xi} \cdot \vec{n}_{12})
 -
  2\,X_2\bigg[1 
 -
  2\,(\vec{e}_{\xi}\cdot\vec{n}_{12})^2
 -
  3\,(\vec{e}_{(0)} \cdot \vec{n}_{12})^2\bigg]\bigg(\frac{r_{12}}{\xi}\bigg)\nonumber\\
&-
  6\,X^2_2\,(\vec{e}_{\xi} \cdot \vec{n}_{12}) \bigg[1
 -
  \frac{4}{3}\,(\vec{e}_{\xi} \cdot \vec{n}_{12})^2
 -
  3\,(\vec{e}_{(0)} \cdot \vec{n}_{12})^2\bigg]\bigg(\frac{r_{12}}{\xi}\bigg)^2\nonumber\\
&+
  2\,X^3_2\, \bigg[1
 -
  4\,(\vec{e}_{\xi} \cdot \vec{n}_{12})^2 
 -
  3\, (\vec{e}_{(0)} \cdot \vec{n}_{12})^2\bigg]\bigg(\frac{r_{12}}{\xi}\bigg)^3
 +
  \mathcal{O}\bigg[\bigg(\frac{r_{12}}{\xi}\bigg)^4\bigg]\Bigg\}e^i_{\xi}\nonumber\\
&-
  \frac{G^2 m_1 m_2}{c^4 r^2_{12}}\Bigg\{2\, X_2\,(\vec{e}_{\xi} \cdot \vec{n}_{12})\bigg(\frac{r_{12}}{\xi}\bigg)
 -
  2\,X^2_2\,\bigg[1 
 -
  2\,(\vec{e}_{\xi}\cdot\vec{n}_{12})^2
    -
  3\,(\vec{e}_{(0)} \cdot \vec{n}_{12})^2\bigg]\bigg(\frac{r_{12}}{\xi}\bigg)^2\nonumber\\
&-
  6\,X^3_2\,(\vec{e}_{\xi} \cdot \vec{n}_{12}) \bigg[1
 -
  \frac{4}{3}\,(\vec{e}_{\xi} \cdot \vec{n}_{12})^2
 -
  3\,(\vec{e}_{(0)} \cdot \vec{n}_{12})^2\bigg]\bigg(\frac{r_{12}}{\xi}\bigg)^3\nonumber\\
&+
  2\,X^3_2\,\bigg[1
 -
  4\,(\vec{e}_{\xi} \cdot \vec{n}_{12})^2 
 -
  3\, (\vec{e}_{(0)} \cdot \vec{n}_{12})^2\bigg]\bigg(\frac{r_{12}}{\xi}\bigg)^4
 +
  \mathcal{O}\bigg[\bigg(\frac{r_{12}}{\xi}\bigg)^5\bigg]\Bigg\}P^i_qn^q_{12}\nonumber\\
&+
  \frac{G^2 m_1 m_2}{c^4 r^2_{12}}\bigg\{-2\,X_2\,(\vec{e}_{\xi} \cdot \vec{n}_{12})\bigg(\frac{r_{12}}{\xi}\bigg)
 +
  X^2_2\,\bigg[1 
 -
  2\,(\vec{e}_{\xi} \cdot \vec{n}_{12})^2\nonumber\\
&-
  3\,(\vec{e}_{(0)} \cdot \vec{n}_{12})^2 \bigg]\bigg(\frac{r_{12}}{\xi}\bigg)^2 
 +
  2\,X^3_2\,(\vec{e}_{\xi} \cdot \vec{n}_{12})\bigg[1
 -
  \frac{4}{3}(\vec{e}_{\xi} \cdot \vec{n}_{12})^2\nonumber\\
&-
  3\,(\vec{e}_{(0)} \cdot \vec{n}_{12})^2 \bigg]\bigg(\frac{r_{12}}{\xi}\bigg)^3
 +\mathcal{O}\bigg[\bigg(\frac{r_{12}}{\xi}\bigg)^4\bigg]\Bigg\}P^i_q n^q_{12}\nonumber\\ 
&-
  8\,\frac{G^2 m_1 m_2}{c^4 r^2_{12}}X_2(\vec{e}_{(0)} \cdot \vec{n}_{12})^2 \Bigg\{\bigg(\frac{r_{12}}{\xi}\bigg)
 +
  2\,X_2\,(\vec{e}_{\xi} \cdot \vec{n}_{12})\bigg(\frac{r_{12}}{\xi}\bigg)^2 
 +
  \mathcal{O}\bigg[\bigg(\frac{r_{12}}{\xi}\bigg)^3\bigg]\Bigg\}e^i_{\xi}\nonumber\\
&+
  8\,\frac{G^2 m_1 m_2}{c^4 r^2_{12}}X^2_2(\vec{e}_{(0)} \cdot \vec{n}_{12})^2 \Bigg\{\bigg(\frac{r_{12}}{\xi}\bigg)^2
 +
  2\,X_2\,(\vec{e}_{\xi} \cdot \vec{n}_{12})\bigg(\frac{r_{12}}{\xi}\bigg)^3 \nonumber\\ 
&+
  \mathcal{O}\bigg[\bigg(\frac{r_{12}}{\xi}\bigg)^4\bigg]\Bigg\}P^i_q n^q_{12}
 +
  (1 \leftrightarrow 2),
\end{align}
where the quantities $\vec{n}_{12}$, $\vec{v}_{12}$ and $r_{12}$ are taken at the time $t^{\ast}$.

\section{The post-linear light deflection and the perturbed light ray trajectory}\label{trajectory}
If we introduce into the equations for the linear perturbations \eqref{Perturbation11} and \eqref{Perturbation12} 
the expression for the perturbed light ray trajectory, we get additional post-linear light deflection terms. 
To compute these terms we have first to find the expression for the perturbation of the photon's trajectory
that is linear in $G$.
We obtain this perturbation by integrating the expression for the total linear perturbation 
given by Eq.~\eqref{Perturbed3} with respect to the parameter $\tau$.
The expression for the total linear perturbation is obtained by summing up the expressions for the linear
perturbations $\delta l^i_{(1)\mathrm{I}}(\tau)$ and $\delta l^i_{(1)\mathrm{II}}(\tau)$.
Considering that in this paper we compute the post-linear light deflection to the order $G^2/c^4$ we do not need
to retain in the expression resulting from the integration of Eq.~\eqref{Perturbed3} the terms of the order
$G/c^4$, since these terms are related to the post-linear light deflection terms of higher order than $G^2/c^4$.
After performing the integration of Eq.~\eqref{Perturbed3} with regard to $\tau$ and retaining only terms of the order
$\mathcal{O}(G/c^2)$ and $\mathcal{O}(G/c^3)$ we obtain

\begin{align}
&\delta z^{i}_{(1)}(\tau)=-2\frac{G}{c^2}\sum^{2}_{a=1} m_a \mathcal{B}_a \big[\xi^i-x^i_a(t^{\ast})\big] 
-
  2 \frac{G}{c^3}\sum^{2}_{a=1} m_a (\vec{e}_{(0)} \cdot \vec{x}_{a}(t^{\ast})) \mathcal{B}_a l^i_{(0)}\nonumber\\
&-
  2\frac{G}{c^3}\sum^{2}_{a=1} m_a \ln \bigg[\frac{c \tau-\vec{e}_{(0)} \cdot \vec{x}_a(t^{\ast})
+ r_a}{r_a(0,t^{\ast})-\vec{e}_{(0)} \cdot \vec{x}_a(t^{\ast})} \bigg]l^i_{(0)}
+4\frac{G}{c^3}\sum^{2}_{a=1} m_a (\vec{e}_{(0)}\cdot\vec{v}_a(t^{\ast})) \mathcal{B}_a\big[\xi^i-x^i_a(t^{\ast})\big]\nonumber\\
&+
  2\frac{G}{c^3}\sum^{2}_{a=1} m_a (\vec{e}_{(0)} \cdot \vec{x}_{a}(t^{\ast}))\mathcal{B}_av^i_a(t^{\ast})
+
  2 \frac{G}{c^3}\sum^{2}_{a=1} m_a \ln \bigg[\frac{c \tau-\vec{e}_{(0)} \cdot \vec{x}_a(t^{\ast})
+ r_a}{r_a(0,t^{\ast})-\vec{e}_{(0)} \cdot \vec{x}_a(t^{\ast})} \bigg]v^i_a(t^{\ast})\nonumber\\
&-
  6\frac{G}{c^3}\sum^{2}_{a=1} m_a \Big[(\vec{e}_{(0)}\cdot\vec{v}_a(t^{\ast}))\mathcal{F}_{a2}
+(\vec{r}_a(0,t^{\ast})\cdot\vec{v}_a(t^{\ast}))\mathcal{F}_{a3}\Big]\big[\xi^i-x^i_a(t^{\ast})\big]
+
  \mathcal{O}\bigg(\frac{G}{c^4}\bigg),\label{Perturbation3}\nonumber\\
\end{align}
where the functions $\mathcal{B}_a$, $\mathcal{F}_{a2}$ and $\mathcal{F}_{a3}$ are given by

\begin{align}
\mathcal{B}_a&=\frac{1}{R_a}\Big[c \tau+r_a\Big],\nonumber\\
\mathcal{F}_{a2}&=\frac{1}{3r_a R^3_a} \Bigg\{(\vec{e}_{(0)}\cdot\vec{x}_a(t^{\ast}))^2\Big[3r^4_a(0,t^{\ast})
+r^2_aR_a+(\vec{e}_{(0)}\cdot\vec{x}_a(t^{\ast}))^2r^2_a(0,t^{\ast})\Big]\nonumber\\
&+r^2_a(0,t^{\ast})\Big[r^4_a(0,t^{\ast})+r^2_aR_a-5(\vec{e}_{(0)}\cdot\vec{x}_a(t^{\ast}))^2r^2_a(0,t^{\ast})\Big]
+\Big[-2(\vec{e}_{(0)}\cdot\vec{x}_a(t^{\ast}))r^4_a(0,t^{\ast})\nonumber\\
&+\Big(r^2_a(0,t^{\ast})+(\vec{e}_{(0)}\cdot\vec{x}_a(t^{\ast}))^2\Big)r_aR_a
+2(\vec{e}_{(0)}\cdot\vec{x}_a(t^{\ast}))^3 \Big(2r^2_a(0,t^{\ast})-(\vec{e}_{(0)}\cdot\vec{x}_a(t^{\ast}))^2\Big)
\Big]c \tau \Bigg\},\nonumber\\
\mathcal{F}_{a3}&=\frac{1}{3r_aR^2_a}\Bigg\{2(\vec{e}_{(0)}\cdot\vec{x}_a(t^{\ast}))r^2_a(0,t^{\ast})
-\Big[R_a-2(\vec{e}_{(0)}\cdot\vec{x}_a(t^{\ast}))r_a+4(\vec{e}_{(0)}\cdot\vec{x}_a(t^{\ast}))^2\Big]c \tau\nonumber\\
&+2(\vec{e}_{(0)}\cdot\vec{x}_a(t^{\ast}))c^2\tau^2\Bigg\}.\nonumber\\
\end{align}
As our integration constant we have chosen in Eq.~\eqref{Perturbation3}
\begin{align*}
K^i&=2 \frac{G}{c^3} \sum_{a=1}^2 m_a\ln \bigg[2 \Big(r_a(0,t^{\ast})-\vec{e}_{(0)} \cdot \vec{x}_a(t^{\ast}) \Big)\bigg] l^i_{(0)}
\nonumber\\
&-2 \frac{G}{c^3} \sum_{a=1}^2 m_a \ln \bigg[2 \Big(r_a(0,t^{\ast})-\vec{e}_{(0)} \cdot \vec{x}_a(t^{\ast}) \Big)\bigg] v^i_a(t^{\ast})
\nonumber\\
&+\mathcal{O}\bigg(\frac{G}{c^4}\bigg),
\end{align*}
because with this integration constant we recover from our expression for the post-linear light deflection the correct expression  
for the post-linear light deflection in the event that the value of one of the masses is equal to zero 
(i.e. the Epstein-Shapiro post-linear light deflection).

It follows from Eq.~\eqref{deflection} that the expression for the linear light deflection for an observer located at infinity
is given by
\begin{align}
\alpha ^{i}_{(1)}&=\lim_{\tau\to\infty}\Bigg\{\frac{1}{c}P^i_q \delta l^{q}_{(1)}(\tau)\Bigg\},\nonumber\\
&=\lim_{\tau\to\infty}\Bigg\{\frac{1}{c}P^i_q\Big[\delta l^{q}_{(1)\mathrm{I}}(\tau)
+\delta l^{q}_{(1)\mathrm{II}}(\tau)\Big]\Bigg\},    \label{linlightdeflection}
\end{align}
where the perturbations $\delta l^{q}_{(1)\mathrm{I}}(\tau)$ and $\delta l^{q}_{(1)\mathrm{II}}(\tau)$ 
are given by Eqs~\eqref{Perturbation11} and \eqref{Perturbation12}.\\
After introducing the perturbation $\delta \vec{z}_{(1)}$ into the equations for $\delta l^q_{(1)\mathrm{I}}(\tau)$ and
$\delta l^q_{(1)\mathrm{II}}(\tau)$ we get a perturbed linear light deflection.
Because the perturbation $\delta \vec{z}_{(1)}$ is a small quantity compared to $\vec{z}(\tau)_{\mathrm{unpert.}}$,
we can resort to a Taylor expansion of the perturbed linear light deflection about $\delta \vec{z}_{(1)}=0$
in order to get the terms of the perturbed linear light deflection that are quadratic in $G$.
We denote these terms by $\alpha^i_{(2)\mathrm{IV}}$ and they are given by

\begin{align}
\alpha ^{i}_{(2)\mathrm{IV}}&=\lim_{\tau\to\infty}\Bigg\{\frac{1}{c}P^i_q
\bigg[\bigg(\frac{\partial \delta l^{q\mathrm{(Pert.)}}_{(1)\mathrm{I}}}{\partial \delta z^m_{(1)}}\bigg)_{\delta \vec{z}_{(1)}=0} \delta z^m_{(1)}
+\bigg(\frac{\partial \delta l^{q\mathrm{(Pert.)}}_{(1)\mathrm{II}}}{\partial \delta z^m_{(1)}}\bigg)_{\delta \vec{z}_{(1)}=0} \delta z^m_{(1)}
\bigg]\Bigg\}\nonumber\\
&=\frac{G}{c} \sum^2_{a=1} m_a \int^{\infty}_{-\infty}
d \tau \,\Bigg\{\bigg[\frac{6}{r^5_a}(\vec{r}_a\cdot \delta \vec{z}_{(1)})
-6\,\Big[\frac{1}{r^5_a}(\vec{v}_a(t^{\ast})\cdot \delta \vec{z}_{(1)}) \nonumber\\
&-
  \frac{5}{r^7_a}(\vec{r}_a\cdot \vec{v}_a(t^{\ast}))(\vec{r}_a\cdot \delta \vec{z}_{(1)})\Big]\tau
-15\,\Big[\frac{2}{r^7_a}(\vec{r}_a\cdot \vec{v}_a(t^{\ast}))(\vec{v}_a(t^{\ast})\cdot \delta \vec{z}_{(1)})\nonumber\\
&+
  \frac{1}{r^7_a}v^2_a(t^{\ast})(\vec{r}_a\cdot \delta \vec{z}_{(1)})
-
  \frac{7}{r^9_a}(\vec{r}_a\cdot \vec{v}_a(t^{\ast}))^2(\vec{r}_a\cdot \delta \vec{z}_{(1)})\Big]\tau^2\bigg]P^i_qr^q_a\nonumber\\
&-
  \bigg[\frac{2}{r^3_a}
+
  \frac{6}{r^5_a}(\vec{r}_a\cdot \vec{v}_a(t^{\ast}))\tau
-
 3\,\Big[\frac{1}{r^5_a}v^2_a(t^{\ast})
-
\frac{5}{r^7_a}(\vec{r}_a\cdot \vec{v}_a(t^{\ast}))^2\Big]\tau^2\bigg]P^i_q \delta z^q_{(1)}\nonumber\\
&-
  \bigg[\frac{6}{r^5_a}(\vec{r}_a\cdot \delta \vec{z}_{(1)})\tau
-
  3\,\Big[\frac{2}{r^5_a}(\vec{v}_a\cdot \delta \vec{z}_{(1)})
-
  \frac{10}{r^7_a}(\vec{r}_a\cdot \vec{v}_a(t^{\ast}))(\vec{r}_a\cdot \delta \vec{z}_{(1)})\Big]\tau^2\bigg]P^i_qv^q_a(t^{\ast})
\Bigg\}_{|(\rightarrow)}\nonumber\\
&+
  \frac{G}{c^2} \sum^2_{a=1} m_a(\vec{e}_{(0)}\cdot\vec{v}_a(t^{\ast}))\int^{\infty}_{-\infty}d \tau \,
\Bigg\{-6\,\bigg[\frac{2}{r^5_a}(\vec{r}_a\cdot \delta \vec{z}_{(1)})
-
  2\,\Big[\frac{1}{r^5_a}(\vec{v}_a(t^{\ast})\cdot \delta \vec{z}_{(1)})\nonumber\\
&-
  \frac{5}{r^7_a}(\vec{r}_a\cdot \vec{v}_a(t^{\ast}))(\vec{r}_a\cdot \delta \vec{z}_{(1)})\Big]\tau
-
  5\,\Big[\frac{1}{r^7_a}v^2_a(t^{\ast})(\vec{r}_a\cdot \delta \vec{z}_{(1)})
+
  \frac{2}{r^7_a}(\vec{r}_a\cdot \vec{v}_a(t^{\ast}))(\vec{v}_a(t^{\ast})\cdot \delta \vec{z}_{(1)})\nonumber\\
&-
  \frac{7}{r^9_a}(\vec{r}_a\cdot \vec{v}_a(t^{\ast}))^2(\vec{r}_a\cdot \delta \vec{z}_{(1)})\Big]\tau^2\bigg]
P^i_qr^q_a
+
  2\, \bigg[\frac{2}{r^3_a}
+
  \frac{6}{r^5_a}(\vec{r}_a\cdot \vec{v}_a(t^{\ast}))\tau \nonumber\\
&-
  3\,\Big[\frac{1}{r^5_a}v^2_a(t^{\ast})-\frac{5}{r^7_a}(\vec{r}_a\cdot \vec{v}_a(t^{\ast}))^2\Big]\tau^2\bigg]P^i_q \delta z^q_{(1)}
+
  12\,\bigg[\frac{1}{r^5_a}(\vec{r}_a\cdot \delta \vec{z}_{(1)})\tau \nonumber\\
&-
  \Big[\frac{1}{r^5_a}(\vec{v}_a(t^{\ast})\cdot \delta \vec{z}_{(1)})
-
  \frac{5}{r^7_a}(\vec{r}_a\cdot \vec{v}_a(t^{\ast}))(\vec{r}_a\cdot \delta \vec{z}_{(1)})\Big]\tau^2\bigg]
P^i_qv^q_a(t^{\ast})\Bigg\}_{|(\rightarrow)}
+
 \frac{G}{c^3}\Big\{...\Big\}.\label{Deflection3}  
\end{align}

In the equation above we need not write out the expression $G/c^3 \big\{...\big\}$ explicitly since it contributes only terms 
of order greater than $G^2/c^4$.
After substituting the perturbation \eqref{Perturbation3} into the preceding equation we obtain the integrals for the
post-linear light deflection $\alpha^i_{(2) \mathrm{IV}}$.
Here, we take into account only the integrals of the order $G^2/c^4$. These integrals are given in an explicit form
in  Appendix \ref{ap:C}.

\section{Light deflection and the centre of mass}\label{centre}
In this section we compute the corrections to the linear and the post-linear light deflection resulting from the introduction of the 
1PN-corrections to the positions of the masses in the equations for the linear perturbations 
$\delta l^q_{(1)\mathrm{I}}(\tau)$ and $\delta l^q_{(1)\mathrm{II}}(\tau)$ given by 
Eqs~\eqref{Perturbation11} and \eqref{Perturbation12}.
It follows from equations \eqref{pos1} and \eqref{pos2} that the 1PN-corrections in the positions are given by
\begin{align}
\delta \vec{x}_1=\delta \vec{x}_2=\frac{1}{c^2}\Bigg[\frac{\nu (m_1-m_2)}{2 M}\Big[v^2_{12}-\frac{GM}{r_{12}}\Big]\Bigg]
\vec{r}_{12}.\label{shift}
\end{align}
From expression \eqref{shift}, it is easy to see that the corrections vanish when $m_1=m_2$. The corrections also vanish
for the case of circular orbits. 
After introducing the 1PN-corrections into the equations for the linear perturbations $\delta l^q_{(1)\mathrm{I}}(\tau)$ 
and $\delta l^q_{(1)\mathrm{II}}(\tau)$ we obtain the expression for the perturbed linear light deflection. 
Because the corrections $\delta \vec{x}_a$ are small quantities compared to $\vec{x}_a$, we can resort to a
Taylor expansion of the perturbed linear light deflection about $\delta \vec{x}_a=0$ in order to find the 
correction terms for the linear and post-linear light deflection. We denote these terms by $\tilde{\alpha}^i_{(1)(2)}$ and they are
given by

\begin{align}
\tilde{\alpha}^{i}_{(1)(2)}&=\lim_{\tau\to\infty}\Bigg\{\frac{1}{c}P^i_q
\bigg[\bigg(\frac{\partial \delta l^{q\mathrm{(Pert.)}}_{(1)\mathrm{I}}}{\partial \delta x^m_{a}}\bigg)_{\delta \vec{x}_{a}=0} \delta x^m_{a}
+\bigg(\frac{\partial \delta l^{q\mathrm{(Pert.)}}_{(1)\mathrm{II}}}{\partial \delta x^m_{a}}\bigg)_{\delta \vec{x}_{a}=0} \delta x^m_{a}
\bigg]\Bigg\}\nonumber\\
&=\frac{G}{c} \sum^2_{a=1} m_a \int^{\infty}_{-\infty}
d \tau \,\Bigg\{\bigg[-\frac{6}{r^5_a}(\vec{r}_a\cdot \delta \vec{x}_{a})
+6\,\Big[\frac{1}{r^5_a}(\vec{v}_a(t^{\ast})\cdot \delta \vec{x}_{a}) \nonumber\\
&-
  \frac{5}{r^7_a}(\vec{r}_a\cdot \vec{v}_a(t^{\ast}))(\vec{r}_a\cdot \delta \vec{x}_{a})\Big]\tau
+15\,\Big[\frac{2}{r^7_a}(\vec{r}_a\cdot \vec{v}_a(t^{\ast}))(\vec{v}_a(t^{\ast})\cdot \delta \vec{x}_{a})\nonumber\\
&+
  \frac{1}{r^7_a}v^2_a(t^{\ast})(\vec{r}_a\cdot \delta \vec{x}_{a})
-
  \frac{7}{r^9_a}(\vec{r}_a\cdot \vec{v}_a(t^{\ast}))^2(\vec{r}_a\cdot \delta \vec{x}_{a})\Big]\tau^2\bigg]P^i_qr^q_a\nonumber\\
&+
  \bigg[\frac{2}{r^3_a}
+
  \frac{6}{r^5_a}(\vec{r}_a\cdot \vec{v}_a(t^{\ast}))\tau
-
 3\,\Big[\frac{1}{r^5_a}v^2_a(t^{\ast})
-
\frac{5}{r^7_a}(\vec{r}_a\cdot \vec{v}_a(t^{\ast}))^2\Big]\tau^2\bigg]P^i_q\delta x^q_{a}\nonumber\\
&+
  \bigg[\frac{6}{r^5_a}(\vec{r}_a\cdot \delta \vec{x}_{a})\tau
-
  3\,\Big[\frac{2}{r^5_a}(\vec{v}_a\cdot \delta \vec{x}_{a})
-
  \frac{10}{r^7_a}(\vec{r}_a\cdot \vec{v}_a(t^{\ast}))(\vec{r}_a\cdot \delta \vec{x}_{a})\Big]\tau^2\bigg]P^i_qv^q_a(t^{\ast})
\Bigg\}_{|(\rightarrow)}\nonumber\\
&+
  \frac{G}{c^2} \sum^2_{a=1} m_a(\vec{e}_{(0)}\cdot\vec{v}_a(t^{\ast}))\int^{\infty}_{-\infty}d \tau \,
\Bigg\{6\,\bigg[\frac{2}{r^5_a}(\vec{r}_a\cdot \delta \vec{x}_{a})
-
  2\,\Big[\frac{1}{r^5_a}(\vec{v}_a(t^{\ast})\cdot \delta \vec{x}_{a})\nonumber\\
&-
  \frac{5}{r^7_a}(\vec{r}_a\cdot \vec{v}_a(t^{\ast}))(\vec{r}_a\cdot \delta \vec{x}_{a})\Big]\tau
-
  5\,\Big[\frac{1}{r^7_a}v^2_a(t^{\ast})(\vec{r}_a\cdot \delta \vec{x}_{a})
+
  \frac{2}{r^7_a}(\vec{r}_a\cdot \vec{v}_a(t^{\ast}))(\vec{v}_a(t^{\ast})\cdot \delta \vec{x}_{a})\nonumber\\
&-
  \frac{7}{r^9_a}(\vec{r}_a\cdot \vec{v}_a(t^{\ast}))^2(\vec{r}_a\cdot \delta \vec{x}_{a})\Big]\tau^2\bigg]
P^i_qr^q_a
-
  2\, \bigg[\frac{2}{r^3_a}
+
  \frac{6}{r^5_a}(\vec{r}_a\cdot \vec{v}_a(t^{\ast}))\tau \nonumber\\
&-
  3\,\Big[\frac{1}{r^5_a}v^2_a(t^{\ast})-\frac{5}{r^7_a}(\vec{r}_a\cdot \vec{v}_a(t^{\ast}))^2\Big]\tau^2\bigg]P^i_q\delta x^q_{a}
-
  12\,\bigg[\frac{1}{r^5_a}(\vec{r}_a\cdot \delta \vec{x}_{a})\tau \nonumber\\
&-
  \Big[\frac{1}{r^5_a}(\vec{v}_a(t^{\ast})\cdot \delta \vec{x}_{a})
-
  \frac{5}{r^7_a}(\vec{r}_a\cdot \vec{v}_a(t^{\ast}))(\vec{r}_a\cdot \delta \vec{x}_{a})\Big]\tau^2\bigg]
P^i_qv^q_a(t^{\ast})\Bigg\}_{|(\rightarrow)}
+
  \frac{G}{c^3}\Big\{...\Big\}.\label{Deflection5}
\end{align}
For the same reason as in Eq.~\eqref{Deflection3}, we do not need to write out the expression $G/c^3 \big\{...\big\}$ explicitly in Eq.~\eqref{Deflection5}.
After substituting the 1PN-corrections given by Eq.~\eqref{shift} into the equation above  and taking into account 
only the terms of the orders $G/c^4$ and $G^2/c^4$ we find,
\begin{align}
\tilde{\alpha}^{i}_{(1)(2)}&=-6\,\frac{G}{c^3}\sum^2_{a=2} m_a \Bigg[\frac{\nu (m_1-m_2)}{2 M}\Big[v^2_{12}(t^{\ast})
-
  \frac{GM}{r_{12}(t^{\ast})}\Big]\Bigg]\nonumber\\
& \int^{\infty}_{-\infty} d \tau \, \frac{1}{r^5_a}\Big[c\,\tau \vec{e}_{(0)} \cdot \vec{r}_{12}(t^{\ast})
-
  \vec{x}_a\cdot \vec{r}_{12}(t^{\ast})\Big]\big[\xi^i-P^i_q x^q_a(t^{\ast})\big]\nonumber\\
&+
  2\,\frac{G}{c^3}\sum^2_{a=2} m_a \Bigg[\frac{\nu (m_1-m_2)}{2 M}\Big[v^2_{12}(t^{\ast})
-
  \frac{GM}{r_{12}(t^{\ast})}\Big]\Bigg] \int^{\infty}_{-\infty} d \tau \, \frac{1}{r^3_a} P^i_q r^q_{12}(t^{\ast})
+
  (1 \leftrightarrow 2).
\end{align}
Here, we have already replaced the photon trajectory by its unperturbed approximation 
$\vec{z}(\tau)_{\mathrm{unpert.}} =\tau \, \vec{l}_{(0)}+\vec{\xi}$. 
 
After performing the integration we obtain,

\begin{align}
\tilde{\alpha}^i_{(1)(2)}&=8\,\frac{G}{c^4}\Bigg[\frac{\nu (m_1-m_2)}{2 M}\Big[v^2_{12}(t^{\ast})-\frac{GM}{r_{12}(t^{\ast})}\Big]\Bigg]
\sum^2_{a=1}\frac{m_a}{R^2_a} \Big[(\vec{x}_a(t^{\ast})\cdot \vec{r}_{12}(t^{\ast}))\nonumber\\
&-
  (\vec{e}_{(0)}\cdot \vec{x}_a(t^{\ast}))\,(\vec{e}_{(0)}\cdot\vec{r}_{12}(t^{\ast}))\Big]\big[\xi^i-P^i_q x^q_a(t^{\ast})\big]\nonumber\\
&+
  4\,\frac{G}{c^4}\Bigg[\frac{\nu (m_1-m_2)}{2 M}\Big[v^2_{12}(t^{\ast})-\frac{GM}{r_{12}(t^{\ast})}\Big]\Bigg]\sum^2_{a=1}\frac{m_a}{R_a}P^i_q r^q_{12}(t^{\ast}),
\end{align}
where the quantity $R_a$ is given by Eq.~\eqref{distanceB}. 
Finally, considering further applications we express the positions of the masses by their centre of mass coordinates 
and expand the preceding expression about the origin of the coordinate system to the second order in $\big(r_{12}/c\big)$ 
to obtain,
\begin{align}
\tilde{\alpha}^i_{(1)(2)}&=8\,X_2\,\frac{G}{c^4} m_1 \Bigg[\frac{\nu (m_1-m_2)}{2 M}\Big[v^2_{12}
-
  \frac{GM}{r_{12}}\Big]\Bigg]\Bigg\{\bigg[1
-
  (\vec{e}_{(0)} \cdot \vec{n}_{12})^2\bigg]\bigg(\frac{r_{12}}{\xi}\bigg)^2
+
  \mathcal{O}\bigg[\bigg(\frac{r_{12}}{\xi}\bigg)^3\bigg]\Bigg\}\frac{e^i}{\xi} \nonumber\\
&+
  4\,\frac{G}{c^4} m_1 \Bigg[\frac{\nu (m_1-m_2)}{2 M}\Big[v^2_{12}-\frac{GM}{r_{12}}\Big]\Bigg]
\Bigg\{\bigg(\frac{r_{12}}{\xi}\bigg)
+
  2\, X_2\,(\vec{e}_{\xi} \cdot \vec{n}_{12})\bigg(\frac{r_{12}}{\xi}\bigg)^2\nonumber\\
&+
  \mathcal{O}\bigg[\bigg(\frac{r_{12}}{\xi}\bigg)^3\bigg]\Bigg\}\frac{1}{\xi} P^i_q n^q_{12}
+(1 \leftrightarrow 2).\label{Defl5}
\end{align}

As in the preceding section, the quantities $\vec{n}_{12}$, $\vec{v}_{12}$ and $r_{12}$ are taken at 
the time $t^{\ast}$.

\section{The total linear Perturbation and the linear light deflection in the gravitational field of two bounded
masses}\label{totlightlin}
To obtain the total linear perturbation $\delta l^i_{(1)}(\tau)$ we have to sum up the expressions for the linear
perturbations $\delta l^i_{(1)\mathrm{I}}(\tau)$ and  $\delta l^i_{(1)\mathrm{II}}(\tau)$ given by Eqs~ 
\eqref{Perturbed1} and \eqref{Perturbed2}. The resulting expression is given by

\begin{align}
&\delta l^i_{(1)}(\tau)=-2\,\frac{G}{c}\sum^2_{a=1}m_a\,B_a \big[\xi^i-x^i_a(t^{\ast})\big]\nonumber\\
&-
 \frac{G}{c^2}\sum^2_{a=1}m_a\,\Bigg\{2\,A_a+\frac{4}{r_a}
+
  \frac{4}{r^3_a}\bigg[(\vec{e}_{(0)}\cdot\vec{v}_a(t^{\ast}))c\, \tau^2+\Big[\vec{\xi}\cdot\vec{v}_a(t^{\ast})
-
 \vec{x}_a(t^{\ast})\cdot\vec{v}_a(t^{\ast})\Big]\tau \bigg]\Bigg\}l^i_{(0)}\nonumber\\
&+
 \frac{G}{c^2}\sum^2_{a=1}m_a\,\Bigg\{4\,(\vec{e}_{(0)}\cdot\vec{v}_a(t^{\ast}))B_a
-
 6\,(\vec{e}_{(0)}\cdot\vec{v}_a(t^{\ast}))F_{a2}
-
 6\,\Big[\vec{\xi}\cdot\vec{v}_a(t^{\ast})
-
 \vec{x}_a(t^{\ast})\cdot\vec{v}_a(t^{\ast})\Big]F_{a3}\Bigg\}\nonumber\\
 &\big[\xi^i-x^i_a(t^{\ast})\big]\nonumber\\
&+
  2\, \frac{G}{c^2}\sum^2_{a=1}m_a\,\Bigg\{A_a+\frac{2}{r_a}+\frac{2}{r^3_a}\,
  \bigg[(\vec{e}_{(0)}\cdot\vec{v}_a(t^{\ast}))c\,\tau^2
+
 \Big[\vec{\xi}\cdot\vec{v}_a(t^{\ast})-\vec{x}_a(t^{\ast})\cdot\vec{v}_a(t^{\ast})\Big]\tau\bigg]\Bigg\}
v^i_a(t^{\ast})\nonumber\\
&+
  \frac{G}{c^3}\sum^2_{a=1}m_a\,\Bigg\{4\,(\vec{e}_{(0)}\cdot\vec{v}_a(t^{\ast}))\bigg[A_a+\frac{1}{r_a}\bigg]
-
 6\,\bigg[(\vec{e}_{(0)}\cdot\vec{v}_a(t^{\ast}))F_{a1}\nonumber\\
&+
 \Big[\vec{\xi}\cdot\vec{v}_a(t^{\ast})-\vec{x}_a(t^{\ast})\cdot\vec{v}_a(t^{\ast})\Big]F_{a2}\bigg]\Bigg\}l^i_{(0)}
 \nonumber\\
&+
  \frac{G}{c^3}\sum^2_{a=1}m_a\,\Bigg\{(\vec{e}_{(0)}\cdot\vec{v}_a(t^{\ast}))^2\Big[15\,F_{a2}-2\,B_a-15\,G_{a2}
\Big]\nonumber\\
&+
 (\vec{e}_{(0)}\cdot\vec{v}_a(t^{\ast}))\,\Big[\vec{\xi}\cdot\vec{v}_a(t^{\ast})
-
 \vec{x}_a(t^{\ast})\cdot\vec{v}_a(t^{\ast})\Big]\Big[18\,F_{a3}-30\,G_{a3}\Big]\nonumber\\
&+
  \Big[\vec{\xi}\cdot\vec{v}_a(t^{\ast})-\vec{x}_a(t^{\ast})\cdot\vec{v}_a(t^{\ast})\Big]^2
  \Big[3\,F_{a4}-15\,G_{a4}\Big]
+
 v^2_a(t^{\ast})\Big[3\,F_{a2}-2\,B_{a}\Big]\Bigg\}\big[\xi^i-x^i_a(t^{\ast})\big]\nonumber\\
&+
  \frac{G}{c^3}\sum^2_{a=1}m_a\,\Bigg\{(\vec{e}_{(0)}\cdot\vec{v}_a(t^{\ast}))\bigg[6\,F_{a1}-6\,A_a
-\frac{4}{r_a}-\frac{4}{r^3_a}\,(\vec{e}_{(0)}\cdot\vec{v}_a(t^{\ast}))c\,\tau^2\nonumber\\
&-
  \frac{4}{r^3_a}\ \Big[\vec{\xi}\cdot\vec{v}_a(t^{\ast})
-\vec{x}_a(t^{\ast})\cdot\vec{v}_a(t^{\ast})\Big]\tau\bigg] 
+
 \Big[\vec{\xi}\cdot\vec{v}_a(t^{\ast})-\vec{x}_a(t^{\ast})\cdot\vec{v}_a(t^{\ast})\Big]
 \Big[6\,F_{a2}-2\,B_a\Big]\Bigg\}v^i_a(t^{\ast})\nonumber\\
&+
 \mathcal{O}\bigg(\frac{G}{c^4}\bigg).\label{Perturbed3}
\end{align}
In the expression above we need only retain the terms of the order $G/c^2$ and $G/c^3$, since the terms
of the order $G/c^4$ are related to linear and post-linear light deflection terms of order higher than the terms
which we compute in this paper.
After introducing the perturbation $\delta l^i_{(1)}(\tau)$ into Eq.~\eqref{deflection} and computing the limit for 
$\tau \rightarrow \infty$, we find
\begin{align}
\alpha^i_{(1)}&=\lim_{\tau\to\infty}\Big[\frac{1}{c}P^i_q \delta l^q_{(1)}(\tau)\Big]\nonumber\\
&=-4\,\frac{G}{c^2}\sum^2_a \frac{m_a}{R_a}\big[\xi^i-P^i_qx^q_a(t^{\ast})\big]\nonumber\\
&+4\,\frac{G}{c^3}\sum_{a=1}^{2} \frac{m_a}{R_a}(\vec{e}_{(0)}\cdot\vec{v}_{a}(t^{\ast}))\big[\xi^i-P^i_qx^q_a(t^{\ast})\big]
\nonumber\\
&+4\,\frac{G}{c^3}\sum_{a=1}^{2} \frac{m_a}{R_a}(\vec{e}_{(0)}\cdot\vec{x}_{a}(t^{\ast}))P^i_q v^q_a(t^{\ast})
\nonumber\\
&-8\,\frac{G}{c^3}\sum_{a=1}^{2} \frac{m_a}{R^2_a}(\vec{e}_{(0)}\cdot\vec{x}_a(t^{\ast}))
\Big[\vec{\xi}\cdot\vec{v}_a(t^{\ast})-\vec{x}_a(t^{\ast})\cdot\vec{v}_a(t^{\ast})
+(\vec{e}_{(0)}\cdot\vec{x}_a(t^{\ast}))(\vec{e}_{(0)}\cdot\vec{v}_a(t^{\ast}))\Big]\nonumber\\
&\big[\xi^i-P^i_qx^q_a(t^{\ast})\big]\nonumber\\
&-2\,\frac{G}{c^4}\sum_{a=1}^{2} \frac{m_a}{R_a}v^2_a(t^{\ast})\big[\xi^i-P^i_qx^q_a(t^{\ast})\big]\nonumber\\
&+4\,\frac{G}{c^4}\sum_{a=1}^{2}\frac{m_a}{R^2_a}v^2_a(t^{\ast})(\vec{e}_{(0)}\cdot\vec{x}_a(t^{\ast}))^2
\big[\xi^i-P^i_qx^q_a(t^{\ast})\big]\nonumber\\
&-16\,\frac{G}{c^4}\sum_{a=1}^{2} \frac{m_a}{R^3_a}(\vec{e}_{(0)}\cdot\vec{x}_a(t^{\ast}))^2
\Big[\vec{\xi}\cdot\vec{v}_a(t^{\ast})-\vec{x}_a(t^{\ast})\cdot\vec{v}_a(t^{\ast})\Big]^2 \nonumber\\
&\big[\xi^i-P^i_qx^q_a(t^{\ast})\big]\nonumber\\
&-32\,\frac{G}{c^4}\sum_{a=1}^{2} \frac{m_a}{R^3_a}(\vec{e}_{(0)}\cdot\vec{x}_a(t^{\ast}))^3
(\vec{e}_{(0)}\cdot\vec{v}_a(t^{\ast}))
\Big[\vec{\xi}\cdot\vec{v}_a(t^{\ast})-\vec{x}_a(t^{\ast})\cdot\vec{v}_a(t^{\ast})\Big]
\big[\xi^i-P^i_qx^q_a(t^{\ast})\big]\nonumber\\
&-16\,\frac{G}{c^4}\sum_{a=1}^{2}\frac{m_a}{R^3_a}(\vec{e}_{(0)}\cdot\vec{x}_a(t^{\ast}))^4
(\vec{e}_{(0)}\cdot\vec{v}_a(t^{\ast}))^2 \big[\xi^i-P^i_qx^q_a(t^{\ast})\big]\nonumber\\
&+8\,\frac{G}{c^4}\sum_{a=1}^{2}\frac{m_a}{R^2_a}(\vec{e}_{(0)}\cdot\vec{x}_a(t^{\ast}))^2
\Big[\vec{\xi}\cdot\vec{v}_a(t^{\ast})-\vec{x}_a(t^{\ast})\cdot\vec{v}_a(t^{\ast})
+(\vec{e}_{(0)}\cdot\vec{x}_a(t^{\ast}))(\vec{e}_{(0)}\cdot\vec{v}_a(t^{\ast}))\Big]P^i_qv^q_a(t^{\ast})\nonumber\\
&-4\,\frac{G}{c^4}\sum_{a=1}^{2}\frac{m_a}{R^2_a}(\vec{e}_{(0)}\cdot\vec{x}_a(t^{\ast}))^2
(\vec{e}_{(0)}\cdot\vec{v}_a(t^{\ast}))^2 \big[\xi^i-P^i_qx^q_a(t^{\ast})\big],\label{Totlineardefl}
\end{align}
where $R_a$ is given by Eq.~\eqref{distanceB}. 

Considering that the expression for the post-linear light deflection computed in this paper is given in
terms of the centre-of-mass-frame coordinates and as an expansion in powers of $(r_{12}/\xi)$, we
must perform the expansion of the expression above about the origin of the coordinate system $\vec{x}_a=0$
and express the positions of the masses in terms of their centre of mass coordinates.
After expressing the positions of the masses by their centre of mass frame coordinates without considering the 
1PN-corrections and expanding the expression given by Eq.~\eqref{Totlineardefl} about the origin 
of the coordinate system to the third order in $\big(r_{12}/\xi\big)$ we finally obtain,

\begin{align}
&\alpha^i_{(1)}=\frac{G m_1}{c^2\,\xi}\,\Bigg\{\bigg[-4 
+ 
  4\,X_2\,\frac{1}{c}\,(\vec{e}_{(0)}\cdot\vec{v}_{12})
- 2\,X^2_2\,\frac{v^2_{12}}{c^2}\bigg]\nonumber\\
&+ 
  \bigg[-8\,X_2\,(\vec{e}_{\xi}\cdot\vec{n}_{12}) 
+ 
  8\,X^2_2\,(\vec{e}_{\xi}\cdot\vec{n}_{12})\,\frac{1}{c}\,(\vec{e}_{(0)}\cdot\vec{v}_{12}) 
- 
  8\,X^2_2\,\frac{1}{c}\,(\vec{e}_{\xi}\cdot\vec{v}_{12})\,(\vec{e}_{(0)}\cdot\vec{n}_{12})\nonumber\\ 
&- 
  4\,X^3_2\,(\vec{e}_{\xi}\cdot\vec{n}_{12})\,\frac{v^2_{12}}{c^2} \bigg]\bigg(\frac{r_{12}}{\xi}\bigg)\nonumber\\  
&+ 
  \bigg[4\,X^2_2 
- 
  16\,X^2_2\,(\vec{e}_{\xi}\cdot\vec{n}_{12})^2
- 
  4\,X^2_2\,(\vec{e}_{(0)}\cdot\vec{n}_{12})^2
- 
  4\,X^3_2\,\frac{1}{c}\,(\vec{e}_{(0)}\cdot\vec{v}_{12})\nonumber\\
&+ 
  16\,X^3_2\,(\vec{e}_{\xi}\cdot\vec{n}_{12})^2\,\frac{1}{c}\,(\vec{e}_{(0)}\cdot\vec{v}_{12})
- 
  32\,X^3_2\,\frac{1}{c}\,(\vec{e}_{\xi}\cdot\vec{v}_{12})\,(\vec{e}_{\xi}\cdot\vec{n}_{12})\,(\vec{e}_{(0)}\cdot\vec{n}_{12})
\nonumber\\
&- 
  4\,X^3_2\,\frac{1}{c}\,(\vec{e}_{(0)}\cdot\vec{v}_{12}(t^{\ast}))\,{(\vec{e}_{(0)}\cdot\vec{n}_{12})}^2
+ 
  8\,X^3_2\,(\vec{e}_{(0)}\cdot\vec{n}_{12})\,\frac{1}{c}\,(\vec{n}_{12}\cdot\vec{v}_{12})\nonumber\\ 
&+ 
  2\,X^4_2\,\frac{v^2_{12}}{c^2}
- 
  8\,X^4_2\,(\vec{e}_{\xi}\cdot\vec{n}_{12})^2\frac{v^2_{12}}{c^2}
+ 
  2\,X^4_2\,(\vec{e}_{(0)}\cdot\vec{n}_{12})^2\frac{v^2_{12}}{c^2}\nonumber\\
&- 
  16\,X^4_2\,\frac{1}{c^2}\,(\vec{e}_{\xi}\cdot\vec{v}_{12})^2\,(\vec{e}_{(0)}\cdot\vec{n}_{12})^2
- 
  4\,X^4_2\,\frac{1}{c^2}\,(\vec{e}_{(0)}\cdot\vec{v}_{12})^2\,(\vec{e}_{(0)}\cdot\vec{n}_{12})^2\bigg] 
  \bigg(\frac{r_{12}}{\xi}\bigg)^2\nonumber\\
&+
  \bigg[16\,X^4_2\,\frac{1}{c}\,(\vec{e}_{\xi}\cdot\vec{v}_{12})\,(\vec{e}_{(0)}\cdot\vec{n}_{12})
- 
  96\,X^4_2\,\frac{1}{c}\,(\vec{e}_{\xi}\cdot\vec{v}_{12})\,(\vec{e}_{\xi}\cdot\vec{n}_{12})^2\,(\vec{e}_{(0)}\cdot\vec{n}_{12})
\nonumber\\
&- 
  32\,X^4_2\,(\vec{e}_{\xi}\cdot\vec{n}_{12})\,\frac{1}{c}\,(\vec{e}_{(0)}\cdot\vec{v}_{12})\,
  (\vec{e}_{(0)}\cdot\vec{n}_{12})^2
- 
  16\,X^4_2\,\frac{1}{c}\,(\vec{e}_{\xi}\cdot\vec{v}_{12})\,(\vec{e}_{(0)}\cdot\vec{n}_{12})^3\nonumber\\
&+ 
  32\,X^4_2\,(\vec{e}_{\xi}\cdot\vec{n}_{12})\,(\vec{e}_{(0)}\cdot\vec{n}_{12})\,\frac{1}{c}\,(\vec{n}_{12}\cdot\vec{v}_{12}) 
+ 
  16\,X^5_2\,(\vec{e}_{\xi}\cdot\vec{n}_{12})\,(\vec{e}_{(0)}\cdot\vec{n}_{12})^2\,\frac{v^2_{12}}{c^2}\nonumber\\
&- 
  96\,X^5_2\,\frac{1}{c^2}\,(\vec{e}_{\xi}\cdot\vec{v}_{12})^2\,(\vec{e}_{\xi}\cdot\vec{n}_{12})\,
(\vec{e}_{(0)}\cdot\vec{n}_{12})^2\nonumber\\
&- 
  16\,X^5_2\,(\vec{e}_{\xi}\cdot\vec{n}_{12})\,\frac{1}{c^2}\,(\vec{e}_{(0)}\cdot\vec{v}_{12})^2\,
(\vec{e}_{(0)}\cdot\vec{n}_{12})^2\nonumber\\
&- 
  32\,X^5_2\,\frac{1}{c^2}\,(\vec{e}_{\xi}\cdot\vec{v}_{12})\,(\vec{e}_{(0)}\cdot\vec{v}_{12})\,
(\vec{e}_{(0)}\cdot\vec{n}_{12})^3\nonumber\\
&+ 
  32\,X^5_2\,\frac{1}{c^2}\,(\vec{e}_{\xi}\cdot\vec{v}_{12})\,(\vec{e}_{(0)}\cdot\vec{n}_{12})^2\,
(\vec{n}_{12}\cdot\vec{v}_{12})\bigg]\bigg(\frac{r_{12}}{\xi}\bigg)^3 
+
  \mathcal{O}\bigg[\bigg(\frac{r_{12}}{\xi}\bigg)^4\bigg]\Bigg\}e^i_{\xi}\nonumber\\
&+
  \frac{G m_1}{c^2\,\xi}\Bigg\{\bigg[4\,X_2 
- 
  4\,X^2_2\,\frac{1}{c}\,(\vec{e}_{(0)}\cdot\vec{v}_{12})
+ 
  2\,X^3_2\,\frac{v^2_{12}}{c^2}\bigg]\bigg(\frac{r_{12}}{\xi}\bigg)\nonumber\\
&+ 
  \bigg[8\,X^2_2\,(\vec{e}_{\xi}\cdot\vec{n}_{12})
- 
  8\,X^3_2\,(\vec{e}_{\xi}\cdot\vec{n}_{12})\,\frac{1}{c}\,(\vec{e}_{(0)}\cdot\vec{v}_{12})
+ 
  8\,X^3_2\,\frac{1}{c}\,(\vec{e}_{\xi}\cdot\vec{v}_{12})\,(\vec{e}_{(0)}\cdot\vec{n}_{12})\nonumber\\
&+ 
  4\,X^4_2\,(\vec{e}_{\xi}\cdot\vec{n}_{12})\,\frac{v^2_{12}}{c^2}\bigg]\bigg(\frac{r_{12}}{\xi}\bigg)^2 
+ 
  \bigg[-4\,X^3_2 
+ 
  16\,X^3_2\,(\vec{e}_{\xi}\cdot\vec{n}_{12})^2
+ 
  4\,X^3_2\,(\vec{e}_{(0)}\cdot\vec{n}_{12})^2\nonumber\\
&+ 
  4\,X^4_2\,\frac{1}{c}\,(\vec{e}_{(0)}\cdot\vec{v}_{12})
- 
  16\,X^4_2\,(\vec{e}_{\xi}\cdot\vec{n}_{12})^2\,\frac{1}{c}\,(\vec{e}_{(0)}\cdot\vec{v}_{12})\nonumber\\
&+ 
  32\,X^4_2\,\frac{1}{c}\,(\vec{e}_{\xi}\cdot\vec{v}_{12})\,(\vec{e}_{\xi}\cdot\vec{n}_{12})\,
  (\vec{e}_{(0)}\cdot\vec{n}_{12})
+ 
  4\,X^4_2\,\frac{1}{c}\,(\vec{e}_{(0)}\cdot\vec{v}_{12})\,(\vec{e}_{(0)}\cdot\vec{n}_{12})^2\nonumber\\
&- 
  8\,X^4_2\,(\vec{e}_{(0)}\cdot\vec{n}_{12})\,\frac{1}{c}\,(\vec{n}_{12}\cdot\vec{v}_{12})
- 
  2\,X^5_2\,\frac{v^2_{12}}{c^2}
+ 
  8\,X^5_2\,(\vec{e}_{\xi}\cdot\vec{n}_{12})^2\,\frac{v^2_{12}}{c^2}\nonumber\\
&- 
  2\,X^5_2\,(\vec{e}_{(0)}\cdot\vec{n}_{12})^2\,\frac{v^2_{12}}{c^2}
+ 
  16\,X^5_2\,\frac{1}{c^2}\,(\vec{e}_{\xi}\cdot\vec{v}_{12})^2\,(\vec{e}_{(0)}\cdot\vec{n}_{12})^2\nonumber\\
&+ 
  4\,X^5_2\,\frac{1}{c^2}\,(\vec{e}_{(0)}\cdot\vec{v}_{12})^2\,(\vec{e}_{(0)}\cdot\vec{n}_{12})^2
\bigg]\bigg(\frac{r_{12}}{\xi}\bigg)^3
+
  \mathcal{O}\bigg[\bigg(\frac{r_{12}}{\xi}\bigg)^4\bigg]\Bigg\} P^i_qn^q_{12}\nonumber\\
&+
  \frac{G m_1}{c^2\,\xi}\Bigg\{4\,X^2_2\,(\vec{e}_{(0)}\cdot\vec{n}_{12})\bigg(\frac{r_{12}}{\xi}\bigg)
+ 
  \bigg[8\,X^3_2\,(\vec{e}_{\xi}\cdot\vec{n}_{12})\,(\vec{e}_{(0)}\cdot\vec{n}_{12})\nonumber\\
&+ 
  8\,X^4_2\,\frac{1}{c}\,(\vec{e}_{\xi}\cdot\vec{v}_{12})\,(\vec{e}_{(0)}\cdot\vec{n}_{12})^2\bigg]
   \bigg(\frac{r_{12}}{\xi}\bigg)^2\nonumber\\
&+ 
  \bigg[-4\,X^4_2\,(\vec{e}_{(0)}\cdot\vec{n}_{12})
+ 
  16\,X^4_2\,(\vec{e}_{\xi}\cdot\vec{n}_{12})^2\,(\vec{e}_{(0)}\cdot\vec{n}_{12})
+ 
  4\,X^4_2\,(\vec{e}_{(0)}\cdot\vec{n}_{12})^3\nonumber\\
&+ 
  32\,X^5_2\,\frac{1}{c}\,(\vec{e}_{\xi}\cdot\vec{v}_{12})\,(\vec{e}_{\xi}\cdot\vec{n}_{12})\,(\vec{e}_{(0)}\cdot\vec{n}_{12})^2
+ 
  8\,X^5_2\,\frac{1}{c}\,(\vec{e}_{(0)}\cdot\vec{v}_{12})\,(\vec{e}_{(0)}\cdot\vec{n}_{12})^3\nonumber\\
&- 
  8\,X^5_2\,(\vec{e}_{(0)}\cdot\vec{n}_{12})^2\,\frac{1}{c}\,(\vec{n}_{12}\cdot\vec{v}_{12})\bigg]\bigg(\frac{r_{12}}{\xi}\bigg)^3 
+
  \mathcal{O}\bigg[\bigg(\frac{r_{12}}{\xi}\bigg)^4\bigg]\Bigg\}\bigg(\frac{v_{12}}{c}\bigg)\,
  P^i_qe^q_{v_{12}}
+
  (1 \leftrightarrow 2),\label{Totlightdefl1}
\end{align}
where the quantities $\vec{n}_{12}$, $\vec{v}_{12}$ and $r_{12}$ are taken at the time $t^{\ast}$.
Notice that in Eq.~\eqref{Totlightdefl1} we chose the order of the expansion in an arbitrary manner
in order to show the structure of the terms belonging to the linear light deflection.
In concrete applications we have to choose the order of the expansion of the linear light deflection
in accordance with the accuracy reached by the post-linear light deflection.

To obtain the total expression for the angle of light deflection linear in $G$ we have to add
to the preceding equation the correction terms arising from the part of $\tilde{\alpha}^i_{(1)(2)}$, which
is linear in $G$. The expression for $\tilde{\alpha}^i_{(1)(2)}$ is given by Eq.~\eqref{Defl5}.

\section{The post-linear light deflection}\label{totlightpostlin}
The final expression for the post-linear light deflection in the gravitational field of two bounded
masses is obtained by summing  up the parts of the light deflection, which are given in the preceding sections 
and in  Appendices A to C. 
The final expression for the angle of light deflection quadratic in $G$ to the first order in $\big(r_{12}/\xi\big)$, 
in which the positions of the masses are expressed in the centre of mass frame coordinates, is given by
\begin{align}
\alpha^i_{(2)}&=\Bigg\{\frac{G^2 m^2_1}{c^4\,\xi^2}\Bigg[-\frac{15}{4}\,\pi 
 - 
  X_2 \bigg(8\,(\vec{e}_{(0)} \cdot \vec{n}_{12})
 +
  \frac{45}{4}\,\pi\,(\vec{e}_{\xi} \cdot \vec{n}_{12})\bigg)\bigg(\frac{r_{12}}{\xi}\bigg)
 +
  \mathcal{O}\bigg[\bigg(\frac{r_{12}}{\xi}\bigg)^2\bigg]\Bigg]\nonumber\\
&+
  \frac{G^2 m^2_2}{c^4\, \xi^2}\Bigg[-\frac{15}{4}\,\pi 
 + 
  X_1\,\bigg(8\,(\vec{e}_{(0)} \cdot \vec{n}_{12})
 +
  \frac{45}{4}\,\pi\,(\vec{e}_{\xi} \cdot \vec{n}_{12})\bigg)\bigg(\frac{r_{12}}{\xi}\bigg)
 +
  \mathcal{O}\bigg[\bigg(\frac{r_{12}}{\xi}\bigg)^2\bigg]\Bigg]\Bigg\}e^i_{\xi}\nonumber\\
&+
  4\,\frac{G^2m_1m_2}{c^4\,\xi\,r_{12}}e^i_{\xi}\nonumber\\
&+ 
  \frac{G^2m_1m_2}{c^4\,\xi^2}\Bigg\{-\frac{15}{2}\,\pi 
 -
  4\,(X_1-X_2)(\vec{e}_{\xi}\cdot\vec{n}_{12})
 +
  \Bigg[-\frac{2}{3}
 -
  (X_1-X_2)^2
 -
  3\,(X^2_1+X^2_2)\nonumber\\
&+
  2\,(X^3_1+X^3_2)
 +
  8\,(X_1-X_2)(\vec{e}_{(0)}\cdot\vec{n}_{12})
 +
  \frac{45}{4}\,\pi\,(X_1-X_2)(\vec{e}_{\xi}\cdot\vec{n}_{12})\nonumber\\
&+
  \bigg[\frac{2}{3}
 +
  \frac{5}{3}\,(X^2_1+X^2_2)
 -
  6\,(X^3_1+X^3_2)
 +
  3\,(X^4_1+X^4_2)
 +
  \frac{4}{3}\,X_1\,X_2
 -
  2\,X^2_1\,X^2_2\nonumber\\
&+
  2\,X_1\,X_2\,(X^2_1+X^2_2)\bigg](\vec{e}_{(0)}\cdot\vec{n}_{12})^2
 +
  \bigg[\frac{44}{3}
 -
  8\,X_1\,X_2\,(3+X_1\,X_2)
 +
  4\,(X^4_1+X^4_2)\bigg](\vec{e}_{\xi}\cdot\vec{n}_{12})^2\nonumber\\
&+
  \frac{16}{3}\,\bigg[X^2_1
 +
  X^2_2
 -
  X_1\,X_2
 -
  3\,(X^3_1+X^3_2)\bigg](\vec{e}_{(0)}\cdot\vec{n}_{12})^2(\vec{e}_{\xi}\cdot\vec{n}_{12})^2
 -
  16\,(X^3_1+X^3_2)(\vec{e}_{\xi}\cdot\vec{n}_{12})^4\nonumber\\
&+
  \frac{2}{3}\,\bigg[2\,(X^2_1+X^2_2)
 -
  2\,X_1\,X_2
 -
  3\,(X^3_1+X^3_2)\bigg](\vec{e}_{(0)}\cdot\vec{n}_{12})^4\Bigg]\bigg(\frac{r_{12}}{\xi}\bigg)
 +
  \Bigg[\mathcal{O}\bigg(\frac{r_{12}}{\xi}\bigg)^2\Bigg]\Bigg\}e^i_{\xi}\nonumber\\ 
&+\frac{G^2 m^2_1}{c^4\,\xi^2}\Bigg[\frac{15}{4}\, \pi\, X_2\,\bigg(\frac{r_{12}}{\xi}\bigg)
+
  X^2_2\,\bigg(8\,(\vec{e}_{(0)} \cdot \vec{n}_{12})
+
  \frac{45}{4}\,\pi\,(\vec{e}_{\xi} \cdot \vec{n}_{12})\bigg)\bigg(\frac{r_{12}}{\xi}\bigg)^2\nonumber\\
&+ 
  \mathcal{O}\bigg[\bigg(\frac{r_{12}}{\xi}\bigg)^3\bigg]\Bigg] P^i_q n^q_{12}\nonumber\\
&+
  \frac{G^2 m^2_2}{c^4\,\xi^2}\Bigg[-\frac{15}{4}\,\pi \, X_1\,\bigg(\frac{r_{12}}{\xi}\bigg)
+
  X^2_1\bigg(8(\vec{e}_{(0)} \cdot \vec{n}_{12})
+
  \frac{45}{4} \pi(\vec{e}_{\xi} \cdot \vec{n}_{12})\bigg)\bigg(\frac{r_{12}}{\xi}\bigg)^2\nonumber\\
&+ 
  \mathcal{O}\bigg[\bigg(\frac{r_{12}}{\xi}\bigg)^3\bigg]\Bigg] P^i_q n^q_{12}\nonumber\\
&+
  \frac{G^2 m_1m_2}{c^4\,\xi^2}\Bigg\{2\,(X_1-X_2)
 +
  \Bigg[-\frac{15}{4}\,\pi\,(X_1-X_2)\nonumber\\
&+
  \bigg[-\frac{22}{3}
 +
  4\,X_1\,X_2\,(3+X_1\,X_2)
 +
  4\,(X^3_1+X^3_2)
 -
  2\,(X^4_1+X^4_2)\bigg](\vec{e}_{\xi} \cdot \vec{n}_{12})\nonumber\\
&+
  -\frac{8}{3}\,\Big[1-3\,X_1\,X_2\Big](\vec{e}_{\xi} \cdot \vec{n}_{12})(\vec{e}_{(0)} \cdot \vec{n}_{12})^2\Bigg]
\bigg(\frac{r_{12}}{\xi}\bigg)
 + 
  \mathcal{O}\bigg[\bigg(\frac{r_{12}}{\xi}\bigg)^2\bigg]\Bigg\} P^i_qn^q_{12}
+
  \alpha^i_{(2)\mathrm{V}},\label{Totlightdefl2}
\end{align}
where $\alpha^i_{(2)\mathrm{V}}$ is given by the part of $\tilde{\alpha}^i_{(1)(2)}$ (i.e. of Eq.~\eqref{Defl5}), which
is quadratic in $G$.
Here, it should be pointed out that some terms which belong to the post-linear light deflection given
by Eq.~\eqref{Totlightdefl2} (e.g. the term $4\frac{G^2m_1m_2}{c^4 \xi r_{12}}e^i_{\xi}$) are related to terms
in $v^2_{12}$ of the linear light deflection given by Eq.~\eqref{Totlightdefl1} through the virial theorem.
In the next section we shall group these terms together before computing the limit $r_{12}\rightarrow 0$ in 
Eqs.~\eqref{Totlightdefl1} and \eqref{Totlightdefl2} in order to remove the formal divergences.
As in the case of the linear light deflection the order of the expansion in Eq.~\eqref{Totlightdefl2} was chosen
in an arbitrary manner in order to show the structure of the terms belonging to the post-linear light
deflection.
In concrete applications the order of the expansion of the post-linear light deflection is to be chosen
in accordance with the accuracy reached by the linear light deflection.

\section{Results}\label{results}
The resulting expressions for the angle of light deflection linear and quadratic in $G$
are given in Sections \ref{totlightlin} and \ref{totlightpostlin}, respectively.
To study the important features of the derived results, let us consider special cases.\\ 
The results take a particularly simple form for the following cases:
 
\subsubsection*{A) the value of one of the two masses is equal to zero}

If in Eqs~\eqref{Totlightdefl1} and \eqref{Totlightdefl2} we put the value of one of the masses equal to zero 
(e.g. $m_1=M$, $m_2=0$) we obtain expressions for the light deflection angle 
for a static point-like mass, 
\begin{align}
\alpha^i_{(1)\mathrm{(E)}}&=-4\,\frac{G M_{\mathrm{ADM}}}{c^2\,\xi}\,e^i_{\xi},\label{Einstein}\\
\alpha^i_{(2)\mathrm{(E-S)}}&= -\frac{15}{4}\,\pi\,\frac{G^2 M^2_{\mathrm{ADM}}}{c^4\,\xi^2}\,e^i_{\xi},\label{Epstein}
\end{align} 
where in this case the ADM mass $M_{\mathrm{ADM}}$ is  equal to the mass $M$.
The deflection angle linear in $G$ is the well known ``Einstein angle'' (see \cite{EIN1916}). 
The light deflection angle given by Eq.~\eqref{Epstein} is the post-post-Newtonian light deflection
for a point-like mass, which was obtained for the first time by Epstein and Shapiro and by other authors in 1980 
(see \cite{ESH80} and \cite{FIFR80}).
\subsubsection*{B) the light deflection when $r_{12}\rightarrow 0$}
In this subsection we are going to compute the limit of the expression for the linear and post-linear
angle of light deflection (Eqs~\eqref{Totlightdefl1} and \eqref{Totlightdefl2}) in the event that the distance $r_{12}$ between the components of the binary goes towards zero (i.e. $r_{12}\rightarrow 0$).
As we explained at the end of the preceding section, we have to group together the terms of
Eqs.~\eqref{Defl5}, \eqref{Totlightdefl1} and \eqref{Totlightdefl2} in an appropriate manner in order 
to remove the formal divergences.
By inspection, it is clear to see that the remaining terms in Eqs.~\eqref{Defl5}, \eqref{Totlightdefl1}
and  \eqref{Totlightdefl2} are given by
\begin{align}
\alpha^i_{(1)}&=\frac{Gm_1}{c^2\,\xi}\Bigg\{\bigg[-4
 +
  4\,X_2\,\frac{1}{c}\,(\vec{e}_{(0)}\cdot\vec{v}_{12})
 -
  2\,X^2_2\,\frac{v^2_{12}}{c^2}\bigg]
 -
  4\,X^3_2\,(\vec{e}_{\xi}\cdot\vec{n}_{12})\,\frac{v^2_{12}}{c^2}\bigg(\frac{r_{12}}{\xi}\bigg)\Bigg\}e^i_{\xi}\nonumber\\
&+
  \frac{Gm_2}{c^2\,\xi}\Bigg\{\bigg[-4
 -
  4\,X_1\,\frac{1}{c}\,(\vec{e}_{(0)}\cdot\vec{v}_{12})
 -
  2\,X^2_1\,\frac{v^2_{12}}{c^2}\bigg]
 +
  4\,X^3_1\,(\vec{e}_{\xi}\cdot\vec{n}_{12})\,\frac{v^2_{12}}{c^2}\bigg(\frac{r_{12}}{\xi}\bigg)\Bigg\}e^i_{\xi}\nonumber\\
&+
  \Bigg\{2\,\frac{Gm_1}{c^2\,\xi}\,X^3_2
 -
  2\,\frac{Gm_2}{c^2\,\xi}\,X^3_1\Bigg\}\frac{v^2_{12}}{c^2}\bigg(\frac{r_{12}}{\xi}\bigg)P^i_qn^q_{12},\label{lim11}
\end{align}

\begin{align}
\tilde{\alpha}^i_{(1)(2)}&=\Bigg[\frac{\nu (m_1-m_2)}{2 M}\Big[v^2_{12}
 -
  \frac{GM}{r_{12}}\Big]\Bigg]\Bigg\{4\,\frac{Gm_1}{c^4\,\xi}
 +
  4\,\frac{Gm_2}{c^4\,\xi}\Bigg\}\bigg(\frac{r_{12}}{\xi}\bigg)P^i_qn^q_{12},\label{lim12}\\
\alpha^i_{(2)}&=\Bigg\{-\frac{15}{4}\,\pi\,\frac{G^2m^2_1}{c^4\,\xi^2}
 -\frac{15}{4}\,\pi\,\frac{G^2m^2_2}{c^4\,\xi^2}
 +
  4\,\frac{G^2m_1m_2}{c^4\,\xi\,r_{12}}\nonumber\\
&+
  \frac{G^2m_1m_2}{c^4\,\xi^2}\bigg[-\frac{15}{2}\,\pi-4\,(X_1-X_2)\,(\vec{e}_{\xi}\cdot\vec{n}_{12})\bigg]\Bigg\}e^i_{\xi}\nonumber\\
&+
  2\,\frac{G^2m_1m_2}{c^4\,\xi^2}\,(X_1-X_2)P^i_qn^q_{12}.\label{lim13}
\end{align}  
After grouping together the terms that are related through the virial theorem in Eqs.~\eqref{lim11}, \eqref{lim12}
and \eqref{lim13} we get
\begin{align}
\alpha^i_{(1)}&=\Bigg\{\bigg[-4\,\frac{G}{c^2\,\xi}(m_1+m_2)
 -
  2\,\frac{G}{c^2\,\xi}\Big[m_1\,X^2_2+m_2\,X^2_1\Big]\frac{v^2_{12}}{c^2}
 +
  4\,\frac{G^2m_1m_2}{c^4\,\xi\,r_{12}}\bigg]\nonumber\\
&+
  \bigg[-4\,\frac{Gm_1}{c^2\,\xi}\,X^3_2\,\frac{v^2_{12}}{c^2}
 +
  4\,\frac{Gm_2}{c^2\,\xi}\,X^3_1\,\frac{v^2_{12}}{c^2} 
 -
  4\,\frac{G^2m_1m_2}{c^4\,\xi\,r_{12}}(X_1-X_2)\bigg](\vec{e}_{\xi}\cdot\vec{n}_{12})
\bigg(\frac{r_{12}}{\xi}\bigg)\Bigg\}e^i_{\xi}\nonumber\\
&+
  2\,\frac{G}{c^2\,\xi}\Bigg\{\bigg[m_1\,X^3_2
 -
  m_2\,X^3_1\bigg]
 +\frac{\nu (m_1-m_2)}{M}(m_1+m_2)\Bigg\}\frac{v^2_{12}}{c^2}\bigg(\frac{r_{12}}{\xi}\bigg)P^i_qn^q_{12},\label{lim21}\\
\alpha^i_{(2)}&=-\frac{15}{4}\,\pi\,\frac{G^2}{c^4\,\xi^2}(m_1+m_2)^2e^i_{\xi}\nonumber\\
&+
  2\,\Bigg\{\frac{G^2m_1m_2}{c^4\,\xi\,r_{12}}(X_1-X_2)
 -
  \frac{\nu(m_1-m_2)}{M}\frac{GM}{r_{12}}\bigg[\frac{Gm_1}{c^4\,\xi}
 +
  \frac{Gm_2}{c^4\,\xi}\bigg]\Bigg\}\bigg(\frac{r_{12}}{\xi}\bigg)P^i_qn^q_{12}.\label{lim22}
\end{align}
After simplifying Eqs~\eqref{lim21} and \eqref{lim22} we finally obtain
\begin{align}
\alpha^i_{(1)}&=-4\,\frac{GM_{\mathrm{ADM}}}{c^2\,\xi}e^i_{\xi}
 +
  4\,\frac{Gm_1m_2}{c^4\,\xi\,M^2}(m_1-m_2)\Big[v^2_{12}
 -
  \frac{GM}{r_{12}}\Big](\vec{e}_{\xi}\cdot\vec{n}_{12})\bigg(\frac{r_{12}}{\xi}\bigg)e^i_{\xi},\label{lim31}\\
\alpha^i_{(2)}&=-\frac{15}{4}\,\pi\frac{G^2M^2}{c^4\,\xi^2}e^i_{\xi},
\end{align}
where 
\begin{align}
M_{\mathrm{ADM}}=M+\frac{1}{2}\,\mu\,\frac{v^2_{12}}{c^2}-\frac{Gm_1m_2}{c^2\,r_{12}}\label{ADM0}
\end{align}
is the ADM mass of the system with $\mu=m_1m_2/M$ and $M=m_1+m_2$.
In Eqs~\eqref{lim11}-\eqref{lim31}, the quantities $\vec{n}_{12}$, $v_{12}$ and $r_{12}$ are taken at the
time $t^{\ast}$.
Note that the second term in Eq.~\eqref{lim31} goes to zero when $r_{12}\rightarrow 0$ since
the expression in brackets remains finite.
At the end, we recover the Einstein-angle (i.e. Eq.~\eqref{Einstein}) with the ADM mass as given by Eq.~\eqref{ADM0} 
as well as the Epstein-Shapiro angle (i.e. Eq.~\eqref{Epstein}).
\subsubsection*{C) the values of the two masses are equal and the light ray is  originally orthogonal to the 
orbital plane of the binary}
In this case we choose $M/2$ for the value of the masses in Eqs~\eqref{Totlightdefl1} and \eqref{Totlightdefl2}
and assume that the light ray is originally propagating orthogonal to the orbital plane of the two bounded
masses (i.e. $\vec{e}_{(0)}\cdot\vec{n}_{12}=0$, $\vec{e}_{(0)}\cdot\vec{v}_{12}=0$). After introducing 
the ADM mass in the resulting expression for the angle of light deflection linear and quadratic 
in $G$ and rearranging the terms, we finally find 
\begin{align}
\alpha^i_{(1)\bot}&=\frac{G M_{\mathrm{ADM}}}{c^2\,\xi}\Bigg\{-4
 +
  \bigg[1
 -
  4\,(\vec{e}_{\xi}\cdot\vec{n}_{12})^2\bigg] \bigg(\frac{r_{12}}{\xi}\bigg)^2
 +
  \bigg[-\frac{1}{4}
 +
  3\,(\vec{e}_{\xi}\cdot\vec{n}_{12})^2\nonumber\\
&-
  4\,(\vec{e}_{\xi}\cdot\vec{n}_{12})^4\bigg] \bigg(\frac{r_{12}}{\xi}\bigg)^4\nonumber
 +
  \bigg[\frac{1}{16}
 -
  \frac{3}{2}\,(\vec{e}_{\xi}\cdot\vec{n}_{12})^2
 +
  5\,(\vec{e}_{\xi}\cdot\vec{n}_{12})^4\nonumber\\
&-
  4\,(\vec{e}_{\xi}\cdot\vec{n}_{12})^6\bigg] \bigg(\frac{r_{12}}{\xi}\bigg)^6
 +
  \bigg[-\frac{1}{64}
 +
  \frac{5}{8}\,(\vec{e}_{\xi}\cdot\vec{n}_{12})^2
 -
  \frac{15}{4}\,(\vec{e}_{\xi}\cdot\vec{n}_{12})^4\nonumber\\
 &+
  7\,(\vec{e}_{\xi}\cdot\vec{n}_{12})^6
 -
  4\,(\vec{e}_{\xi}\cdot\vec{n}_{12})^8\bigg] \bigg(\frac{r_{12}}{\xi}\bigg)^8
 +
  \bigg[\frac{1}{256}
 -
  \frac{15}{64}\,(\vec{e}_{\xi}\cdot\vec{n}_{12})^2\nonumber\\
&+
  \frac{35}{16}\,(\vec{e}_{\xi}\cdot\vec{n}_{12})^4
 -
  7\,(\vec{e}_{\xi}\cdot\vec{n}_{12})^6+9\,(\vec{e}_{\xi}\cdot\vec{n}_{12})^8
 -
  4\,(\vec{e}_{\xi}\cdot\vec{n}_{12})^{10}\bigg] \bigg(\frac{r_{12}}{\xi}\bigg)^{10}\nonumber\\
&+
  \bigg[-\frac{1}{1024}
 +
  \frac{21}{256}\,(\vec{e}_{\xi}\cdot\vec{n}_{12})^2
 -
  \frac{35}{32}\,(\vec{e}_{\xi}\cdot\vec{n}_{12})^4
 +
  \frac{21}{4}\,(\vec{e}_{\xi}\cdot\vec{n}_{12})^6\nonumber\\
&-
  \frac{45}{4}\,(\vec{e}_{\xi}\cdot\vec{n}_{12})^8
 +
  11\,(\vec{e}_{\xi}\cdot\vec{n}_{12})^{10}-4\,(\vec{e}_{\xi}\cdot\vec{n}_{12})^{12}\bigg] 
\bigg(\frac{r_{12}}{\xi}\bigg)^{12}\nonumber\\
&+
  \mathcal{O}\Bigg[\bigg(\frac{r_{12}}{\xi}\bigg)^{14}\Bigg]\Bigg\}e^i_{\xi}\nonumber\\
&+
  \frac{G M_{\mathrm{ADM}}}{c^2\,\xi}\,(\vec{e}_{\xi}\cdot\vec{n}_{12})\,\Bigg\{2\,\bigg(\frac{{r}_{12}}{\xi}\bigg)^2
 +
  \bigg[-1
 +
  2\,(\vec{e}_{\xi}\cdot\vec{n}_{12})^2\bigg] \bigg(\frac{{r}_{12}}{\xi}\bigg)^4\nonumber\\
&+
  \bigg[\frac{3}{8}
 -
  2\,(\vec{e}_{\xi}\cdot\vec{n}_{12})^2
 +
  2\,(\vec{e}_{\xi}\cdot\vec{n}_{12})^4\bigg] \bigg(\frac{{r}_{12}}{\xi}\bigg)^6
 +
  \bigg[-\frac{1}{8}
 +
  \frac{5}{4}\,(\vec{e}_{\xi}\cdot\vec{n}_{12})^2\nonumber\\
&-
  3\,(\vec{e}_{\xi}\cdot\vec{n}_{12})^4
 +
  2\,(\vec{e}_{\xi}\cdot\vec{n}_{12})^6\bigg] \bigg(\frac{{r}_{12}}{\xi}\bigg)^8
 +
  \bigg[\frac{5}{128}
 -
  \frac{5}{8}\,(\vec{e}_{\xi}\cdot\vec{n}_{12})^2\nonumber\\
&+
  \frac{21}{8}\,(\vec{e}_{\xi}\cdot\vec{n}_{12})^4
 -
  4\,(\vec{e}_{\xi}\cdot\vec{n}_{12})^6
 +
  2\,(\vec{e}_{\xi}\cdot\vec{n}_{12})^8\bigg] \bigg(\frac{{r}_{12}}{\xi}\bigg)^{10}\nonumber\\
 &+
  \bigg[-\frac{3}{256}
 +
  \frac{35}{128}\,(\vec{e}_{\xi}\cdot\vec{n}_{12})^2
 -
  \frac{7}{4}\,(\vec{e}_{\xi}\cdot\vec{n}_{12})^4
 +
  \frac{9}{2}\,(\vec{e}_{\xi}\cdot\vec{n}_{12})^6
 -
  5\,(\vec{e}_{\xi}\cdot\vec{n}_{12})^8\nonumber\\
&+
  2\,(\vec{e}_{\xi}\cdot\vec{n}_{12})^{10}\bigg] \bigg(\frac{{r}_{12}}{\xi}\bigg)^{12}
 +
  \mathcal{O}\Bigg[\bigg(\frac{{r}_{12}}{\xi}\bigg)^{14}\Bigg]\Bigg\}P^i_qn^q_{12}\label{OrthLightdefl1}
\end{align}
and
\begin{align}
\alpha^i_{(2)\bot}&=\frac{G^2 M^2_{\mathrm{ADM}}}{c^4\,\xi^2}\Bigg\{-\frac{15}{4}\,\pi
-
  \frac{1}{6}\,\bigg[1-7\,(\vec{e}_{\xi}\cdot\vec{n}_{12})^2+6\,(\vec{e}_{\xi}\cdot\vec{n}_{12})^4\bigg] \bigg(\frac{r_{12}}{\xi}\bigg)\nonumber\\
&+
  \frac{75}{256}\,\pi\,\bigg[10-31\,(\vec{e}_{\xi}\cdot\vec{n}_{12})^2\bigg] \bigg(\frac{r_{12}}{\xi}\bigg)^2
 +
  \mathcal{O}\Bigg[\bigg(\frac{r_{12}}{\xi}\bigg)^3\Bigg]\Bigg\}e^i_{\xi}\nonumber\\
&+
  \frac{G^2 M^2_{\mathrm{ADM}}}{c^4\,\xi^2}\,(\vec{e}_{\xi}\cdot\vec{n}_{12})\,\Bigg\{-\frac{1}{3}\,\bigg(\frac{r_{12}}{\xi}\bigg)
 +
  \frac{465}{128}\,\pi\,\bigg(\frac{r_{12}}{\xi}\bigg)^2
 +
  \mathcal{O}\Bigg[\bigg(\frac{r_{12}}{\xi}\bigg)^3\Bigg]\Bigg\}P^i_qn^q_{12}\label{OrthLightdefl2}
\end{align}
where in this case the ADM mass is given by
\begin{align}
M_{\mathrm{ADM}}=M\bigg[1+\frac{1}{4}\,\Big(\frac{v^2_{12}}{2 c^2}-\frac{G M}{c^2 r_{12}}\Big)\bigg].\label{ADM}
\end{align}
The expression for the linear light deflection given by Eq.~\eqref{OrthLightdefl1} was expanded to the order $(r_{12}/\xi)^{12}$ in order
to reach the accuracy of the post-linear light deflection given in Eq.~\eqref{OrthLightdefl2}. Note that an expansion
to the order $(r_{12}/\xi)^{12}$ in the linear part of the light deflection which is of the same accuracy as an expansion to the
second order in $(r_{12}/\xi)$ in the post-linear part implies that  $(r_{12}/\xi) \sim (G M_{\mathrm{ADM}}/c^2 \xi)^{1/10}$.
In Eqs~\eqref{OrthLightdefl1}-\eqref{ADM}, the quantities $\vec{n}_{12}$, $v_{12}$ and $r_{12}$ are taken at the time $t^{\ast}$. 
Note, that in this case the correction arising from the shift of the 1PN-centre of mass with respect to the Newtonian 
centre of mass (see Eq.~\eqref{Defl5}) vanishes.

\subsubsection*{D) the values of the two masses are equal and the light ray is originally parallel 
to the orbital plane of the binary}
In this case we choose the value of the masses equal to $M/2$ in Eqs.~\eqref{Totlightdefl1} and \eqref{Totlightdefl2} 
and assume that the light ray is originally propagating parallel to the orbital plane of the binary 
(i.e. $\vec{e}_{\xi}\cdot \vec{n}_{12}=0$). After introducing the ADM mass as given by Eq.~\eqref{ADM} and rearranging the terms, we finally find: 

\begin{align}
\alpha^i_{(1)\|}&=\frac{G M_{\mathrm{ADM}}}{c^2\,\xi}\Bigg\{-4
 +
  \bigg[1
 -
  (\vec{e}_{(0)}\cdot\vec{n}_{12})^2\bigg]\bigg(\frac{r_{12}}{\xi}\bigg)^2
 +
  \bigg[-\frac{1}{4}
 +
  \frac{1}{2}\,(\vec{e}_{(0)}\cdot\vec{n}_{12})^2\nonumber\\
&-
  \frac{1}{4}\,(\vec{e}_{(0)}\cdot\vec{n}_{12})^4\bigg]\bigg(\frac{r_{12}}{\xi}\bigg)^4
 +
  \bigg[\frac{1}{16}
 -
  \frac{3}{16}\,(\vec{e}_{(0)}\cdot\vec{n}_{12})^2
 +
  \frac{3}{16}\,(\vec{e}_{(0)}\cdot\vec{n}_{12})^4\nonumber\\
&-
  \frac{1}{16}\,(\vec{e}_{(0)}\cdot\vec{n}_{12})^6\bigg]\bigg(\frac{r_{12}}{\xi}\bigg)^6
 +
  \bigg[-\frac{1}{64}
 +
  \frac{1}{16}\,(\vec{e}_{(0)}\cdot\vec{n}_{12})^2
 -
  \frac{3}{32}\,(\vec{e}_{(0)}\cdot\vec{n}_{12})^4\nonumber\\
&+
  \frac{1}{16}\,(\vec{e}_{(0)}\cdot\vec{n}_{12})^6
 -
  \frac{1}{64}\,(\vec{e}_{(0)}\cdot\vec{n}_{12})^8\bigg]\bigg(\frac{r_{12}}{\xi}\bigg)^8
 +
  \bigg[\frac{1}{256}
 -
  \frac{5}{256}\,(\vec{e}_{(0)}\cdot\vec{n}_{12})^2\nonumber\\
&+
  \frac{5}{128}\,(\vec{e}_{(0)}\cdot\vec{n}_{12})^4
 -
  \frac{5}{128}\,(\vec{e}_{(0)}\cdot\vec{n}_{12})^6
 +
  \frac{5}{256}\,(\vec{e}_{(0)}\cdot\vec{n}_{12})^8\nonumber\\
&-
  \frac{1}{256}\,(\vec{e}_{(0)}\cdot\vec{n}_{12})^{10}\bigg]\bigg(\frac{r_{12}}{\xi}\bigg)^{10}
 +
  \bigg[-\frac{1}{1024}
 +
  \frac{3}{512}\,(\vec{e}_{(0)}\cdot\vec{n}_{12})^2\nonumber\\
&-
  \frac{15}{1024}\,(\vec{e}_{(0)}\cdot\vec{n}_{12})^4
 +
  \frac{5}{256}\,(\vec{e}_{(0)}\cdot\vec{n}_{12})^6
 -
  \frac{15}{1024}\,(\vec{e}_{(0)}\cdot\vec{n}_{12})^8\nonumber\\
&+
  \frac{3}{512}\,(\vec{e}_{(0)}\cdot\vec{n}_{12})^{10}
 -
  \frac{1}{1024}\,(\vec{e}_{(0)}\cdot\vec{n}_{12})^{12}\bigg]\bigg(\frac{r_{12}}{\xi}\bigg)^{12}
 +
  \mathcal{O}\Bigg[\bigg(\frac{r_{12}}{\xi}\bigg)^{14}\Bigg]\Bigg\}e^i_{\xi}\nonumber\\
&+
  \frac{G M_{\mathrm{ADM}}}{c^3\,\xi}\,\Bigg\{-(\vec{e}_{(0)}\cdot\vec{v}_{12})\,\bigg(\frac{r_{12}}{\xi}\bigg)
 +
  \bigg[\frac{1}{4}\,(\vec{e}_{(0)}\cdot\vec{v}_{12})\,\Big[1
 +
  (\vec{e}_{(0)}\cdot\vec{n}_{12})^2\Big]\nonumber\\
&-
  \frac{1}{2}\,(\vec{n}_{12}\cdot\vec{v}_{12})\,(\vec{e}_{(0)}\cdot\vec{n}_{12})\bigg]\bigg(\frac{r_{12}}{\xi}\bigg)^3
 +
  \bigg[-\frac{1}{16}\,(\vec{e}_{(0)}\cdot\vec{v}_{12})\,\Big[1
 +
  2\,(\vec{e}_{(0)}\cdot\vec{n}_{12})^2\nonumber\\
&-
  3\,(\vec{e}_{(0)}\cdot\vec{n}_{12})^4\Big]
 +
  \frac{1}{4}\,(\vec{n}_{12}\cdot\vec{v}_{12})\,(\vec{e}_{(0)}\cdot\vec{n}_{12})\Big[1
 -
  (\vec{e}_{(0)}\cdot\vec{n}_{12})^2\Big]\bigg]\bigg(\frac{r_{12}}{\xi}\bigg)^5\nonumber\\
&+
  \bigg[\frac{1}{64}\,(\vec{e}_{(0)}\cdot\vec{v}_{12})\,\Big[1
 +
  3\,(\vec{e}_{(0)}\cdot\vec{n}_{12})^2
 -
  9\,(\vec{e}_{(0)}\cdot\vec{n}_{12})^4
 +
  5\,(\vec{e}_{(0)}\cdot\vec{n}_{12})^6\Big]\nonumber\\
&-
  \frac{3}{32}\,(\vec{n}_{12}\cdot\vec{v}_{12})\,(\vec{e}_{(0)}\cdot\vec{n}_{12})\Big[1
 -
  2\,(\vec{e}_{(0)}\cdot\vec{n}_{12})^2
 +
  (\vec{e}_{(0)}\cdot\vec{n}_{12})^4\Big]\bigg]\bigg(\frac{r_{12}}{\xi}\bigg)^7\nonumber\\
&+
  \mathcal{O}\Bigg[\bigg(\frac{r_{12}}{\xi}\bigg)^{9}\Bigg]\Bigg\}P^i_qn^q_{12}\nonumber\\
&+
  \frac{G M_{\mathrm{ADM}}}{c^3\, \xi}\,(\vec{e}_{(0)}\cdot\vec{n}_{12})\,\Bigg\{\frac{1}{2}\,\bigg(\frac{r_{12}}{\xi}\bigg)
 -
  \frac{1}{8}\,\bigg[1
 -
  (\vec{e}_{(0)}\cdot\vec{n}_{12})^2\bigg] \bigg(\frac{r_{12}}{\xi}\bigg)^3\nonumber\\
&+
  \frac{1}{32}\,\bigg[1
 -
  2\,(\vec{e}_{(0)}\cdot\vec{n}_{12})^2
 +
  (\vec{e}_{(0)}\cdot\vec{n}_{12})^4\bigg]\bigg(\frac{r_{12}}{\xi}\bigg)^5\nonumber\\
&-
  \frac{1}{128}\,\bigg[1
 -
  3\,(\vec{e}_{(0)}\cdot\vec{n}_{12})^2
 +
  3\,(\vec{e}_{(0)}\cdot\vec{n}_{12})^4
 -
  (\vec{e}_{(0)}\cdot\vec{n}_{12})^6\bigg]\bigg(\frac{r_{12}}{\xi}\bigg)^7
 +
  \mathcal{O}\Bigg[\bigg(\frac{r_{12}}{\xi}\bigg)^9\Bigg]\Bigg\}P^i_qv^q_{12}\nonumber\\
&+
  \frac{G M_{\mathrm{ADM}}}{c^4\, \xi}(\vec{e}_{(0)}\cdot\vec{n}_{12})^2
\Bigg\{\frac{1}{4}\bigg[v^2_{12}
 -
  (\vec{e}_{(0)}\cdot \vec{v}_{12})^2\bigg]\bigg(\frac{r_{12}}{\xi}\bigg)^2
 +
 \mathcal{O}\Bigg[\bigg(\frac{r_{12}}{\xi}\bigg)^4\Bigg]\Bigg\}e^i_{\xi} \label{ParLightdefl1}
\end{align}
and

\begin{align}
\alpha^i_{(2)\|}&=\frac{G^2 M^2_{\mathrm{ADM}}}{c^4\,\xi^2}\Bigg\{-\frac{15}{4}\,\pi
 -
  \frac{1}{24}\,\bigg[4
 +
  (\vec{e}_{(0)}\cdot\vec{n}_{12})^2
 +
  (\vec{e}_{(0)}\cdot\vec{n}_{12})^4\bigg]\bigg(\frac{r_{12}}{\xi}\bigg)\nonumber\\
&+
  \frac{3}{256}\,\pi\,\bigg[250
 -
  797\,(\vec{e}_{(0)}\cdot\vec{n}_{12})^2\bigg]\bigg(\frac{r_{12}}{\xi}\bigg)^2
 +
  \mathcal{O}\Bigg[\bigg(\frac{r_{12}}{\xi}\bigg)^3\Bigg]\Bigg\}e^i_{\xi}\nonumber\\
&+
  \frac{G^2 M^2_{\mathrm{ADM}}}{c^4\,\xi^2}\,
\Bigg\{4\,(\vec{e}_{(0)}\cdot\vec{n}_{12})\bigg(\frac{r_{12}}{\xi}\bigg)^2
 +
  \mathcal{O}\Bigg[\bigg(\frac{r_{12}}{\xi}\bigg)^3\Bigg]\Bigg\}P^i_qn^q_{12}\label{ParLightdefl2}
\end{align}
where, the quantities $\vec{n}_{12}$, $\vec{v}_{12}$ and $r_{12}$ are taken at the time $t^{\ast}$.  

In Eq.~\eqref{ParLightdefl1} the components $e^i_{\xi}$, $P^i_qn^q_{12}$ and $P^i_qv^q_{12}$ of the linear
light deflection were expanded to the order $(r_{12}/\xi)^{12}$, $(r_{12}/\xi)^{7}$ and $(r_{12}/\xi)^{7}$ 
respectively in order to reach the accuracy of the post-linear light deflection given in Eq.~\eqref{ParLightdefl2}.
As in the preceding subsection here the correction arising from the shift of the 1PN-centre of mass with respect
to the Newtonian centre of mass (see Eq.~\eqref{Defl5}) vanishes.
\end{widetext}

Finally, we apply the formulae for the angle of light deflection \eqref{Einstein}-\eqref{ParLightdefl2} 
to the double pulsar PSR J0737-3039.
The parameters of the pulsar PSR J0737-3039 (e.g. see \cite{KRA04}) are given in table \ref{tab:1}.

\begin{table}
\tabcolsep3em
\caption{\label{tab:1}The parameters of PSR J0737-3039.}
\begin{tabular}{ll}
\hline\hline
Orbital period $P_b$ (day)                  & 0.102251563(1)\\ 
Eccentricity $e$                            & 0.087779(5)\\ 
Total mass $m_A+m_B$ ($M_{\odot}$)   & 2.588(3)\\
Mass ratio $R\equiv m_A/m_B$                & 1.069(6)\\
\hline\hline
\end{tabular}
\end{table}

We compute the angle of light deflection for the cases when the distance between the two stars $r_{12}$
is maximal and minimal. In our computations we assume that the masses of the binary's components are equal, 
i.e. that the mass ratio $R$ is equal to 1. 
For the impact parameter, we choose $\xi=5\,r_{12}$.
In order to compute the angle of light deflection we have first to calculate  $r_{12}$ and $v_{12}$.
Note that we only use Newtonian relations, since the uncertainties in observational data, although small,
are nonetheless greater than the corrections that post-Newtonian corrections would yield.

To compute $r_{12}$ we have to calculate the semi-major axis of the elliptical orbit by means of the following
equation (because of the low accuracy of the observational data, only Newtonian relations can be used),
\begin{align}
a=\sqrt[3]{\frac{G (m_1+m_2) T^2}{4\,\pi^2}},
\end{align} 
where $a$ denotes the semi-major axis and $T$ the orbital period. The preceding equation follows from Kepler's third law
(e.g. see \cite{GOLD02}).
The relationships between the the distances $r_{12\mathrm{max.}}$, $r_{12\mathrm{min.}}$ and the semi-major axis
$a$ are given by,
\begin{align}
r_{12\mathrm{max}}&=a(1+e),\\
r_{12\mathrm{min}}&=a(1-e).
\end{align}
We obtain the corresponding velocities to $r_{12\mathrm{max}}$ and $r_{12\mathrm{min}}$ from the
equations given by
\begin{align}
v_{12\mathrm{min}}=8\,\pi^3 \sqrt{\frac{G (m_1+m_2)}{a}\frac{(1-e)}{(1+e)}},
\end{align}
and
\begin{align}
v_{12\mathrm{max}}=8\, \pi^3 \sqrt{\frac{G (m_1+m_2)}{a}\frac{(1+e)}{(1-e)}}
\end{align}
respectively.

After introducing the parameters of PSR J0737-3039 into the preceding equations and into the  formulae for the light 
deflection given in this section we find the results presented in table \ref{tab:2}.

\begin{table}[h]
\begin{ruledtabular}
\caption{The angles of light deflection linear and quadratic in $G$ are given in arcsec. We denote by  $\alpha^i_{(1)\mathrm{(E)}}$
and $\alpha^i_{(2)\mathrm{(E-S)}}$ the Einstein angle and the Epstein-Shapiro angle. For the light ray
originally orthogonal to the orbital plane we assume that $\vec{e}_{\xi}\cdot\vec{n}_{12}=1$. For the light
ray originally parallel to the orbital plane we assume that $\vec{e}_{(0)}\cdot\vec{n}_{12}=1$.}\label{tab:2}
\begin{tabular}{p{0.05\textwidth}ll}
   & $r_{12\mathrm{max}}=9.56\cdot10^{10}$ cm & $r_{12\mathrm{min}}=8.02\cdot10^{10}$ cm \\ 
   & $v_{12\mathrm{min}}=5.72\cdot10^7⁷$ cm/s & $v_{12\mathrm{max}}=6.83\cdot 10^7$ cm/s \\  
   & $\xi= 5\,r_{12\mathrm{max}}$ & $\xi= 5\,r_{12\mathrm{min}}$ \\ \hline
   
$\alpha^i_{(1)\mathrm{(E)}}$ & $-0.659\,e^i_{\xi}$ & $-0.785\, e^i_{\xi}$ \\ 
 
$\alpha^i_{(2)\mathrm{(E-S)}}$ & $-1.55\cdot10^{-6}\,e^i_{\xi}$ & $-2.20\cdot10^{-6}\,e^i_{\xi}$ \\[3mm] 

$\alpha^i_{(1)\bot}$ &  $-0.679\,e^i_{\xi}+0.013\,P^i_qn^q_{12}$ & $-0.809\, e^i_{\xi}+0.016\,P^i_qn^q_{12}$\\
$\alpha^i_{(2)\bot}$ & $-1.65\cdot10^{-6}\,e^i_{\xi}+5\cdot10^{-8}\,P^i_qn^q_{12}$ & $-2.35\cdot10^{-6}\,e^i_{\xi}
                                                                                         +7\cdot10^{-8}\,P^i_qn^q_{12}$  \\[3mm]

$\alpha^i_{(1)\|}$  & $-0.659\,e^i_{\xi}$ & $-0.785 \, e^i_{\xi}$ \\
$\alpha^i_{(2)\|}$ & $-1.66\cdot10^{-6}\,e^i_{\xi}+2\cdot10^{-8}\,P^i_qn^q_{12}$ & $-2.36\cdot10^{-6}\, e^i_{\xi}+3\cdot10^{-8}\,P^i_qn^q_{12}$ \\ 
\end{tabular}
\end{ruledtabular}
\end{table}
Table \ref{tab:2} shows the angles of light deflection computed by means of Eqs~\eqref{OrthLightdefl2}
and \eqref{ParLightdefl2} corresponding to the Epstein-Shapiro angles of $\alpha^i_{(E-S)}=-1.55\cdot10^{-6}\,e^i_{\xi}$ arcsec 
and $\alpha^i_{(E-S)}=-2.20\cdot10^{-6}\,e^i_{\xi}$ arcsec.
If we define the corrections to the Epstein-Shapiro angle that we calculated in this paper by 
$\delta \alpha^i_{(2)\bot,\|}=\alpha^i_{(2)\bot,\|}-\alpha^i_{(E-S)}$, we find:
\begin{enumerate}
\item light ray originally orthogonal to the orbital plane\\
$\delta \alpha^i_{(2)\bot}=(-1.0\cdot10^{-7}\,e^{i}_{\xi}+5\cdot10^{-8}\,P^i_qn^q_{12})$ arcsec for $\xi=5\,r_{12\mathrm{max}}$,\\
$\delta \alpha^i_{(2)\bot}=(-1.5\cdot10^{-7}\,e^i_{\xi}+7\cdot10^{-8}\,P^i_qn^q_{12})$ arcsec for $\xi=5\,r_{12\mathrm{min}}$.\\
\item light ray originally parallel to the orbital plane\\
$\delta \alpha^i_{(2)\|}=-1.1\cdot10^{-7}\,e^{i}_{\xi}+2\cdot10^{-8}\,P^i_qn^q_{12}$ arcsec for $\xi=5\,r_{12\mathrm{max}}$,\\
$\delta \alpha^i_{(2)\|}=-1.6\cdot10^{-7}\,e^i_{\xi}+3\cdot10^{-8}\,P^i_qn^q_{12}$ arcsec for $\xi=5\,r_{12\mathrm{min}}$.
\end{enumerate}
From the values of $\delta \alpha^i_{(2)\bot}$ and  $\delta \alpha^i_{(2)\|}$ we see that the corrections to the 
Epstein-Shapiro light deflection angle are greater for the case when the light ray is
originally orthogonal to the orbital plane.
Details related to the measurement of the corrections to the Epstein-Shapiro angle will be discussed in the next section.

\section{Discussion and Conclusions}\label{discussion}
The angle of light deflection in the post-linear gravitational field of two bounded point-like masses has been
computed to the second order in $G/c^2$.
Both the light source and the observer were assumed to be located at infinity in an asymptotically flat space.
The light deflection linear in $G$ has been exactly computed. It was shown that the expression obtained
for the linear light deflection is fully equivalent to the expression given by Kopeikin and Sch\"afer
in \cite{KSCH99} in the event that the velocities of the masses are small with respect to the velocity of light and
the retarded times in the expression of Kopeikin and Sch\"afer are close to the time of closest approach of the
unperturbed light ray to the origin of the coordinate system.
To evaluate the integrals related to the light deflection quadratic in $G$, which could not be integrated by
means of elementary functions we resorted to a series expansion of the integrands. For this reason the resulting 
expressions for the angle of light deflection quadratic in $G$ are only valid for the case when the distance between
the two masses $r_{12}$ is smaller than the impact parameter $\xi$ (i.e. $r_{12}/\xi<1$).
The final result is given as a power series in $r_{12}/\xi$.
The expression for the angle of light deflection in terms of the ADM mass to the order $G^2/c^4$ including a power expansion to the second order in $(r_{12}/\xi)$, in which $(r_{12}/\xi) \sim (G M_{\mathrm{ADM}}/c^2 \xi)^{1/10}$ is being
assumed, is given in an explicit form for a binary with equal masses in the event that the light ray is originally 
orthognal to the orbital plane of the binary. The expression for the angle of light deflection in terms of the ADM mass 
to the same order is also given for a binary with equal masses in the event that the light ray is originally parallel 
to the orbital plane of the binary.
For a light ray originally propagating orthogonal to the orbital plane of a binary with equal masses the
deflection angle takes a particularly simple form.
In the case when one of the masses is equal to zero, we obtain the ``Einstein angle'' 
and the ``Shapiro-Epstein light deflection angle'', as we do when $r_{12} \rightarrow 0$. 
Application of the derived formulae for the deflection angle to the double pulsar PSR J0737-3039 has
shown that the corrections to the ``Einstein angle'' are of the order $10^{-2}$ arcsec for the case
when $r_{12}/\xi=0.2$, see table \ref{tab:2}. The corrections to the ``Epstein-Shapiro light deflection angle'' lie between
$10^{-7}$ and $10^{-8}$ arcsec, see table \ref{tab:2}.
We conclude that the corrections to the ``Epstein-Shapiro light deflection angle'' are beyond the sensitivity of the
current astronomical interferometers.
Nevertheless, taking into account that the interferometer for the planned mission
LATOR \cite{TUSHNO04} will be able to measure light deflection angles of the order $10^{-8}$ arcsec, we believe
that the corrections to the ``Epstein-Shapiro light deflection angle'' computed in the present work might be
measured by space-borne interferometers in the foreseeable future.

\begin{acknowledgments}
The author is grateful to G. Sch\"afer for helpful and enlightening discussions
as well as to G. Faye for discussions and useful hints on the use of Mathematica IV.
The author is pleased to thank D. Petroff for the careful reading of the manuscript and for important remarks.
This work was supported in part by a Th\"uringer Landesgraduiertenstipendium and 
by the Deutsche Forschungsgemeinschaft (DFG) through SFB/TR 7 Gravitationswellenastronomie. 
Part of the computations were performed with the help of Mathematica IV.
\end{acknowledgments}

\begin{widetext}
\begin{appendix}
\section{The post-linear light deflection $\alpha^{i}_{(2) \rm{I}}$.}\label{ap:A}
In the integrals that are given in this appendix as well as in Appendices \ref{ap:B} and \ref{ap:C}  we already replaced the photon trajectory by its unperturbed approximation $\vec{z}(\tau)=\tau \,\vec{l}_{(0)}+\vec{\xi}$, where $\vec{l}_{(0)}$ is given by
$\vec{l}_{(0)}= c\,\vec{e}_{(0)}$.
The distances $r_1$ and $r_2$ are given by
\begin{align*}
r_a=\bigg[c^2 \tau^2+\xi^2-2\,c\,\tau \,\vec{e}_{(0)}\cdot \vec{x}_{a}(t^{\ast})- 2\,\vec{\xi}\cdot \vec{x}_{a}(t^{\ast})
+x^{2}_{a}(t^{\ast})\bigg]^{1/2},
\end{align*}
with $a=1$ and $a=2$ for the distances $r_1$ and $r_2$.
The distance $S$ is defined by
\begin{align*}
S=r_1+r_2+r_{12}.
\end{align*}
Here, $r_{12}=|\vec{x}_1(t^{\ast})-\vec{x}_{2}(t^{\ast})|$ is the distance between the two
masses $m_1$ and $m_2$ at the time $t^{\ast}$ and the unit vector $\vec{n}_{12}$ is given by
\begin{align*}
\vec{n}_{12}=\frac{1}{r_{12}}\,\big[\vec{x}_{1}(t^{\ast})-\vec{x}_2(t^{\ast})\big].
\end{align*}
The positions of the masses in the centre of mass frame without considering the 1PN-corrections are given by
\begin{align*}
\vec{x}_1=X_2\,\vec{r}_{12}(t^{\ast})
\end{align*}
and
\begin{align*}
\vec{x}_2=-X_1\,\vec{r}_{12}(t^{\ast}).
\end{align*}

The integrals resulting from the introduction of the post-linear metric coefficients \eqref{Metric2} and
\eqref{Metric3} into the expression for $\alpha^i_{(2) \mathrm{I}}$ given by Eq.~\eqref{deflection1} are:

\begin{align}
\alpha^{i}_{(2)\mathrm{I}}(\vec{\xi}\,)&=2\,\frac{G^2 m^{2}_{1}}{c^3} \int^{\infty}_{-\infty}
d \tau \,\frac{1}{r^6_1}\,\big[c\,\tau - X_2\,\vec{e}_{(0)} \cdot \vec{r}_{12}(t^{\ast})\big]^2 
\big[\xi^i-X_2\,P^i_q r^q_{12}(t^{\ast})\big]\nonumber\\
&-4\,\frac{G^2 m_1 m_2}{c^3}\int^{\infty}_{-\infty}
d \tau \, \frac{1}{r_{12}r_1 S^2}\,\big[\xi^i-X_2\,P^i_q r^q_{12}(t^{\ast})\big]\nonumber\\
&+\frac{G^2 m_1 m_2}{c^3}\int^{\infty}_{-\infty}
d \tau \,\frac{1}{r^{3}_{12}}\,\Big[\frac{1}{r_2}-\frac{1}{r_1}\Big]\,\big[\xi^i-X_2\,P^i_q r^q_{12}(t^{\ast})\big]\nonumber\\   
&+\frac{1}{2}\,\frac{G^2 m_1 m_2}{c^3}\int^{\infty}_{-\infty}
d \tau \,\frac{1}{r^{3}_{12}}\,\Big[\frac{1}{r_1}-\frac{r^{2}_{2}}{r^{3}_{1}}\Big]\,\big[\xi^i-X_2\,P^i_q r^q_{12}(t^{\ast})\big]
\nonumber\\
&+16\,\frac{G^2 m_1 m_2}{c^3}\int^{\infty}_{-\infty}
d \tau \,\frac{1}{r_1r_2 S^{3}}\,(\vec{e}_{(0)}\cdot\vec{n}_{12})\,\big[c\,\tau + X_1\,\vec{e}_{(0)}\cdot \vec{r}_{12}(t^{\ast})\big]
\nonumber\\
&\big[\xi^i-X_2\,P^i_q r^q_{12}(t^{\ast})\big]\nonumber\\
&+8\,\frac{G^2 m_1 m_2}{c^3} \int^{\infty}_{-\infty} 
d \tau \,\frac{1}{r_1 S^3}\,(\vec{e}_{(0)} \cdot \vec{n}_{12})^2\,\big[\xi^i-X_2\,P^i_q r^q_{12}(t^{\ast})\big]\nonumber\\
&+4\,\frac{G^2 m_1 m_2}{c^3} \int^{\infty}_{-\infty} 
d \tau \,\frac{1}{r_{12}r_1 S^2}\,(\vec{e}_{(0)} \cdot \vec{n}_{12})^2\,\big[\xi^i-X_2 P^i_q r^q_{12}(t^{\ast})\big]\nonumber\\
&+\frac{5}{2}\,\frac{G^2 m_1 m_2}{c^3} \int^{\infty}_{-\infty} 
d \tau \,\frac{1}{r_{12}r^{3}_{1}}\,\big[\xi^i-X_2\,P^i_q r^q_{12}(t^{\ast})\big]\nonumber\\
&-16\,\frac{G^2 m_1 m_2}{c^3} \int^{\infty}_{-\infty}
d \tau \, \frac{1}{r_{12}r_1 S^3}\,(\vec{e}_{(0)} \cdot \vec{n}_{12})\,\big[c\,\tau-X_2\,\vec{e}_{(0)} 
\cdot \vec{r}_{12}(t^{\ast})\big]\,P^i_qr^q_{12}(t^{\ast})\nonumber\\
&-8\,\frac{G^2 m_1 m_2}{c^3} \int^{\infty}_{-\infty}
d \tau \, \frac{1}{r^2_{12}r_1S^2}\,(\vec{e}_{(0)} \cdot \vec{n}_{12})\, 
\big[c\,\tau -X_2\,\vec{e}_{(0)} \cdot \vec{r}_{12}(t^{\ast})\big]\,P^i_qr^q_{12}(t^{\ast})\nonumber\\
&-4\,\frac{G^2 m_1 m_2}{c^3} \int^{\infty}_{-\infty}
d \tau \, \frac{1}{r_1r_2S^2}\,\big[\xi^i-X_2\,P^i_q r^q_{12}(t^{\ast})\big]\nonumber\\
&+4\,\frac{G^2 m_1 m_2}{c^3} \int^{\infty}_{-\infty}
d \tau \,\frac{1}{r_1r^3_2S^2}\,\big[c\,\tau+X_1\,\vec{e}_{(0)} \cdot \vec{r}_{12}(t^{\ast})\big]^2\,
\big[\xi^i-X_2\,P^i_q r^q_{12}(t^{\ast})\big] \nonumber\\
&+8\,\frac{G^2 m_1 m_2}{c^3}\int^{\infty}_{-\infty}
d \tau \, \frac{1}{r_1r^2_2S^3}\,\big[c\,\tau+X_1\,\vec{e}_{(0)} \cdot \vec{r}_{12}(t^{\ast})\big]^2\,
\big[\xi^i-X_2\,P^i_q r^q_{12}(t^{\ast})\big]\nonumber\\
&-4\,\frac{G^2 m_1 m_2}{c^3} \int^{\infty}_{-\infty}
d \tau \, \Big[\frac{1}{r_{12}r_1S^2}-\frac{1}{r_{12}r^3_1S^2}\,
\big[c\,\tau-X_2\,\vec{e}_{(0)}\cdot \vec{r}_{12}(t^{\ast})\big]^2\,\Big]\,P^i_qr^q_{12}(t^{\ast})\nonumber\\
&+8 \frac{G^2 m_1 m_2}{c^3}\int^{\infty}_{-\infty}
d \tau \, \frac{1}{r_{12}r^2_1S^3}\big[c \tau-X_2\vec{e}_{(0)}\cdot \vec{r}_{12}(t^{\ast})\big]^2
P^i_qr^q_{12}(t^{\ast})\nonumber\\
&+8\,\frac{G^2 m_1 m_2}{c^3}\int^{\infty}_{-\infty}
d \tau \,\frac{1}{r_{12}r_1r_2S^3}\,\big[c\,\tau -X_2\,\vec{e}_{(0)} \cdot \vec{r}_{12}(t^{\ast})\big]\nonumber\\
&\big[c\,\tau+X_1\,\vec{e}_{(0)}\cdot \vec{r}_{12}(t^{\ast})\big]\,P^i_qr^q_{12}(t^{\ast})\nonumber\\
&+(1 \leftrightarrow 2).  
\end{align}

\section{The post-linear light deflection $\alpha^{i}_{(2) \mathrm{II}}$}\label{ap:B}

After introducing the expressions for the perturbations $\delta l^i_{(1)}(\tau)$ given
by Eq.~\eqref{Perturbed3} and the metric coefficients \eqref{Metric4} and \eqref{Metric5}
into the expression for $\alpha^i_{(2)\mathrm{II}}$ given by Eq.~\eqref{deflection2} we find,

\begin{align}
\alpha^{i}_{(2)\rm{II}}(\vec{\xi}\,)&=12\,\frac{G^2 m^2_1}{c^3}\int^{\infty}_{-\infty}
d \tau \, \frac{1}{r^4_1}\,\big[\xi^i-X_2\,P^i_q r^q_{12}(t^{\ast})\big]  \nonumber\\
&+12\,\frac{G^2 m_1 m_2}{c^3}\int^{\infty}_{-\infty}
d \tau \, \frac{1}{r^3_1 r_2}\,\big[\xi^i-X_2\,P^i_q r^q_{12}(t^{\ast})\big] \nonumber\\
&+4\,\frac{G^2 m^2_1}{c^3}\int^{\infty}_{-\infty}
d \tau \,\frac{1}{r^3_1}\,A_1\,\big[\xi^i-X_2\,P^i_q r^q_{12}(t^{\ast})\big] \nonumber\\
&+4\,\frac{G^2 m_1 m_2}{c^3}\int^{\infty}_{-\infty}
d \tau \, \frac{1}{r^3_1}\,A_2\,\big[\xi^i-X_2\,P^i_q r^q_{12}(t^{\ast})\big] \nonumber\\
&-4\,\frac{G^2 m^2_1}{c^3} \int^{\infty}_{-\infty}
d \tau \,\frac{1}{r^3_1}\,B_1\,\big[\vec{e}_{(0)} \cdot (X_2\,\vec{r}_{12}(t^{\ast}))\big]\, 
\big[\xi^i-X_2\,P^i_q r^q_{12}(t^{\ast})\big] \nonumber\\ 
&+4\,\frac{G^2 m_1 m_2}{c^3} \int^{\infty}_{-\infty}
d \tau \,\frac{1}{r^3_1}\,B_2\,\big[\vec{e}_{(0)} \cdot (X_1 \vec{r}_{12}(t^{\ast}))\big]\,
\big[\xi^i-X_2\,P^i_q r^q_{12}(t^{\ast})\big] \nonumber\\ 
&-8\,\frac{G^2 m^2_1}{c^3}\int^{\infty}_{-\infty}
d \tau \, \frac{1}{r^3_1}\,B_1\,\big[c\,\tau-X_2\,\vec{e}_{(0)} \cdot \vec{r}_{12}(t^{\ast})\big]\,
\big[\xi^i-X_2\, P^i_q r^q_{12}(t^{\ast})\big]\nonumber\\
&-8\,\frac{G^2 m_1 m_2}{c^3}\int^{\infty}_{-\infty}
d \tau \, \frac{1}{r^3_2}\,B_1\,\big[c\,\tau+X_1\,\vec{e}_{(0)} \cdot \vec{r}_{12}(t^{\ast})\big]\,
\big[\xi^i-X_2 P^i_q r^q_{12}(t^{\ast})\big] \nonumber\\
&+(1\leftrightarrow 2).
\end{align}
The functions $A_1$, $A_2$, $B_1$ and $B_2$ are given in Section \ref{lightlingrav}.

\section{The post-linear light deflection $\alpha^{i}_{(2) \mathrm{IV}}$}\label{ap:C}
In this appendix we give the integrals of the order $G^2/c^4$ resulting from the introduction of the perturbation
$\delta z^i_{(1)}(\tau)$ given by Eq.~\eqref{Perturbation3} into the expression for $\alpha^{i}_{(2)\mathrm{III}}$
given by Eq.~\eqref{Deflection3}.
\begin{align}
&\alpha^{i}_{(2) \mathrm{IV}}(\vec{\xi}\,)=-12 \frac{G^2 m^{2}_{1}}{c^3} r^2_1(0,t^{\ast}) \int^{\infty}_{-\infty} d \tau \,
  \frac{1}{r^5_1}\mathcal{B}_1 \big[\xi^i-X_2\,P^i_q r^q_{12}(t^{\ast})\big]\nonumber\\
&-
  12 \frac{G^2 m_1 m_2}{c^3}(\vec{r}_1(0,t^{\ast})\cdot \vec{r}_2(0,t^{\ast})) \int^{\infty}_{-\infty} d \tau \,
  \frac{1}{r^5_1}\mathcal{B}_2 \big[\xi^i-X_2\, P^i_q r^q_{12}(t^{\ast})\big]\nonumber\\
&+
  12 \frac{G^2 m^2_1}{c^3}(X_2\, \vec{e}_{(0)}\cdot\vec{r}_{12}(t^{\ast}))^2 \int^{\infty}_{-\infty} d \tau \,
\frac{1}{r^5_1} \mathcal{B}_1 \big[\xi^i-X_2\, P^i_q r^q_{12}(t^{\ast})\big]\nonumber\\
&-
  12 \frac{G^2 m_1m_2}{c^3}(X_1\,X_2)(\vec{e}_{(0)}\cdot\vec{r}_{12}(t^{\ast}))^2 \int^{\infty}_{-\infty} d \tau \,
\frac{1}{r^5_1} \mathcal{B}_2 \big[\xi^i-X_2\, P^i_q r^q_{12}(t^{\ast})\big]\nonumber\\
&-
  12 \frac{G^2 m^2_1}{c^3} \int^{\infty}_{-\infty} d \tau \,
  \frac{1}{r^5_1} \ln \bigg[\frac{c \tau - X_2\,\vec{e}_{(0)} \cdot \vec{r}_{12}(t^{\ast})
 +
  r_1(\tau,t^{\ast})}{r_1(0,t^{\ast})-X_2\,\vec{e}_{(0)} \cdot \vec{r}_{12}(t^{\ast})}\bigg]\nonumber\\
& \big[c \tau -X_2\,\vec{e}_{(0)} \cdot \vec{r}_{12}(t^{\ast})\big]\big[\xi^i-X_2\,P^i_q r^q_{12}(t^{\ast})\big] \nonumber\\
&-
  12 \frac{G^2 m_1m_2}{c^3} \int^{\infty}_{-\infty} d \tau \,
  \frac{1}{r^5_1} \ln \bigg[\frac{c \tau+ X_1\,\vec{e}_{(0)} \cdot \vec{r}_{12}(t^{\ast})) 
 +
  r_2(\tau,t^{\ast})}{r_2(0,t^{\ast})+X_1\,\vec{e}_{(0)}\cdot \vec{r}_{12}(t^{\ast})}\bigg] \nonumber\\ 
& \big[c \tau -X_2\, \vec{e}_{(0)} \cdot \vec{r}_{12}(t^{\ast})\big]\big[\xi^i-X_2\, P^i_q r^q_{12}(t^{\ast})\big]\nonumber\\
&+
  4 \frac{G^2 m^2_1}{c^3}\int^{\infty}_{-\infty} d \tau \,
  \frac{1}{r^3_1} \mathcal{B}_1\big[\xi^i-X_2\, P^i_q r^q_{12}(t^{\ast})\big] \nonumber\\
&+
  4 \frac{G^2 m_1m_2}{c^3}\int^{\infty}_{-\infty} d \tau \,
  \frac{1}{r^3_2} \mathcal{B}_1\big[\xi^i-X_2\, P^i_q r^q_{12}(t^{\ast})\big] \nonumber\\
&+(1 \leftrightarrow 2),
\end{align}
where the functions $\mathcal{B}_1$ and $\mathcal{B}_2$ are given in Section \ref{trajectory}.

\section{The linear light deflection terms arising from the terms of $h^{(1)}_{00}$ and
$h^{(1)}_{pq}$ which contain the accelerations of the masses}\label{ap:D}
As we mentioned in Section \ref{gravfield}, the terms of the metric coefficients $h^{(1)}_{00}$ and
$h^{(1)}_{pq}$ which contain the accelerations of the masses were introduced into the metric
quadratic in $G$ after substituting the accelerations by explicit functionals of coordinate positions
of the masses by means of the Newtonian equations of motion.
To get the light deflection terms arising from these terms in a form suitable for the comparison of our
computations with the linear light deflection computed by Kopeikin and Sch\"afer \cite{KSCH99},
we compute here the light deflection resulting from these terms before performing the substitution
of the accelerations. The terms of $h^{(1)}_{00}$ and $h^{(1)}_{pq}$ which contains the accelerations
are given by
\begin{align}
&\tilde{h}^{(1)}_{00}=-\frac{G}{c^4} \sum_{a=1}^{2} m_a (\vec{n}_a \cdot \vec{a}_a)
=-\frac{G}{c^4} \sum_{a=1}^{2} m_a \frac{(\vec{r}_a \cdot \vec{a}_a)}{r_a},\nonumber\\
&\tilde{h}^{(1)}_{pq}=\tilde{h}^{(1)}_{00} \delta_{pq}.\label{metricacc}
\end{align}
From Eq.~\eqref{deflection} it follows that the linear light deflection is given by
\begin{align}
\alpha^i_{(1)}=\lim_{\tau\to\infty}\Bigg\{\frac{1}{c}P^i_q\delta l^q_{(1)}(\tau)\Bigg\},
\end{align}
where $\delta l^q_{(1)}(\tau)$ is given by Eq.~\eqref{Perturbation11}.
After introducing the metric coefficients \eqref{metricacc} into the equation above we obtain
\begin{align}
\tilde\alpha^i_{(1)}&=\frac{G}{c^3}\sum_{a=1}^{2} m_a \int_{-\infty}^{\infty} d \tau \,
\frac{1}{r^3_a} \Big[c \tau (\vec{e}_{(0)}\cdot \vec{a}_a(t^{\ast}))+\vec{\xi}\cdot \vec{a}_a(t^{\ast})
-\vec{x}_a(t^{\ast})\cdot\vec{a}_a(t^{\ast})\Big]\big[\xi^i-P^i_qx^q_a(t^{\ast})\big]\nonumber\\
&-\frac{G}{c^3}\sum_{a=1}^{2} m_a \int_{-\infty}^{\infty} d \tau \,\frac{1}{r_a}P^i_qa^q_a(t^{\ast}).
\label{acceleration1}
\end{align}
As in Eq.~\eqref{motion2} the second integral in the preceding equation diverges. After performing the Taylor 
expansion of the second integrand about the origin of the coordinate system $\vec{x}_a=0$ up to the second order
and taking into account the Newtonian centre of mass theorem we perform the integration of 
Eq.~\eqref{acceleration1}. As result we find
\begin{align}
\tilde\alpha^i_{(1)}&=2 \frac{G}{c^4} \sum_{a=1}^{2} \frac{m_a}{R_a}(\vec{e}_{(0)}\cdot\vec{x}_a(t^{\ast}))
(\vec{e}_{(0)}\cdot\vec{a}_a(t^{\ast}))\big[\xi^i-P^i_qx^q_a(t^{\ast})\big]\nonumber\\
&+2\frac{G}{c^4} \sum_{a=1}^{2} \frac{m_a}{R_a}\Big[\vec{\xi}\cdot\vec{a}_a(t^{\ast})
-\vec{x}_a(t^{\ast})\cdot\vec{a}_a(t^{\ast})\Big]\big[\xi^i-P^i_qx^q_a(t^{\ast})\big]\nonumber\\
&+\frac{G}{c^4}\sum_{a=1}^{2} m_a\Bigg\{-2\frac{(\vec{\xi}\cdot\vec{x}_a(t^{\ast}))}{\xi^2}
-\frac{(\vec{e}_{(0)}\cdot\vec{x}_a(t^{\ast}))^2}{\xi^2}
+\frac{x^2_a(t^{\ast})}{\xi^2}-2\frac{(\vec{\xi}\cdot\vec{x}_a(t^{\ast}))^2}{\xi^4}\Bigg\}P^i_qa^q_a(t^{\ast}).
\label{acceleration2}
\end{align} 

\section{The angle of light deflection in the linear gravitational field of two arbitrarily moving point-like 
masses}\label{ap:E}
In this appendix we show that the expression for the light deflection angle in the linear gravitational field of two
arbritrarily moving point-like masses computed by Kopeikin and Sch\"afer \cite{KSCH99} is fully equivalent to the linear light deflection computed in this paper in the event that the velocities of the masses
are small with respect to the velocity of light and that the retarded times in the Kopeikin and Sch\"afer expression are close to the time of closest approach $t^{\ast}$.
The light deflection angle given in \cite{KSCH99} in the case that the source is located at infinity and the observer 
at $\vec{z}(\tau)$ is given by
\begin{align}
\alpha^i_{(1)}(\tau)&=-2 \frac{G}{c^2} \sum_{a=1}^{2} \frac{m_a 
\big[1-\frac{\vec{e}_{(0)} \cdot \vec{v}_a(s_a)}{c}\big]^2 \big[r_a(\tau,s_a)+(\vec{e}_{(0)}\cdot 
\vec{r}_a(\tau,s_a))\big]P^i_qr^q_a(\tau,s_a)}
{\sqrt{1-\frac{v^2(s_a)}{c^2}}
\Big[r^2_a(\tau,s_a)-(\vec{e}_{(0)}\cdot \vec{r}_a(\tau,s_a))^2\Big]
\Big[r_a(\tau,s_a)-\frac{\vec{v}_a(s_a)\cdot \vec{r}_a(\tau,s_a)}{c}\Big]}\nonumber\\
&-4\frac{G}{c^3} \sum_{a=1}^{2} \frac{m_a \big[1-\frac{\vec{e}_{(0)} \cdot \vec{v}_a(s_a)}{c}\big]}
{\sqrt{1-\frac{v^2(s_a)}{c^2}}\Big[r_a(\tau,s_a)-\frac{\vec{v}_a(s_a)\cdot \vec{r}_a(\tau,s_a)}{c}\Big]}
P^i_qv^q_a(s_a),\label{KSCH1}
\end{align}
where $\vec{v}_a(s_a)$ is the velocity of the a-th mass, $\vec{e}_{(0)}$ is the unit vector
tangent to the unperturbed light ray, $\vec{r}_a(\tau,s_a)$ is given by 
$\vec{r}_a(\tau,s_a)=\vec{z}(\tau)-\vec{x}_a(s_a)$ and $r_a(\tau,s_a)$ is the Euclidean norm
of $\vec{r}_a(\tau,s_a)$.
Here, $s_a$ is the retarded time for the a-th mass defined by the light cone equation for 
$s_a$. The light-cone equation for $s_a$ is given by
\begin{align}
s_a+r_a(\tau,s_a)=\tau+t^{\ast}.\label{KSCH2}
\end{align}
From Eq.~\eqref{KSCH1} it follows that for an observer located at infinity, the angle of light deflection
is given by,
\begin{align}
\alpha^i_{(1)}&=\lim_{\tau\to\infty} \alpha^i_{(1)}(\tau)\nonumber\\ 
&=-4 \frac{G}{c^2}\sum_{a=1}^{2} \frac{m_a \big[1-\frac{\vec{e}_{(0)} \cdot \vec{v}_a(s_a)}{c}\big]}
{\sqrt{1-\frac{v^2(s_a)}{c^2}}R_a(s_a)}\big[\xi^i-P^i_qx^q_a(s_a)\big],\label{KSCH3}
\end{align}
where the quantity $R_a(s_a)$ is given by
\begin{align}
R_a(s_a)=r^2_a(0,s_a)-(\vec{e}_{(0)}\cdot \vec{x}_a(s_a))^2.
\end{align}
Here, it is worthwhile to note that the preceding expression for the light deflection angle is
equivalent to the expression given by Eq.~(139) in \cite{KSCH99}.

With the help of the light-cone equation \eqref{KSCH2} and the relationship for the time of closest
approach from \cite{KSCH99},
\begin{align}
t^{\ast}&=t-\frac{1}{c}(\vec{e}_{(0)}\cdot \vec{z}(t))\nonumber\\
&\simeq t-\frac{1}{c} \vec{e}_{(0)}\cdot \vec{r}_a(t,s_a)-\frac{1}{c}\vec{e}_{(0)}\cdot \vec{x}_a(s_a),
\end{align}
it was shown that
\begin{align}
s_a-t^{\ast}&=\frac{1}{c}\vec{e}_{(0)}\cdot \vec{x}_a(s_a)-\frac{d^2_a}{2r_a}\nonumber\\
&\simeq \frac{1}{c}\vec{e}_{(0)}\cdot \vec{x}_a(s_a),\label{KSCH4}
\end{align}
where $d_a$ is given by $d_a=|\vec{z}_a(s_a)-\vec{x}_a(s_a)|$.

If the velocities of the masses are small with respect to the velocity of light and the retarded times do not
differ significantly from the time of closest approach $t^{\ast}$, we are allowed to use the Taylor expansion of the quantity
\begin{align}
x^i_a(s_a) \simeq x^i_a(t^{\ast})+v^i_a(t^{\ast})(s_a-t^{\ast})+\frac{1}{2}a^i_a(t^{\ast}))(s_a-t^{\ast})^2 
\label{KSCH5}
\end{align}
After substituting into Eq.~\eqref{KSCH4}, we find
\begin{align}
s_a-t^{\ast} \simeq \frac{1}{c} (\vec{e}_{(0)} \cdot \vec{x}_a(t^{\ast}))
+\frac{1}{c}(\vec{e}_{(0)}\cdot \vec{v}_a(t^{\ast}))(s_a-t^{\ast})
+\frac{1}{2c}(\vec{e}_{(0)}\cdot \vec{a}_a(t^{\ast}))(s_a-t^{\ast})^2.\label{KSCH6}
\end{align}
Now we solve Eq.~\eqref{KSCH6} iteratively with respect to $(s_a-t^{\ast})$ to obtain
\begin{align}
s_a-t^{\ast} \simeq \frac{1}{c} \vec{e}_{(0)}\cdot\vec{x}_a(t^{\ast})
+\frac{1}{c^2}(\vec{e}_{(0)}\cdot\vec{x}_a(t^{\ast}))(\vec{e}_{(0)}\cdot\vec{v}_a(t^{\ast}))
+\mathcal{O}\Big(\frac{1}{c^3}\Big). \label{KSCH7}
\end{align}
After performing the Taylor expansion of the expression for the light deflection angle given by
Eq.~\eqref{KSCH3} and taking into account Eq.~\eqref{KSCH7}, we finally obtain  

\begin{align}
&\alpha^i_{(1)}=-4\frac{G}{c^2} \sum_{a=1}^{2} \frac{m_a}{R_a}\big[\xi^i-P^i_qx^q_a(t^{\ast})\big]\nonumber\\
&+4\frac{G}{c^3}\sum_{a=1}^{2} \frac{m_a}{R_a}(\vec{e}_{(0)}\cdot\vec{v}_{a}(t^{\ast}))\big[\xi^i-P^i_qx^q_a(t^{\ast})\big]
\nonumber\\
&+4 \frac{G}{c^3}\sum_{a=1}^{2} \frac{m_a}{R_a}(\vec{e}_{(0)}\cdot\vec{x}_{a}(t^{\ast}))P^i_q v^q_a(t^{\ast})
\nonumber\\
&-8 \frac{G}{c^3}\sum_{a=1}^{2} \frac{m_a}{R^2_a}(\vec{e}_{(0)}\cdot\vec{x}_a(t^{\ast}))
\Big[\vec{\xi}\cdot\vec{v}_a(t^{\ast})-\vec{x}_a(t^{\ast})\cdot\vec{v}_a(t^{\ast})
+(\vec{e}_{(0)}\cdot\vec{x}_a(t^{\ast}))(\vec{e}_{(0)}\cdot\vec{v}_a(t^{\ast}))\Big]\nonumber\\
&\big[\xi^i-P^i_qx^q_a(t^{\ast})\big]\nonumber\\
&-2 \frac{G}{c^4}\sum_{a=1}^{2} \frac{m_a}{R_a}v^2_a(t^{\ast})\big[\xi^i-P^i_qx^q_a(t^{\ast})\big]\nonumber\\
&+4 \frac{G}{c^4}\sum_{a=1}^{2}\frac{m_a}{R^2_a}v^2_a(t^{\ast})(\vec{e}_{(0)}\cdot\vec{x}_a(t^{\ast}))^2
\big[\xi^i-P^i_qx^q_a(t^{\ast})\big]\nonumber\\
&-16\frac{G}{c^4}\sum_{a=1}^{2} \frac{m_a}{R^3_a}(\vec{e}_{(0)}\cdot\vec{x}_a(t^{\ast}))^2
\Big[\vec{\xi}\cdot\vec{v}_a(t^{\ast})-\vec{x}_a(t^{\ast})\cdot\vec{v}_a(t^{\ast})\Big]^2 \nonumber\\
&\big[\xi^i-P^i_qx^q_a(t^{\ast})\big]\nonumber\\
&-32 \frac{G}{c^4}\sum_{a=1}^{2} \frac{m_a}{R^3_a}(\vec{e}_{(0)}\cdot\vec{x}_a(t^{\ast}))^3
(\vec{e}_{(0)}\cdot\vec{v}_a(t^{\ast}))
\Big[\vec{\xi}\cdot\vec{v}_a(t^{\ast})-\vec{x}_a(t^{\ast})\cdot\vec{v}_a(t^{\ast})\Big]
\big[\xi^i-P^i_qx^q_a(t^{\ast})\big]\nonumber\\
&-16\frac{G}{c^4}\sum_{a=1}^{2}\frac{m_a}{R^3_a}(\vec{e}_{(0)}\cdot\vec{x}_a(t^{\ast}))^4
(\vec{e}_{(0)}\cdot\vec{v}_a(t^{\ast}))^2 \big[\xi^i-P^i_qx^q_a(t^{\ast})\big]\nonumber\\
&+8\frac{G}{c^4}\sum_{a=1}^{2}\frac{m_a}{R^2_a}(\vec{e}_{(0)}\cdot\vec{x}_a(t^{\ast}))^2
\Big[\vec{\xi}\cdot\vec{v}_a(t^{\ast})-\vec{x}_a(t^{\ast})\cdot\vec{v}_a(t^{\ast})
+(\vec{e}_{(0)}\cdot\vec{x}_a(t^{\ast}))(\vec{e}_{(0)}\cdot\vec{v}_a(t^{\ast}))\Big]P^i_qv^q_a(t^{\ast})\nonumber\\
&-4\frac{G}{c^4}\sum_{a=1}^{2}\frac{m_a}{R^2_a}(\vec{e}_{(0)}\cdot\vec{x}_a(t^{\ast}))^2
(\vec{e}_{(0)}\cdot\vec{v}_a(t^{\ast}))^2 \big[\xi^i-P^i_qx^q_a(t^{\ast})\big]\nonumber\\
&-4 \frac{G}{c^4}\sum_{a=1}^{2}\frac{m_a}{R^2_a}(\vec{e}_{(0)}\cdot\vec{x}_a(t^{\ast}))^2
\Big[\vec{\xi}\cdot\vec{a}_a(t^{\ast})-\vec{x}_a(t^{\ast})\cdot\vec{a}_a(t^{\ast})
+(\vec{e}_{(0)}\cdot\vec{x}_a(t^{\ast}))(\vec{e}_{(0)}\cdot\vec{a}_a(t^{\ast}))\Big]\nonumber\\
&\big[\xi^i-P^i_qx^q_a(t^{\ast})\big]\nonumber\\
&+2\frac{G}{c^4}\sum_{a=1}^{2}\frac{m_a}{R_a}(\vec{e}_{(0)}\cdot\vec{x}_a(t^{\ast}))^2P^i_qa^q_a(t^{\ast})
\nonumber\\
&+4\frac{G}{c^4}\sum_{a=1}^{2}\frac{m_a}{R_a}(\vec{e}_{(0)}\cdot\vec{x}_a(t^{\ast}))(\vec{e}_{(0)}\cdot\vec{a}_a(t^{\ast}))
\big[\xi^i-P^i_qx^q_a(t^{\ast})\big].\label{KSCH8}
\end{align}
It is easy to see that the expression above is equal to the sum of Eqs \eqref{Defl4a}, \eqref{Totlineardefl} and
\eqref{acceleration2}. Note, that to obtain the term in $P^i_a a^q_a(t^{\ast})$ of Eq.~\eqref{KSCH8}, it is better 
to sum up the corresponding terms in Eqs~\eqref{motion2} and \eqref{acceleration1} before performing the 
integration in order to remove the formal divergences. After summing up and performing the integration of these terms we get:
\begin{align}
&\frac{G}{c}\sum_{a=1}^{2} m_a \int_{-\infty}^{\infty} d \tau \,\frac{\tau^2}{r^3_a}P^i_qa^q_a(t^{\ast})
 -
  \frac{G}{c^3}\sum_{a=1}^{2} m_a \int_{-\infty}^{\infty} d \tau \,\frac{1}{r_a}P^i_qa^q_a(t^{\ast})\nonumber\\
&=\frac{G}{c^3}\sum_{a=1}^{2} m_a \Bigg\{\int_{-\infty}^{\infty} d \tau \,\frac{1}{r^3_a}\Big[c^2\tau^2-r^2_a\Big]
P^i_qa^q_a(t^{\ast})\Bigg\}\nonumber\\
&=
  -2\, \frac{G}{c^4}\sum_{a=1}^{2} m_a P^i_qa^q_a(t^{\ast})
+2\,\frac{G}{c^4}\sum_{a=1}^{2}\frac{m_a}{R_a}(\vec{e}_{(0)}\cdot\vec{x}_a(t^{\ast}))^2P^i_qa^q_a(t^{\ast}),
\end{align}
where the first term on the last line vanishes as consequence of the Newtonian centre of mass theorem.
\end{appendix}
\end{widetext}


\bibliographystyle{apsrev.bst}
\end{document}